\documentclass[aps,pra,twocolumn,showpacs,preprintnumbers,amsmath,amssymb]{revtex4-1}
\usepackage{graphicx}
\usepackage[usenames,dvipsnames]{xcolor}
\usepackage[normalem]{ulem}
\usepackage{cancel}
\usepackage{float}
\usepackage{lineno}
\usepackage[caption=false]{subfig}
\newcommand{\Tr}{\textrm{Tr}}

\newcommand{\bra}[1]{\ensuremath{\langle #1 |}}
\newcommand{\ket}[1]{\ensuremath{| #1 \rangle}}
\newcommand{\braket}[2]{\ensuremath{\langle #1 | #2 \rangle}}

\begin{document}
\title{Quantum Joule Expansion of One-dimensional Systems}

\author{Jin Zhang$^{1,2}$}
\author{Y. Meurice$^3$}
\author{S.-W. Tsai$^1$}
\affiliation{$^1$Department of Physics and Astronomy, University of California, Riverside, CA 92521, USA}
\affiliation{$^2$Physics Department, University of Massachusetts, Amherst, Massachusetts 01003, USA}
\affiliation{$^3$Department of Physics and Astronomy, University of Iowa, Iowa City, IA 52242, USA }
\definecolor{burnt}{cmyk}{0.2,0.8,1,0}
\def\lt{\lambda ^t}
\def\note{note}
\def\beq{\begin{equation}}
\def\enq{\end{equation}}

\date{\today}
\begin{abstract}
We investigate the Joule expansion of nonintegrable quantum systems that contain bosons or spinless fermions in one-dimensional lattices. A barrier initially confines the particles to be in half of the system in a thermal state described by the canonical ensemble and is removed at time $t = 0$. We investigate the properties of the time-evolved density matrix, the diagonal ensemble density matrix and the corresponding canonical ensemble density matrix with an effective temperature determined by the total energy conservation using exact diagonalization. The weights for the diagonal ensemble and the canonical ensemble match well for high initial temperatures that correspond to negative effective final temperatures after the expansion. At long times after the barrier is removed, the time-evolved R\'enyi entropy of subsystems bigger than half can equilibrate to the thermal entropy with exponentially small fluctuations. The time-evolved reduced density matrix at long times can be approximated by a thermal density matrix for small subsystems. Few-body observables, like the momentum distribution function, can be approximated by a thermal expectation of the canonical ensemble with strongly suppressed fluctuations. The negative effective temperatures for finite systems go to nonnegative temperatures in the thermodynamic limit for bosons, but is a true thermodynamic effect for fermions, which is confirmed by finite temperature density matrix renormalization group calculations. We propose the Joule expansion as a way to dynamically create negative temperature states for fermion systems with repulsive interactions.
\end{abstract}


\maketitle

\section{Introduction}\label{sec:introduction}

With the remarkable advances in efficient computing algorithms and cold atom experiments, nonequilibrium dynamics has been extensively studied both theoretically and experimentally in recent years. Quantum thermalization is one of the most important topics in this area. In 1929, the classical theory of statistical mechanics was reformulated quantum-mechanically by von Neumann \cite{von2010proof}, which opened the door to the study of quantum thermalization through the unitary dynamics of quantum systems. Later studies \cite{jensen1985,deutsch1991quantum, goldstein2010long} show that quantum thermalization is closely related to  random matrix theory \cite{mehta2004random, guhr1998random} and the eigenstate thermalization hypothesis (ETH) \cite{srednicki1999approach, rigol2012alternatives} was proposed to understand it in finite quantum systems. For nonintegrable quantum systems with Wigner-Dyson level spacing distribution \cite{santos2010onset}, ETH asserts that every eigenstate is thermal and, as a result, few-body observables thermalize \cite{kim2014testing, yoshizawa2018numerical}. ETH has been verified for a variety of nonintegrable quantum systems \cite{rigol2008thermalization, rigol2009quantum, santos2010localization, biroli2010effect, genway2012thermalization, khatami2012quantum, neuenhahn2012thermalization, khatami2013fluctuation, ikeda2013finite, kim2014testing, beugeling2014finite, sorg2014relaxation, beugeling2015off, mondaini2016eigenstate, mondaini2017eigenstate, yoshizawa2018numerical, hikida2018eigenstate}, with  exceptions when little entanglement in the eigenbasis results in equilibration without thermalization\cite{gogolin2011absence}. A modified version of ETH for subdiffusive thermalization \cite{luitz2016anomalous} and strong forms of ETH for reduced density matrix \cite{garrison2018does, dymarsky2018subsystem} have been formulated recently.

The quantum quench protocol is often used in studies of nonequilibrium dynamics, where an initial state of the system is prepared at time $t = 0$,
some parameters are then suddenly or time-dependently changed and the state is let to evolve under the new Hamiltonian. This operation involves a global quench \cite{calabrese2006time, calabrese2007quantum, sotiriadis2009quantum, rigol2009quantum, rigol2011initial, he2012initial, he2013initial, sorg2014relaxation, Mistakidis_2014, PhysRevA.91.033611, cardy2016quantum, PhysRevA.95.053610, PhysRevA.95.013625, Pla_mann_2018, MISTAKIDIS2018106}, local quench \cite{calabrese2007entanglement, eisler2007evolution} or sudden expansion \cite{rigol2004emergence, minguzzi2005exact, rigol2005fermionization, rigol2005free, camalet2008joule, heidrich2009quantum, kajala2011expansion, vidmar2013sudden, ronzheimer2013expansion, sorg2014relaxation, xia2015quantum, campbell2015sudden, vidmar2015dynamical, vidmar2017emergent, vidmar2017emergentpra, xu2017expansion, herbrych2017efficiency, PhysRevA.95.013617, noh2018heating, scherg2018nonequilibrium, PhysRevA.97.053626}.
The initial states can be either pure states or any mixed states.
The dynamics of sudden expansion has rich results including dynamical fermionization in the expansion of a Tonks-Girardeau gas \cite{minguzzi2005exact, vidmar2017emergentpra}, hard-core bosons \cite{rigol2005fermionization, rigol2005free, xu2017expansion} or a Bose gas \cite{campbell2015sudden}, quasi-condensation at finite momentum in hard-core bosons \cite{rigol2004emergence, rigol2005fermionization, rigol2005free, vidmar2013sudden, vidmar2015dynamical, xu2017expansion}, ballistic and diffusive expansion for bosons and fermions in one and two dimensions \cite{kajala2011expansion, ronzheimer2013expansion, vidmar2013sudden}, quantum distillation of singlons and doublons during the expansion \cite{heidrich2009quantum, xia2015quantum, herbrych2017efficiency, scherg2018nonequilibrium}, expansions of atomic clouds and interwell tunneling dynamics of a Bose-Fermi mixture \cite{PhysRevA.97.053626} and self-trapping of bosons initially confined in an harmonic trap \cite{PhysRevA.95.013617}. A time-dependent emergent local Hamiltonian can be constructed to analytically describe these nonequilibrium states as its eigenstates \cite{vidmar2017emergent, vidmar2017emergentpra, xu2017expansion}.

A celebrated experiment in the context of classical statistical mechanics is the Joule expansion. The free expansion of an ideal gas from an initial volume $V$ to a final volume $2V$ does not change the temperature of the gas and the increase in entropy is $nR\ln{2}$. For interacting gases, the temperature decreases for attractive interactions, such as for the van der Waals gas, and increases for repulsive interactions. But what happens for quantum systems? The Joule expansion of an isolated perfect  quantum gas is discussed in \cite{camalet2008joule}, where the time evolution of the particle number density displays periodicity with period proportional to the square of the size of the system, consistent with the recurrence time for free bosons and free fermions obtained in \cite{kaminishi2015recurrence}, and much smaller than the classical Poincar\'e recurrence time, which scales exponentially with the size of the system. For generic quantum systems, the recurrence time \cite{bocchieri1957quantum} is much longer, typically scaling as a double exponential of the system size \cite{kaminishi2015recurrence,mori2018thermalization}.

Sudden expansion of an initially confined thermal state for hard-core bosons was studied in Ref. \cite{xu2017expansion}, where the dynamical fermionization and the quasi-condensation during the expansion are observed. These studies focus on integrable systems where analytic tools can be used. But discussions of the question of thermalization at long times are still missing, which is a prerequisite to answer the Joule question in the quantum regime.

In this article, we discuss the Joule expansion for two interacting but nonintegrable one-dimensional quantum Hamiltonians. We use thermal initial states for a range of temperatures. We discuss the thermalization after a sudden expansion where the particles are kept confined within a length which is twice the initial one. We discuss two important problems  which to the best of our knowledge have not been discussed in the context of sudden expansion in the ETH literature. The first one is to establish if the long-time evolution can be characterized by the diagonal ensemble or the canonical ensemble for a final temperature and to which extent  these ensemble descriptions are compatible.  The second one is the long-time behavior of the R\'enyi entropies for subsystems of various sizes. If the subsystem is large enough, can these entropies stabilize to thermal values? In addition, we will discuss the momentum distributions before and after the expansion as these can be measured experimentally by time-of-flight experiments with cold atoms trapped into one-dimensional optical lattices often called ``tubes".
 
 The paper is arranged as follows. In Section \ref{sec:model} we introduce the boson and spinless fermion models that we study and discuss the time evolution following the removal of the barrier (expansion from size $L/2$ to size $L$), the diagonal and the canonical ensemble descriptions for observables in the equilibrated regime, and the R\'enyi entropy. In Section \ref{sec:wightsensembles}, the weights of the eigenstates for the diagonal ensemble (DE) and the canonical ensemble (CE) descriptions are calculated and compared. We find that, above a certain initial temperature, the final effective temperature obtained from the canonical ensemble description is negative. Results for the von Neumann entropy and the second order R\'enyi entropy are presented in Section \ref{sec:entropy}. In Section \ref{sec:reduceddm} we investigate the full long-time evolved solution, the diagonal ensemble description and the canonical ensemble with an effective final temperature at the level of the reduced density matrices and their eigenvalues and eigenstates in these three cases. In Section \ref{sec:momentum}, we calculate the momentum distribution function, which at long times shows the expected population inversion for the negative effective temperature cases. In Section \ref{sec:effectiveT} we present the answer to the Joule question, that is, the final effective temperature after expansion from $L/2$ to $L$ as function of the temperature of the thermal state at $t=0$. Finite-size analysis shows that effective negative temperatures survive in the thermodynamic limit for the case of spinless fermions, suggesting that Joule expansion can be used as a way to dynamically create negative temperature fermionic states. Finally, Section \ref{sec:conclusion} contains a summary and conclusions.

\section{Model and Methods}\label{sec:model}
We consider two 
models: the one-dimensional Bose-Hubbard model and one-dimensional spinless fermions with nearest-neighbor (NN) and next-nearest-neighbor (NNN) hopping and interaction. Open boundary conditions are used for both models. The Hamiltonian for the Bose Hubbard model is:
\beq \label{eq:BHhamiltonian}
\hat{H}^b = - J \sum_{l=1}^{L-1} (a_l^{\dagger} a_{l+1} + H.c.) + \frac{U}{2} \sum_{l=1}^{L} n^b_l (n^b_l - 1) \ ,
\enq
and the Hamiltonian for spinless fermions is:
\begin{eqnarray} \label{eq:SFhamiltonian}
\nonumber \hat{H}^f &=& - J_1 \sum_{l=1}^{L-1} (c_l^{\dagger} c_{l+1} + H.c.) + V_1 \sum_{l=1}^{L-1} n^f_l n^f_{l+1}\\ && - J_2 \sum_{l=1}^{L-2} (c_l^{\dagger} c_{l+2} + H.c.) + V_2 \sum_{l=1}^{L-2} n^f_l n^f_{l+2} \ ,
\end{eqnarray}
where  $L$ is the total number of sites,  $a^\dagger_l$ ($a_l$) is the creation (annihilation) operator of bosons at site $l$, satisfying the commutation relations $[a_l, a^\dagger_{l^\prime}] = \delta_{l l^\prime}$, and $c^\dagger_l$ ($c_l$) is the creation (annihilation) operator of spinless fermions at site $l$, satisfying the anitcommutation relations $\{c_l, c^\dagger_{l^\prime}\} = \delta_{l l\prime}$. $n^b_l = a^\dagger_l a_l$ ($n^f_l = c^\dagger_l c_l$) is the particle number operators for bosons (fermions) at site $l$. Repulsive interactions $U, V_1, V_2 > 0$ are considered for both systems.
For nonintegrable systems, the energy levels are repulsive and the level spacing distribution is given by the Wigner-Dyson distribution, which has been found for both $\hat{H}^b$ \cite{kollath2010statistical} and $\hat{H}^f$ \cite{santos2010onset}. The fermionic model $\hat{H}^f$ contains next-nearest-neighbor terms $J_2$ and $V_2$ that may seem prohibitively difficult to achieve experimentally. We note, however, that this model can also be realized with a zigzag ladder, with one leg containing the odd sites of Eq. (\ref{eq:SFhamiltonian}) and the other leg containing the even sites. The $J_1$ and $V_1$ in Eq. (\ref{eq:SFhamiltonian}) are terms connecting odd and even sites, and therefore are the inter-leg zig-zag terms in the ladder system. In the ladder system, the $J_2$ and $V_2$ terms then become the intra-leg nearest-neighbor terms. We set $\hbar = k_B = 1$ in the following and fix the hopping energy $J_1 = V_1 = J = 1$ in our calculations. The results we present are for spinless fermions with $J_2 = V_2 = 1$ and bosons with $U = 3$, unless otherwise specified.

In the Joule expansion, a volume of gas is initially prepared with a thermal state inside one half of a container via a partition, with the other half of the container empty. The partition is then suddenly removed and the gas is let to expand freely to occupy the whole volume. For particles on a lattice, a similar initial setup can be achieved by adding a high potential throughout the right half of the lattice, $\hat{H}_W = W \sum_{l=L/2+1}^{L} n_l$, where $W \gg J, U, J_1, J_2, V_1, V_2$. In the numerical calculations presented here, we set $W = 10^6$ to make sure that $\beta W \gg 1$ in all calculations, so particles won't hop to the right half due to high temperatures. At time $t=0$, the high potential is suddenly removed ($W$ is quenched to zero) and particles can then move on a twice bigger lattice. For fermions (bosons), we consider the initial state to be a $N_p = L/4$-particle ($L =20 $ unless otherwise specified) thermal state $\hat{\rho}^f_0$ ($\hat{\rho}^b_0$) at a finite temperature $T^f_0 = 1/\beta^f_0$ ($T^b_0 = 1/\beta^b_0$) in the left half of the lattice.

The general language describing the time evolution of a quantum system is the following. Let $\{\ket{i}\}$ be the eigenstates of the initial Hamiltonian before expansion, $\hat{H}_0 \ket{i} =  e_i \ket{i}$, where $\hat{H}_0 = \hat{H}^{b(f)} + \hat{H}_{W}$ for bosons (fermions). The initial mixed thermal state $\hat{\rho}_0$ with inverse temperature $\beta_0$ in this eigenbasis is given by
\beq \label{eq:initrho}
\hat{\rho}^0 = \sum_{i} \rho^0_{i} \ket{i}\bra{i}
\enq
where $\rho^0_{i} = e^{-\beta_0 e_i}/Z_0(\beta_0)$, and $Z_0(\beta_0) = \Tr{(e^{-\beta_0 \hat{H}_0})}$ is the partition function.

Let $\{\ket{m}\}$ be the eigenstates of the final Hamiltonian $\hat{H}_F$ that drives the time evolution after quenching (removing) the high wall, that is, $\hat{H}_{F} \ket{m} = E_{m} \ket{m}$, with $\hat{H}_{F} = \hat{H}^{b(f)}$ for bosons (fermions). The initial state, written in this eigenbasis is given by,
\begin{eqnarray}
\label{eq:initdenop}
\nonumber \hat{\rho}^0 &=& \sum_{m, n} \rho^0_{m n} \ket{m} \bra{n} \\ &=& \sum_{m, n} \left( \sum_{i} \rho^0_{i} \braket{m}{i} \braket{i}{n} \right) \ket{m} \bra{n} \ ,
\end{eqnarray}
and contains diagonal ($m=n$) and off-diagonal ($m\ne n$) terms. The time-dependent density operator at time $t$ during the expansion is
\begin{eqnarray}
\label{eq:tevoldenop}
\hat{\rho}^t = \sum_{m, n} \rho^0_{m n} \exp[-i t (E_{m} - E_{n})] \ket{m} \bra{n}.
\end{eqnarray}
If there is no degeneracy in the spectrum, which is generally true for nonintegrable systems \cite{bohigas1984characterization}, the long-time average of the density operator above will give the diagonal ensemble (DE) \cite{rigol2008thermalization} density matrix
\begin{eqnarray} \label{eq:diagen}
\nonumber \hat{\bar{\rho}} = \hat{\rho}^d &=&  \lim_{\tau \rightarrow \infty}  \frac{1}{\tau}  \int_{0}^{\tau} d\tau \hat{\rho}^t \\ \nonumber &=& \sum_{m} \rho^0_{mm} \ket{m}\bra{m} \\ &=& \sum_{m} W^d_m \ket{m}\bra{m}.
\end{eqnarray}
where $\rho^0_{m m}$ is the weight $W^d_m$ that the projector $\hat{P}_m = \ket{m}\bra{m}$ has in the DE.

For an arbitrary observable $\hat{O} = O_{m^\prime n^\prime} \ket{m^\prime} \bra{n^\prime}$, the time-dependent expectation value is,
\begin{eqnarray}
\label{eq:tevolexpectation}
\nonumber \langle \hat{O} \rangle_t &=& Tr(\hat{O} \hat{\rho}^t) \\
\nonumber &=& \sum_{m,n} O_{mn} \rho^0_{nm} \exp[-i t (E_n - E_m)] \\
&=& \langle \hat{O} \rangle_d + \sum_{m \neq n} O_{mn} \rho^0_{nm} \exp[-i t (E_n - E_m)],
\end{eqnarray}
where the first term $\langle \hat{O} \rangle_d = \sum_{m} O_{mm} \rho^0_{mm}$ is the expectation value in the diagonal ensemble, or the long-time average value. The second term, for nonintegrable systems, is expected to be very small at long times for two reasons: firstly, according to ETH, the off-diagonal matrix elements for a few-body observable $\hat{O}$ is exponentially small in the size of the system \cite{deutsch1991quantum,srednicki1999approach}, secondly the long time temporal dephasing \cite{rigol2008thermalization} guarantees the canceling of oscillations at long enough times. It seems problematic at first that one may need to wait an exponentially long time for the cancellation due to dephasing to occur because of the exponentially small level spacings. It was however pointed out  \cite{dalessio2016} that since the off-diagonal matrix elements of $\hat{O}$ are exponentially small, the phase coherence between only a finite fraction of the eigenstates with a significant contribution to the expectation value needs to be destroyed. If the observable $\hat{O}$ thermalizes, the expectation in the diagonal ensemble must agree with the expectation value obtained in statistical mechanics. We use the canonical ensemble (CE) to describe this thermalized state. We find the effective inverse temperature $\beta_{eff}$ of this CE by matching the total energy of the system, which is a time-independent conserved quantity,
\begin{eqnarray}
\label{eq:CETbyenergy}
\nonumber \langle \hat{H}_{F} \rangle &=& Tr(\hat{H}_{F} \hat{\rho}^t) \\
\nonumber &=& \sum_{m} E_{m} W^d_m \\
\label{eq:tevolexpE} &=& Tr(\hat{H}_{F} \exp[-\beta_{eff} \hat{H}_{F}] / Z(\beta_{eff}))
\end{eqnarray}
where $Z(\beta_{eff}) = \Tr{(e^{-\beta_{eff} \hat{H}_F})}$ is the partition function for the CE at temperature $T_{eff} = 1/\beta_{eff}$. The function $E(\beta) = \Tr{ ( \hat{H} e^{-\beta \hat{H}} ) } / Z(\beta)$ is monotonically decreasing with $\beta$ because
\begin{eqnarray}
\label{eq:dE/dbeta}
 \frac{dE(\beta)}{d\beta} = \sum_{m, n} -\frac{e^{-\beta(E_n + E_m)} (E_n - E_m)^2}{2 Z^2} < 0,
 \end{eqnarray}
and therefore there is only one solution for the effective temperature. For the CE we then have 
\begin{eqnarray}
\hat{\rho}^c = \sum_m \frac{e^{-\beta_{eff}E_m}}{Z(\beta_{eff})} \ket{m}\bra{m} = \sum_m W^c_m \ket{m}\bra{m}.
\end{eqnarray}
Like in a classical gas, the average energy of the initial Hamiltonian $H_0$ (particles confined to $L/2$) with {\it repulsive} interactions at a certain temperature is higher than that of the final  Hamiltonian $H_F$ (system size $L$) with the same temperature. So a higher effective temperature is needed for Eq. (\ref{eq:CETbyenergy}) to be satisfied. Conversely, for attractive interactions (such as in the classical van der Waals gas), the system is expected to cool down upon expansion.
For repulsive interactions, such as the systems studied here, an interesting possibility can occur.
If the initial average energy is higher than the infinite-temperature energy for the final Hamiltonian, a negative temperature is needed to compensate for the difference. It is convenient to describe this in terms of inverse temperature: the system is prepared in some positive inverse temperature that decreases upon expansion for repulsive interactions. There will be some initial positive inverse temperature for which it decreases to zero (infinite temperature) after expansion. Any positive initial inverse temperature that is smaller than that will then lead to even smaller final inverse temperatures, leading to negative values.

We have the initial thermal state $\hat{\rho}^0$, the time-evolved state $\hat{\rho}^t$, the DE density matrix $\hat{\rho}^d$ and the corresponding CE density matrix $\hat{\rho}^c$. To better compare these states, we investigate their reduced density matrices. Note that for an arbitrary density operator $\hat{\rho}$, if the observable $\hat{O}$ only resides on a subsystem $A$ with a size $l_A$, whose complement is denoted as $B$ with a size $l_B$, the expectation value only depends on the reduced density operator of the subsystem $\hat{\rho}_A = \Tr_B{\hat{\rho}}$ by
\begin{eqnarray}
\label{eq:expecreduced}
\langle \hat{O} \rangle = \Tr_A{\left( \hat{O} \hat{\rho}_A \right)}.
\end{eqnarray}
If the reduced density matrix is similar to the CE reduced density matrix, every observable that resides in subsystem $A$ has the same expectation value as in the canonical ensemble.

Another characterization of the reduced density matrix is the R\'enyi entropy. The $p$-th order bipartite R\'enyi entropy for $\hat{\rho}_A$ is defined as
\begin{eqnarray}
\label{eq:renyi}
S_{p A} = \frac{1}{1-p} \ln{\Tr_A{\left(\hat{\rho}_A\right)^p}}.
\end{eqnarray}
The von Neumann entanglement entropy $S_{1A} = - \Tr{\left(\hat{\rho}_A \ln{\hat{\rho}_A}\right)}$ is obtained by taking the limit $p \rightarrow 1^+$. For pure states, the R\'enyi entropy probes the entanglement between subsystems $A$ and $B$, and $S_{pA} = S_{pB}$. The conformal field theory (CFT) prediction of R\'enyi entropy for the ground state of critical one-dimensional systems is discussed in \cite{holzhey1994geometric, vidal2003entanglement, jin2004quantum, calabrese2004entanglement, calabrese2009entanglement}. For excited states of chaotic many-body Hamiltonians that have finite energy density in the sense that $\lim_{V \rightarrow \infty} (E_n-E_0)/V \neq 0$ \cite{garrison2018does}, Ref. \cite{lu2017renyi} argues that the von Neumann entanglement entropy is proportional to the subsystem size for $l_A / L < 1/2$, while $S_{pA}$ $(p>1)$ are convex functions of the subsystem size, only linear at $\beta=0$, more convex at larger $|\beta|$ and larger than those of thermal mixed states at the same energy density. This means that $\Tr_A{\left(\hat{\rho}_A\right)^p}$ is exponentially small with the subsystem size. The states we are dealing with here are mixed states, and generally $S_{pA} \neq S_{pB}$, unless the partition is in the middle. For a thermal state at temperature $\beta^{-1}$, the R\'enyi entropy for subsystem $A$ contains contributions from both entanglement and thermal mixture. Refs.~\cite{korepin2004universality, calabrese2004entanglement, calabrese2009entanglement} discuss the crossover of R\'enyi entropy from the logarithmic scaling ($l_A / \beta \ll 1$) to the volume law ($l_A / \beta \gg 1$) using finite temperature CFT. This finite temperature 
behavior can place an additional challenge in the preparation of ground states of cold atom systems in optical lattices \cite{unmuth2017probing}. If $A$ equals the whole system, $S_{pA}$ is the diagonal (thermal) R\'enyi entropy for the DE (CE). In particular, for the time-evolved density matrix $\hat{\rho}^t$, taking the second order R\'enyi entropy as an example, it can be expressed by
\begin{eqnarray}
\label{eq:reduceds2}
\nonumber S^t_{2A} = && - \ln (
\sum_{\{m,n,i\}}
\rho^0_{mn}\rho^0_{m^\prime n^\prime} e^{-it(E_{m} - E_{n} + E_{m^\prime} - E_{n^\prime})} \\&& \braket{n}{i_B} \braket{i_B^\prime}{m^\prime} \braket{n^\prime}{i_B^\prime} \braket{i_B}{m} ) ,
\end{eqnarray}
where
$\sum_{\{m,n,i\}}$ is a sum over $m$, $m^\prime$, $n$, $n^\prime$, $i_B$, and $i^\prime_B$, and where $i_B$ and $i_B^\prime$ run over all the states of a basis for subsystem $B$. If $A$ is the whole system (size $L$), then $B$ has size zero, and $n=m^\prime$ and $n^\prime=m$, and the phase is therefore always zero. The same argument holds for all orders of the R\'enyi entropy. The R\'enyi entropy for the whole system is therefore conserved during the time evolution.

\section{Weights in Ensembles}\label{sec:wightsensembles}
\begin{figure}[ht]
  \centering
    \includegraphics[width=0.5\textwidth]{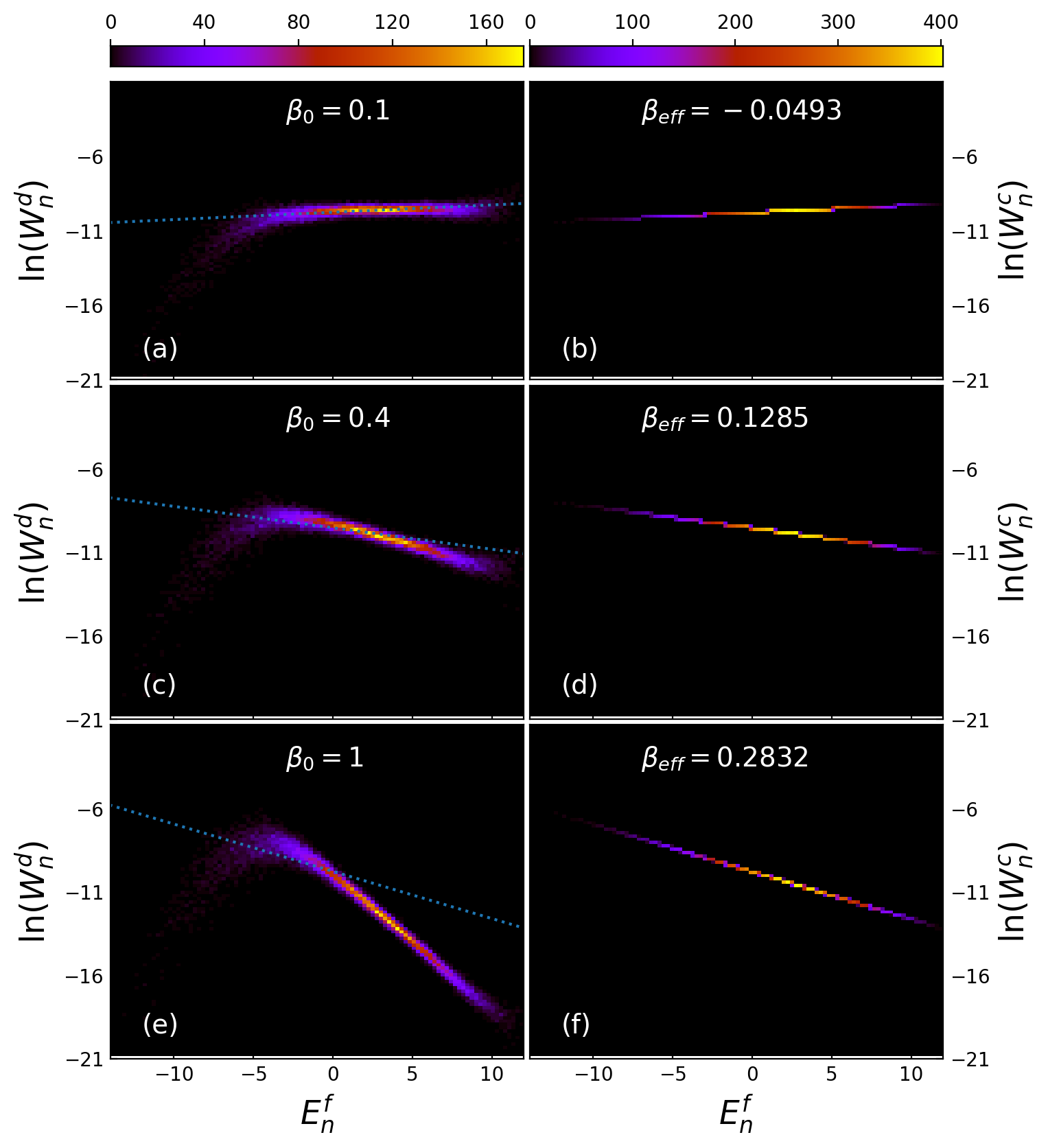}
    \caption{\label{fig:spinlessfdece} (Color online) Two-dimensional histograms for the weights of eigenstates in the DE, $W^d_n$ (a,c,e) and those in the corresponding CE, $W^c_n$ (b,d,f). Results are for spinless fermions with initial inverse temperatures $\beta_0 = 0.1$ (a), $\beta_0 = 0.4$ (c) and $\beta_0 = 1$ (e). The corresponding CE with effective inverse temperatures are shown in (b,d,f). The color scale represents the number of states per unit area. The dashed lines in the left column are translations of the lines from the right column, indicating the slopes in the right column.
    }
  \end{figure} 
\begin{figure}[t!]
  \centering
    \includegraphics[width=0.5\textwidth]{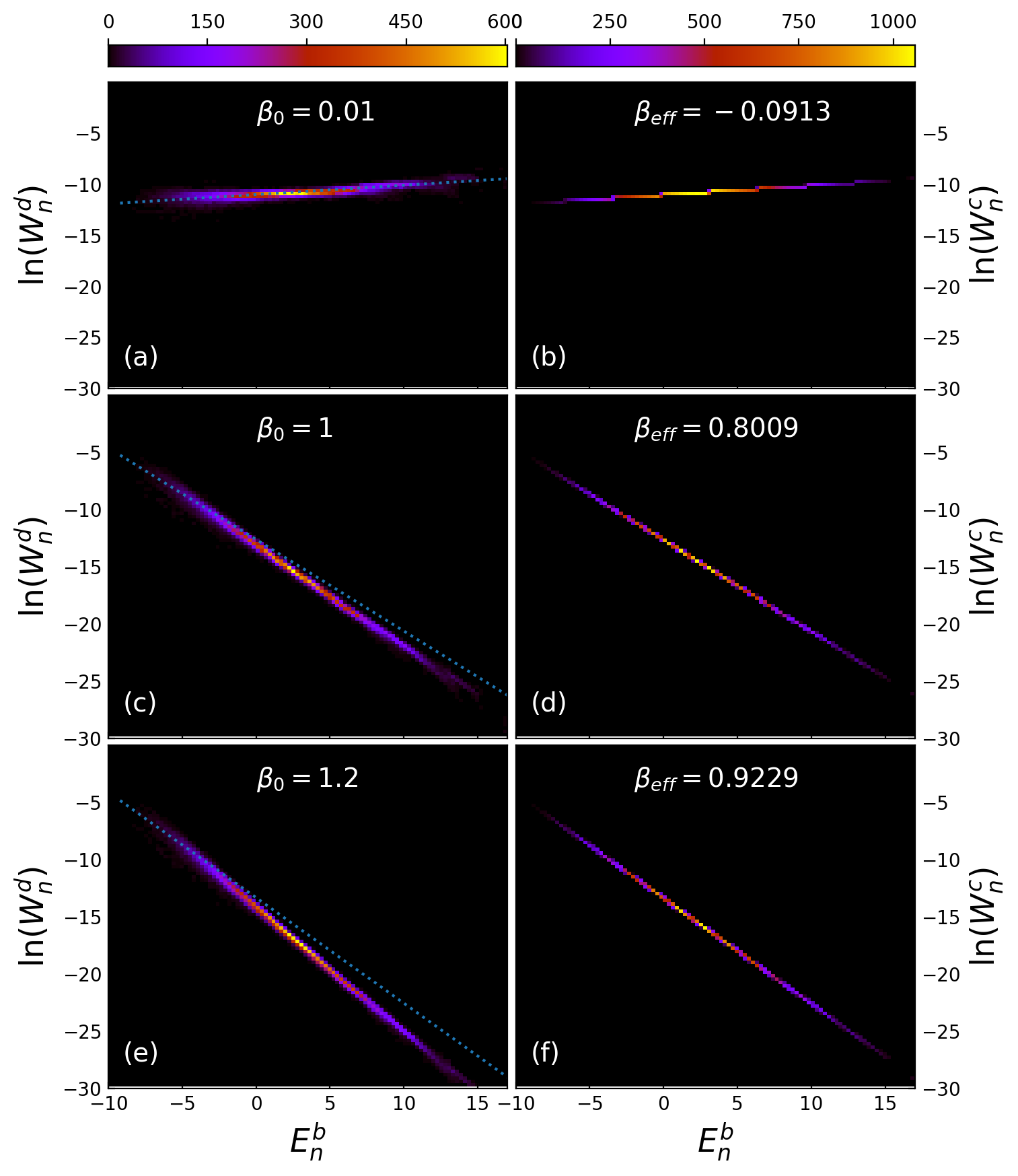}
    \caption{\label{fig:bosondece} (Color online) Same as Fig. \ref{fig:spinlessfdece}, but for bosons with 20 sites, 5 particles and $U = 3$.
    }
  \end{figure} 
In global quench protocols, the energy uncertainty $\Delta E / \bar{E}$, which only depends on the DE, is algebraically small with the system size ($1/\sqrt{L}$) for pure initial states \cite{rigol2008thermalization}, which is also numerically verified for our case. To show this explicitly, writing the Hamiltonian after quench as $\hat{H}_F = \hat{H}_0 - \hat{H}_W$ and following the derivation in Ref.~\cite{rigol2008thermalization}, we calculate the energy width of the diagonal ensemble, which is equal to the variance of energy in the initial thermal state,
\begin{eqnarray}
\label{eq:energywidth}
\nonumber \Delta E &=& \sqrt{\Tr{(\hat{H}^2_F \hat{\rho}_0)} - [\Tr{(\hat{H}_F \hat{\rho}_0)}]^2} \\ \nonumber &=& \bigg\{\Tr{(\hat{H}^2_0 \hat{\rho}_0)} - [\Tr{(\hat{H}_0 \hat{\rho}_0)}]^2 - 2\Tr{[\hat{H}_W (\hat{H}_0 - \bar{E})\hat{\rho}_0]} \\ &+& \Tr{(\hat{H}_W^2 \hat{\rho}_0)} - [\Tr{(\hat{H}_W \hat{\rho}_0)}]^2 \bigg\}^{\frac{1}{2}}.
\end{eqnarray}
If $\hat{H}_W$ is a sum of local operators with finite matrix elements, and in the absence of long-range correlations, $\Delta E \sim \sqrt{L}$ still holds using the same argument in \cite{rigol2008thermalization}. In particular, in our case, $\hat{H}_W$ is a large chemical potential on the right half of the lattice, and the particles are confined in the left half in the initial state, so the last three traces are zero in Eq.~\eqref{eq:energywidth}. So $\Delta E$ equals the variance of the initial energy before quench in a thermal state, which  obviously scales as $\sqrt{L}$.
The small energy window $[\bar{E}-\Delta E, \bar{E}+\Delta E]$ defines a microcanonical ensemble which is equivalent to the CE in the thermodynamic limit. As our initial states already have the Boltzmann weights, it is interesting to explore and compare the  weights in the whole spectrum in the DE, $W^d_n$, and those in the corresponding CE, $W^c_n$, and see if these spectra will match by changing the initial temperature. Similar comparisons have been conducted in studies of global quantum quenches from pure states \cite{sorg2014relaxation} and thermal states \cite{zhang2011quantum, he2012initial}.

In Fig. \ref{fig:spinlessfdece}, we show the 2-d histograms for natural logarithms of these weights in the DE [Figs. \ref{fig:spinlessfdece}(a), \ref{fig:spinlessfdece}(c) and \ref{fig:spinlessfdece}(e)] and the CE [Figs. \ref{fig:spinlessfdece}(b), \ref{fig:spinlessfdece}(d) and \ref{fig:spinlessfdece}(f)]. We choose $100 \times 100$ bins, and the color scale on each bin represents the number of states inside it. The results are for spinless fermions expanding from thermal states with three different initial temperatures. We see that in the DE, the weights lie in a narrow band for all temperatures. There is a tail bending down for the spectrum at low energies, but with a much smaller density of states. The weights in the CE are all straight lines as should be the case by definition. We have included these as dashed lines in the left column of Fig. \ref{fig:spinlessfdece} to compare the slopes. At low initial temperature, e.g. $\beta_0 = 1$, the slopes of the narrow straight band in the DE and the line for the CE are very different. As we increase the initial temperature, they match better and better. It can be understood in the following way. Note that the summation in Eq. (\ref{eq:CETbyenergy}), $\sum_{m} E_{m} W^d_m$, can be divided into two parts: one is the summation in the bending tail of the weights in the DE which is close to the bottom of the energy spectrum, the other one is the summation in the remaining narrow straight band. Although the tail has small density of states, it can still dominate the summation at very low initial temperatures, because the weight in the DE, $W^d_m = \sum_i \rho^0_{i} \braket{m}{i} \braket{i}{m}$, contains the Boltzmann weight $\rho^0_{i} = e^{-\beta_0 e_i}/Z_0(\beta_0)$ given by the initial temperature. As the tail is bending down, the line of the CE must rotate anticlockwise from the straight narrow band of the DE to fit the data in it at low initial temperatures [see Fig. \ref{fig:spinlessfdece}(e)]. For higher initial temperatures, the narrow straight band dominates the summation, and Eq. (\ref{eq:CETbyenergy}) is effectively a linear fit for this band. Note that for the highest initial temperature $\beta_0=0.1$, the effective temperature is negative, and the weights in the DE and the CE match perfectly in the densest part of the spectrum [see Fig. \ref{fig:spinlessfdece}(a)]. This match between the DE and the CE weights results in agreements for expectations of almost all equithermal operators \cite{garrison2018does} in the two ensembles and the model has strong thermalization \cite{banuls2011strong} in the negative temperature regime.

The results for bosons are depicted in Fig. \ref{fig:bosondece}. The weights in the DE also lie in a straight narrow band. The tail in the low energy spectrum only bends down slightly. So the difference in slopes for the DE and the CE is much smaller than in the case of fermions. We still see that the higher the temperature, the better is the match between the DE and the CE. The slopes also agree accurately at very high initial temperature $\beta_0 = 0.01$. Note that a much higher initial temperature is needed for the effective temperature to be negative. We will see that the negative effective temperature actually does not exist in the thermodynamic limit for bosons, but it does for fermions. This is discussed in Section \ref{sec:effectiveT}.

\section{Entropy For Ensembles}\label{sec:entropy}

In this section we investigate the entropy profiles for the initial state, the state after long-time expansion, the DE density matrix and the CE density matrix. We calculate the quantity in Eq. (\ref{eq:renyi}) for $n = 1$ (the von Neumann entropy) and $n = 2$ (the second order R\'enyi entropy), and subsystems containing left $l_A = 1, 2, ..., 20$ sites. We choose the time evolved density matrix to be at a long time after expansion, $tJ = 6000$, where thermalization has occured for a long time. Numeric results show the equilibration of entropy after $tJ = 50$ for all cases considered here. We compare the results for different temperatures for both spinless fermions and bosons.

\subsection{Von Neumann Entropy}\label{sec:vnentropy}

\begin{figure}[t!]
  \centering
    \includegraphics[width=0.5\textwidth]{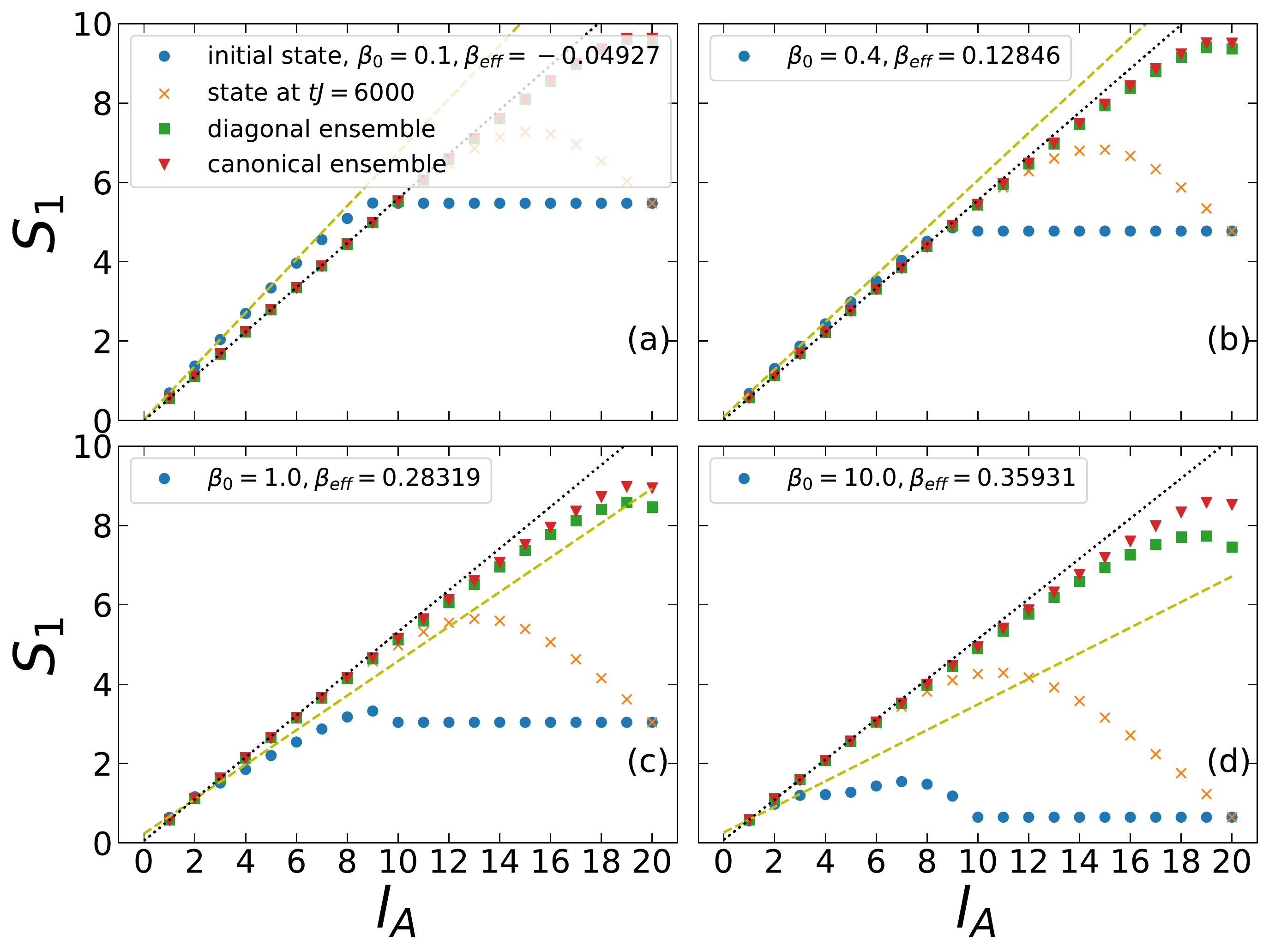}
    \caption{\label{fig:splessfententropy4bs} (Color online)  The von Neumann entropy of the reduced density matrix containing left $l_{A}$ sites for the initial thermal state (blue circle), the state at long time $tJ=6000$ (orange cross), the diagonal ensemble (green square) and the canonical ensemble (red inverted triangle). Results are for spinless fermions with initial inverse temperatures $\beta_0 =0.1$ (a), $\beta_0 =0.4$ (b), $\beta_0 =1$ (c) and $\beta_0 =10$ (d). The green dashed lines are linear fits of the first 3 subsystems for the initial states. The black dotted lines are linear fits of the first 3 subsystems for the canonical ensemble density matrices.
    }
  \end{figure} 

\begin{figure}[t!]
  \centering
    \includegraphics[width=0.50\textwidth]{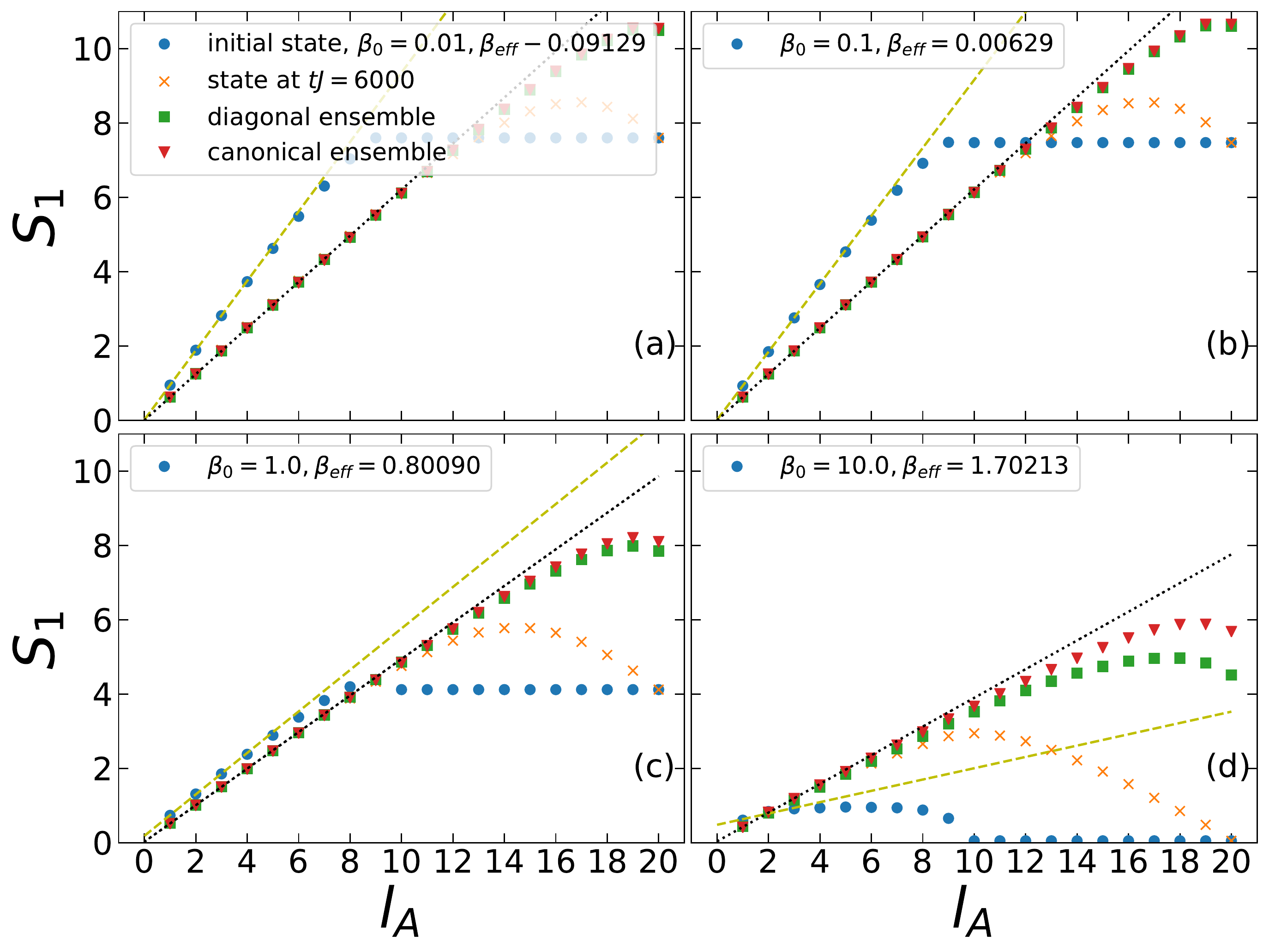}
    \caption{\label{fig:bhententropy4bs} (Color online) Same as Fig. \ref{fig:splessfententropy4bs} but for bosons with initial inverse temperatures $\beta_0 =0.01$ (a), $\beta_0 =0.1$ (b), $\beta_0 =1$ (c) and $\beta_0 =10$ (d).
    }
  \end{figure} 
  
\begin{figure}[t!]
  \centering
    \includegraphics[width=0.50\textwidth]{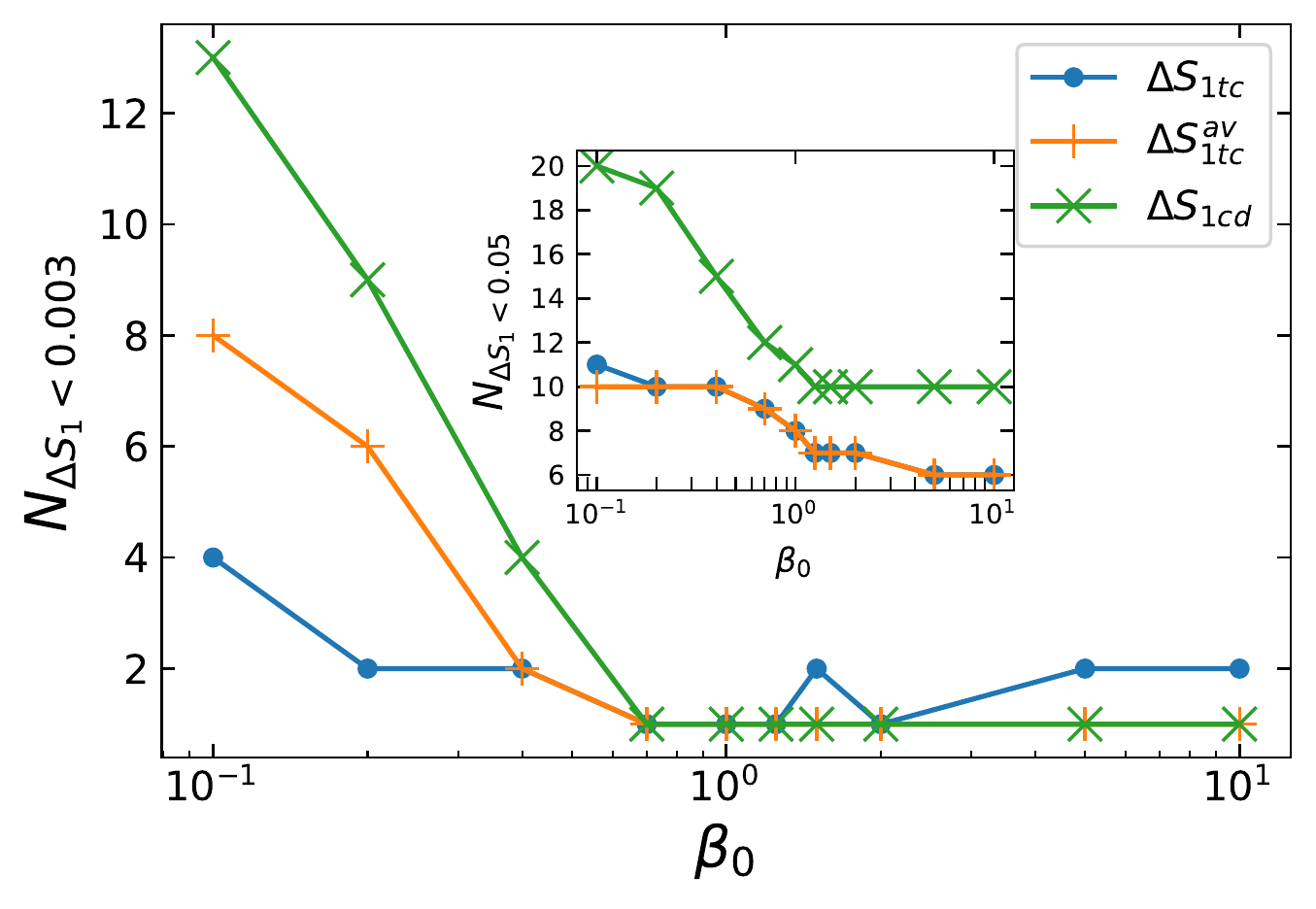}
    \caption{\label{fig:ndeltaeevsbeta} (Color online) The maximal size of the subsystem in which the difference between $S_1$ for the time-evolved state at $tJ = 6000$, or the average $S_1$ for time-evolved states at $tJ = 81, 82, ..., 200$, or $S_1$ for the DE and $S_1$ for the CE is smaller than $0.003$, as a function of initial inverse temperature $\beta_0$. The results are for spinless fermions. The inset shows the results for $\Delta S_1 < 0.05$.
    }
  \end{figure} 

Fig. \ref{fig:splessfententropy4bs} shows the von Neumann entropy of the reduced density matrix for the subsystem containing left $l_A$ sites. The results are for spinless fermions expanding from thermal states with four different initial temperatures. For initial thermal states, the particles are initially confined in the left half part, so the R\'enyi entropy is identical for all subsystems bigger than half. We can clearly see from Fig. \ref{fig:splessfententropy4bs}(a) the volume law in $S_1$ for up to six subsystems at high temperature $\beta_0 = 0.1$. In Fig. \ref{fig:splessfententropy4bs}(b) a lower temperature $\beta_0=0.4$ results in five subsystems satisfying the volume law. As we decrease the initial temperature further, there are only three subsystems satisfying volume law in Fig. \ref{fig:splessfententropy4bs}(c) for $\beta_0=1$ and no sign of volume law in Fig. \ref{fig:splessfententropy4bs}(d) for $\beta_0=10$. These results show the crossover from the volume law for subsystems at high temperatures to the logarithmic scaling in the ground state. The CE density matrices all show volume law for many subsystems. Because the effective temperature is always high even for very low initial temperature (see Fig. \ref{fig:splessfententropy4bs}(d)). Still, the higher the effective temperature is, the more subsystems satisfying the volume law. The von Neumann entropy increases linearly up to 11 sites in Fig. \ref{fig:splessfententropy4bs}(a) where the negative effective temperature should be considered to be ``larger" than the infinite temperature. And it increases linearly up to 6 sites in Fig. \ref{fig:splessfententropy4bs}(d). Intuitively, as the R\'enyi entropy has contributions from both entanglement and thermal mixture, the former increases at first with increasing subsystem size, but decreases for subsystem sizes bigger than half of the whole system. So it is reasonable that the entropy bends down from the linearly increasing line. For a detailed study of properties of thermal R\'enyi entropy, see \cite{bonnes2013entropy}.

In Fig. \ref{fig:spinlessfdece} we have shown that the weights in the DE are very similar to the weights in the CE for high initial temperatures. So the diagonal entropy and thermal entropy should be very close. This is confirmed in Fig. \ref{fig:splessfententropy4bs}(a), (b) and Fig. \ref{fig:splessfre4bs}(a), (b), where the R\'enyi entropy of reduced density matrices for $\hat{\rho}_d$ and $\hat{\rho}_c$ agrees for almost all subsystems. At lower temperatures, on one hand, the deviation for the full systems is consistent with the results in Fig. \ref{fig:spinlessfdece}(e)(f) where it shows that the distribution of weights is very different for the DE and the CE. On the other hand, $S_{1A}$ still has no visible difference for subsystems bigger than half even if the distribution of weights is very different for the whole system. This can be seen in the inset of Fig.~\ref{fig:ndeltaeevsbeta}, where the maximal size of the subsystem in which the von Neumann entropy difference between the DE and the CE, $\Delta S_{1cd} = |S_1^c - S_1^d|$, is smaller than $0.05$ as a function of $\beta_0$ is plotted. For $0.2 < \beta_0 < 1.25$, roughly we have $N_{\Delta S_1 < 0.05} \sim \log_{10}{T_0}$. In the main plot of Fig.~\ref{fig:ndeltaeevsbeta}, we plot $N_{\Delta S_1 < 0.003}$ as a function of $\beta_0$, where $N_{\Delta S_1 < 0.003} \sim \log_{10}{T_0}$ for $0.1 < \beta_0 < 0.7$. Note that it can be proved that $S_1$ of the DE for the whole system only has non-extensive difference from that of the microcanonical ensemble, no matter whether the initial state is pure or mixed \cite{dalessio2016}. And this non-extensive difference can be decreased by increasing the initial temperature in our case.

Finally, we check the reduced von Neumann entropy for the time-evolved density matrix in Fig. \ref{fig:splessfententropy4bs}. We see that at the high temperature $\beta_0=0.1$, the time-evolved density matrix has the same $S_1$ as the CE and the DE density matrices for subsystems up to eleven sites. And ten-, eight- and six-site subsystems have no visible difference in von Neumann entropy for temperatures $\beta_0=0.4$, $\beta_0=1$ and $\beta_0=10$, respectively. The maximal size of the subsystem in which the difference between $S_1$ for $\hat{\rho}_t$ at $tJ = 6000$ and $S_1$ for the CE, $\Delta S_{1tc} = |S_1^t - S_1^c|$, is smaller than $0.003$ ($0.05$) as a function of $\beta_0$ is plotted in the main (inset) figure of Fig.~\ref{fig:ndeltaeevsbeta}. The results for $\Delta S^{av}_{1tc} = |\bar{S_1^t} - S_1^c|$ with average von Neumann entropy $\bar{S_1^t}$ over $tJ = 81, 82, ..., 200$, are also plotted. The plots for $\Delta S_{1tc}$ and $\Delta S^{av}_{1tc}$ are almost the same in the inset figure, indicating that the time fluctuations are small enough that the instantaneous entropy agrees with the time-averaged entropy at the precision of $0.05$. It can also be seen that roughly $N_{\Delta S_1 < 0.05} \sim \log_{10}{T_0}$ for $0.4 < \beta_0 < 5$. Nevertheless, the plots for $\Delta S_{1tc}$ and $\Delta S^{av}_{1tc}$ in the main figure are different because $0.003$ is smaller than the fluctuations of $S_1^t$. In this case $N_{\Delta S^{av}_{1tc} < 0.003} \sim \log_{10}{T_0}$ only for $0.1 < \beta_0 < 0.7$. It is concluded that when the initial temperature is higher, the maximum subsystem size that has the same von Neumann entropy in $\hat{\rho}_t$ and $\hat{\rho}_d$ as that in $\hat{\rho}_c$ is bigger and is expected to scale as $\log_{10}{T_0}$. And it is always smaller for $\hat{\rho}_t$ than for $\hat{\rho}_d$ because $\hat{\rho}_t$ preserves the initial entropy. Note that the reduced $S_1$ increases first to the highest point and then bends down until the full von Neumann entropy goes back to the initial full von Neumann entropy. This is because the full von Neumann entropy is invariant under the unitary change of basis, which is the unitary time evolution here.

The results for bosons depicted in Fig. \ref{fig:bhententropy4bs} are similar to those for spinless fermions. They are summarized as follows. For initial thermal state, there are six, six, five and three sites having volume law von Neumann entropy at initial inverse temperature $\beta_0=0.01$ [Fig. \ref{fig:bhententropy4bs}(a)], $\beta_0=0.1$ [Fig. \ref{fig:bhententropy4bs}(b)], $\beta_0=1$ [Fig. \ref{fig:bhententropy4bs}(c)] and $\beta_0=10$ [Fig. \ref{fig:bhententropy4bs}(d)] respectively. And for the canonical ensemble density matrix, the von Neumann entropy in the left eleven, eleven, ten and seven sites show volume law at these four initial temperatures. The DE von Neumann entropy is identical to the CE von Neumann entropy for all subsystems at initial inverse temperatures $\beta_0=0.01, 0.1$. They are equal in the left fifteen sites and left eight sites at initial inverse temperatures $\beta_0=1$ and $\beta_0=10$, respectively. These results are also consistent with those in Fig. \ref{fig:bosondece}. The time-evolved density matrix at $tJ=6000$ has the same von Neumann entropy with  both $\hat{\rho}^d$ and $\hat{\rho}^c$ in the left eleven, eleven, nine and six sites in the four graphs respectively.

\subsection{The second order R\'enyi entropy}\label{sec:s2entropy}
\begin{figure}[t!]
  \centering
    \includegraphics[width=0.5\textwidth]{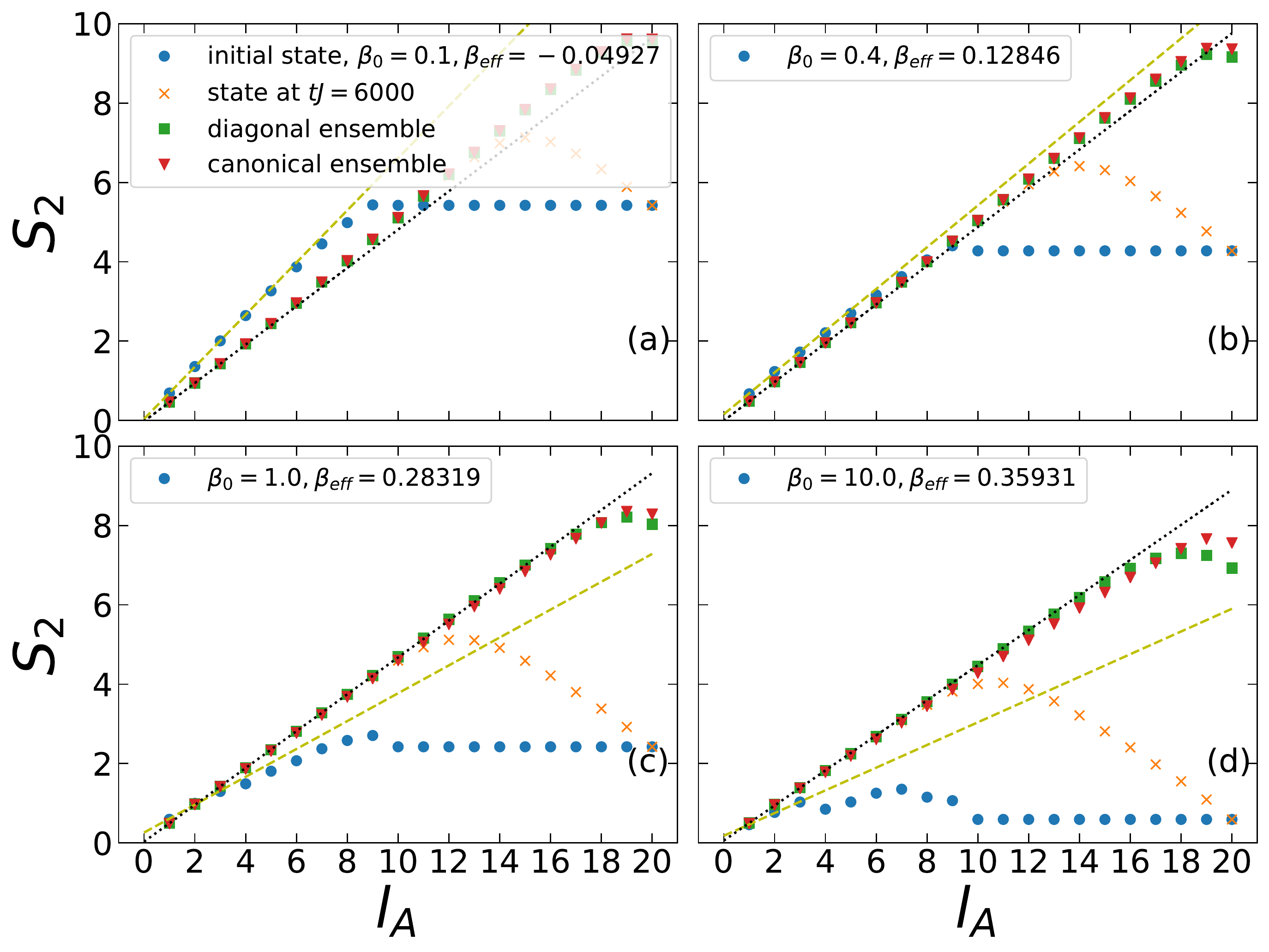}
    \caption{\label{fig:splessfre4bs} (Color online) Same as Fig. \ref{fig:splessfententropy4bs}, but for the second order R\'enyi entropy.
    }
  \end{figure} 

Since the results for the two models are very similar, we only consider  spinless fermions in this section. As shown in Fig. \ref{fig:splessfre4bs}, the second order R\'enyi entropy also has high agreement for the CE and the DE, that is, nineteen, eighteen, eleven and eight sites have the same value for $\beta_0=0.1, 0.4, 1, 10$, respectively. And the time-evolved state has twelve, eleven, ten and eight sites with $S^t_{2A}$ equal to them. The main difference from Fig. \ref{fig:splessfententropy4bs} and Fig. \ref{fig:bhententropy4bs} is that, unlike the von Neumann entropy where the higher the initial temperature is, the bigger the subsystems having linearly increasing entropy are, there are five, six, twelve (sixteen for the DE) and six sites having linearly increasing $S^{d(c)}_{2A}$, respectively. We find that the entropy bends up for high initial temperatures [the data of the DE and the CE lie on or above the black dotted linear fit in Fig. \ref{fig:splessfre4bs}(a)]. This convexity is weakened by decreasing the initial temperature and finally the entropy becomes concave again [the data of the DE and the CE lie on or below the black dotted linear fit in Fig. \ref{fig:splessfre4bs}(d)]. As mentioned in Ref. \cite{lu2017renyi}, $S_{pA}$ $(p>1)$ are convex functions of the subsystem size, which is verified in our models with twenty sites and quarter filling, but just in a small energy window around the infinite temperature eigenstate. Moreover, more obvious convexity is observed in high energy levels near the top of the spectrum while the levels near the bottom of the spectrum are all concave, which explains why $S^{c(d)}_{2A}$ with negative effective temperature is more convex in Fig. \ref{fig:splessfre4bs} even if $|\beta_{eff}|$ is smaller. Note that at low temperatures, the second order R\'enyi entropy of the DE can be bigger than that of the CE for big subsystems, while the von Neumann entropy of the DE is always no greater than that of the CE.

As the second order R\'enyi entropy can be measured experimentally by a beam-splitter operation implemented via a controlled tunneling operation between the two copies of a many-body state \cite{daley2012measuring, islam2015measuring}, we investigate the details of time evolution of $S^t_{2A}$ in this section. Note that for twenty sites and quarter filling, the numerical results show small time fluctuations in R\'enyi entropy for all subsystems at long times, even for the lowest temperature considered here $\beta_0 = 10$, which indicates that the time-dependent terms are cancelled out due to dephasing. In Eq. (\ref{eq:reduceds2}), the only time-independent terms are those with $m=n, m^\prime=n^\prime$ or $m=n^\prime, m^\prime=n$. So for long times, 
\begin{eqnarray}
\nonumber && S^t_{2A} \approx \\ \nonumber&&- \ln (\sum_{m, m^\prime,i_B,i_B^\prime} \rho^0_{m m}\rho^0_{m^\prime m^\prime} \braket{m}{i_B} \braket{i_B^\prime}{m^\prime} \braket{m^\prime}{i_B^\prime} \braket{i_B}{m} \\ \nonumber &&+ \sum_{m \neq m^\prime,i_B,i_B^\prime} \rho^0_{m m^\prime}\rho^0_{m^\prime m} \braket{m^\prime}{i_B} \braket{i_B^\prime}{m^\prime} \braket{m}{i_B^\prime} \braket{i_B}{m} ) \\ \nonumber &&= - \ln (\sum_{m, m^\prime} \rho^0_{m m}\rho^0_{m^\prime m^\prime} T_{AB}^{m m^\prime} \\&&+ \sum_{m \neq m^\prime} \rho^0_{m m^\prime}\rho^0_{m^\prime m} T_{BA}^{m^\prime m} )
\end{eqnarray}
where $T_{AB}^{m m^\prime} = \Tr_A{\left[(\Tr_B{\ket{m}\bra{m}})(\Tr_B{\ket{m^\prime}\bra{m^\prime}})\right]}$ and $T_{BA}^{m^\prime m} = \Tr_B{\left[(\Tr_A{\ket{m^\prime}\bra{m^\prime}})(\Tr_A{\ket{m}\bra{m}})\right]}$. The second summation contributes to the difference between the equilibrated $S_{2A}$ and the $S_{2A}$ for the DE. If it is negligible compared to the first summation, the time-evolved R\'enyi entropy equilibrates to the value for the DE which agrees with the value for the CE. For small subsystem sizes $l_A \ll l_B$, consider any $m \neq m^\prime$, we expect $T_{AB}^{m m^\prime} \gg T_{BA}^{m^\prime m}$ and $S^t_{2A} \approx S^d_{2A}$ because the overlap of $\Tr_B(\ket{m}\bra{m})$ for two $m$ values is much larger than the corresponding overlap for $\Tr_A(\ket{m}\bra{m})$. Furthermore, as $\sum_{m, m^\prime} \rho^0_{m m}\rho^0_{m^\prime m^\prime} = 1$, the first summation is equal to some typical value of $T_{AB}^{m m^\prime}$, denoted as $T_{AB}^{o}$, while the second summation can be written as
\begin{eqnarray}
\label{eq:s2difference}
T_{BA}^{o} \sum_{m \neq m^\prime} \rho^0_{m m^\prime}\rho^0_{m^\prime m} = T_{BA}^{o} \left( e^{-S_2^{\beta_0}} - e^{-S_2^d} \right)
\end{eqnarray}
where $T_{BA}^{o}$ is some typical value of $T_{BA}^{m^\prime m}$, $S_2^{\beta_0}$ is the second order R\'enyi entropy for the initial state, $S_2^d$ is the second order R\'enyi entropy for the DE. $S_2^d$ is extensive with $L$ because it is close to $S_2^c$ (see Fig. \ref{fig:splessfre4bs}), while $S_2^{\beta_0}$ is extensive with $L/2$. Because the average entropy per site for quarter filling is bigger than that for half filling and the effective temperature is higher than the initial temperature after the expansion, the second term in Eq. (\ref{eq:s2difference}) is negligible for large $L$. So as long as the initial state is thermal, Eq. (\ref{eq:s2difference}) is at least exponentially small with $L/2$. 

\begin{figure}[t!]
  \centering
    \includegraphics[width=0.50\textwidth]{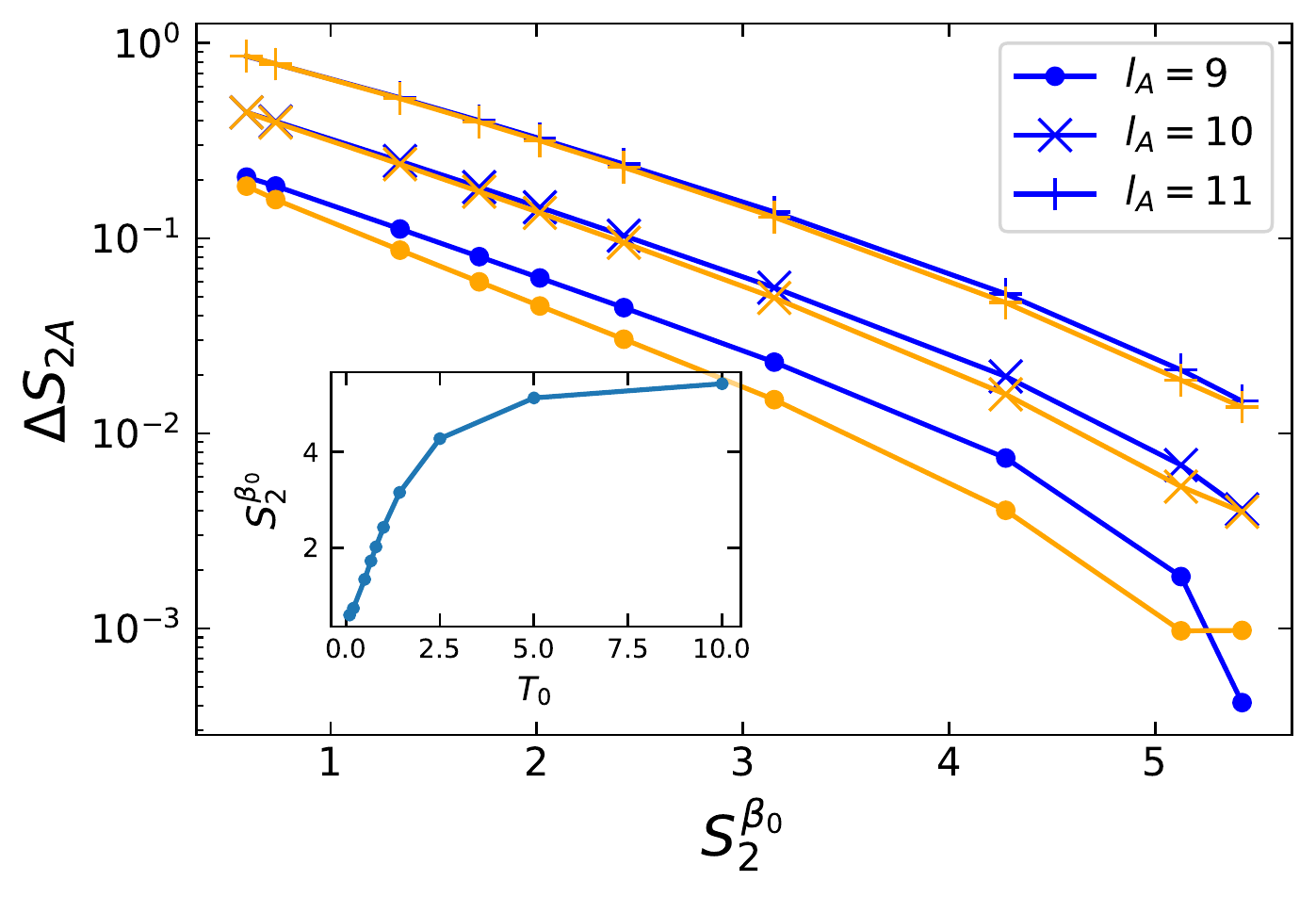}
    \caption{\label{fig:sf20s5pdREvsInitRE} (Color online) The difference between time-averaged R\'enyi entropy $\bar{S}_{2A}$ and the DE R\'enyi entropy $S_{2A}^d$ as a function of initial second order R\'enyi entropy. The results are for spinless fermions, three subsystems with $l_A = 9, 10, 11$. The inset shows the initial second order R\'enyi entropy as a function of initial temperature $T_0$. The yellow markers depict the results for $tJ=6000$.
    }
  \end{figure}

\begin{figure}[t!]
  \centering
    \includegraphics[width=0.50\textwidth]{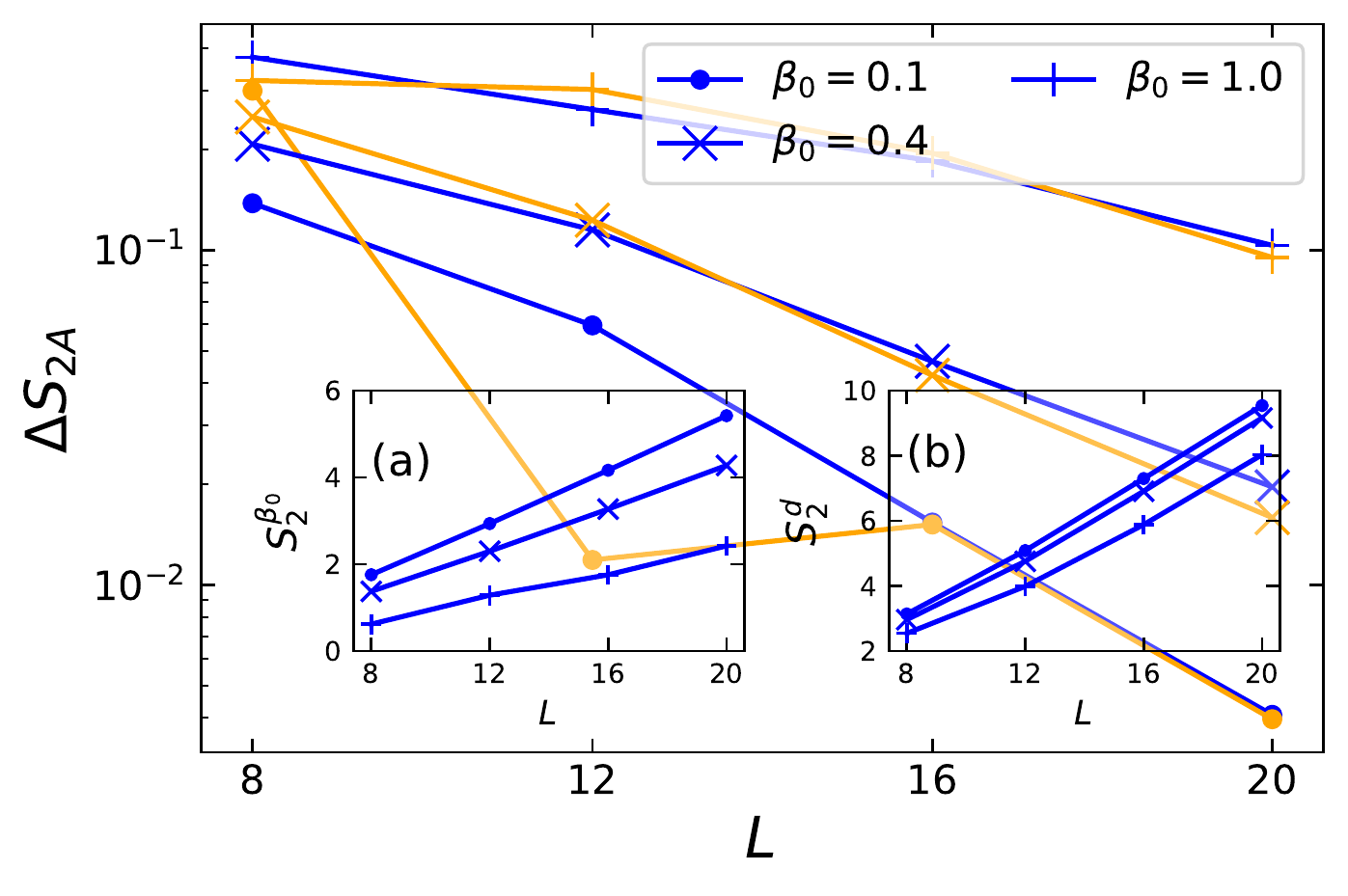}
    \caption{\label{fig:sffssdREvsL} (Color online) The difference between time-averaged R\'enyi entropy $\bar{S}_{2A}$ and the DE R\'enyi entropy $S_{2A}^d$ for $l_A/L=1/2$ as a function of the system size. The results are for spinless fermions, three initial inverse temperatures $\beta_0 = 0.1, 0.4, 1.0$. The insets show the initial second order R\'enyi entropy (a) and the diagonal second order R\'enyi entropy (b) as a function of the system size. The yellow markers depict the results for $tJ=6000$.
    }
  \end{figure}

Note that the dependence of $T_{AB}^{m m^\prime}$ and $T_{BA}^{m^\prime m}$ on the initial temperature and the system size also contributes to the difference between the time-evolved R\'enyi entropy and the DE R\'enyi entropy. If Eq. (\ref{eq:s2difference}) is small, 
\begin{eqnarray}
\label{eq:s2tapproxs2d}
S^t_{2A} \approx S^d_{2A} - T^o_{BA} \  e^{S^d_{2A}-S_2^{\beta_0}}
\end{eqnarray}
where $T^o_{BA} < \Tr_B\left[\left(\Tr_A \ket{m_o}\bra{m_o} \right)^2 \right]$ is also exponentially small with the subsystem size for $\ket{m_o}$ being a finite-energy density eigenstate of a chaotic system \cite{lu2017renyi}.
We plot Figs. \ref{fig:sf20s5pdREvsInitRE} and  \ref{fig:sffssdREvsL} to quantitatively investigate the second term in Eq. (\ref{eq:s2tapproxs2d}). To plot Fig. \ref{fig:sf20s5pdREvsInitRE}, we calculate the time evolution of $S_{2A}^t$ for subsystems containing the left $9, 10, 11$ sites respectively to $tJ=200$, with time step size $\tau J = 1$. To avoid large deviations due to the fluctuations in time and the short-time far-from-equilibrium states, we average $S_{2A}^t$ using only the last $120$ steps of the time evolution data. In Fig. \ref{fig:sf20s5pdREvsInitRE}, we plot the difference between $S_{2A}^d$ for the DE and the time-averaged value $\Delta S_{2A} = S_{2A}^d - \bar{S}_{2A}$ as a function of the second order R\'enyi entropy for the initial thermal state $S^{\beta_0}_{2}$. Varying the number of long-time data points in the time-average  calculation does not change the plot. The exponential decay of $\Delta S_{2A}$ with the initial thermal entropy can be clearly seen from these numerical results. The decay rates are almost the same for three subsystems at least for $\beta_0 > 0.2$. The results at $tJ=6000$ are also plotted, which are very close to the results for time-averaged entropy. The inset of Fig. \ref{fig:sf20s5pdREvsInitRE} shows $S^{\beta_0}_2$ as a function of initial temperature $T_0$, where we see $S^{\beta_0}_2$ is firstly linear with $T_0$ at low initial temperature and saturates at high initial temperature. So $\Delta S_{2A}$ also exponentially decays with the initial temperature when $T_0 < 1$. In Fig. \ref{fig:sffssdREvsL}, the same quantity for $l_A / L = 1/2$ is plotted as a function of the system size, where we see that $\Delta S_{2A}$ is exponentially small with the system size. The decay rate is higher for higher initial temperatures. The inset (a) confirms that $S_{2A}^{\beta_0}$ is proportional to the size of the system. So $\Delta S_{2A}$ also exponentially decays with $S_{2A}^{\beta_0}$ across different system sizes. The inset (b) shows the linearity of $S^d_2$ with the system size, where the weak convexity is due to the increase of effective temperature for larger system sizes. The two insets also confirm that $e^{-S_2^d}$ is negligible in Eq. \ref{eq:s2difference}, at least for $L \ge 12$ where the data of $S^d_{2}$ is more linear with $L$. The results at $tJ=6000$ are not close to the time-averaged ones due to the large time fluctuations for small systems.
  
\begin{figure}[t!]
  \centering
    \includegraphics[width=0.50\textwidth]{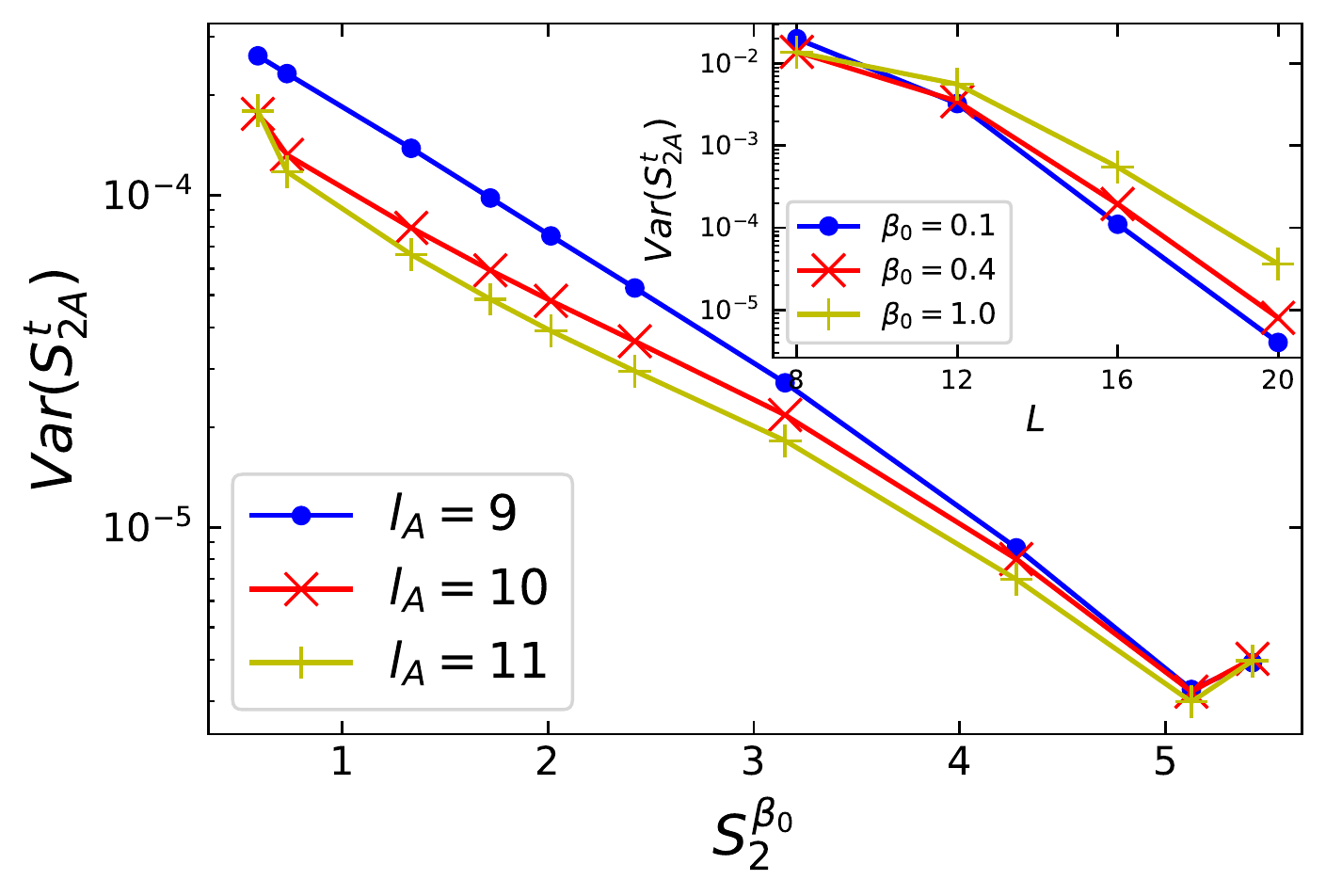}
    \caption{\label{fig:sffssflucvsL} (Color online) Variance of $S^t_{2A}$ in time for $tJ=81, 82, ..., 200$, as a function of $S^{\beta_0}_2$. The results are for spinless fermions, subsystems with $l_A = 9, 10, 11$. The inset shows the same quantity with $l_A = 10$ but as a function of system size at temperatures $\beta_0 = 0.1, 0.4, 1.0$.
    }
  \end{figure}

Finally, we investigate the time fluctuations of the R\'enyi entropy. We calculate the variance of $S^t_{2A}$ in time, 
\begin{eqnarray}
\nonumber Var(S^t_{2A}) &=& \lim_{\tau \rightarrow \infty}  \frac{1}{\tau}  \int_{0}^{\tau} d\tau \left( S^{\tau}_{2A} - \bar{S}_{2A} \right) ^2 \\ &\approx& \frac{1}{N_{\tau}} \sum_{i=1}^{N_{\tau}} \left( S^{\tau_i}_{2A} - \bar{S}_{2A} \right) ^2 ,
\end{eqnarray}
where we take $N_\tau = 120$ and $\{\tau_i\} = \{81, 82, ..., 200\}$. The variance as a function of $S^{\beta_0}_2$ is depicted in Fig. \ref{fig:sffssflucvsL}, where we see that the fluctuations also exponentially decay with $S^{\beta_0}_2$, at least for not very high temperatures ($\beta_0 = 0.1$ for the largest $S^{\beta_0}_2$ in the figure). The inset also shows that the variance is exponentially small with the system size, with a bigger decay rate for higher temperatures.
  
\section{Reduced Density Matrices}\label{sec:reduceddm}

\begin{figure}[t!]
  \centering
    \includegraphics[width=0.48\textwidth]{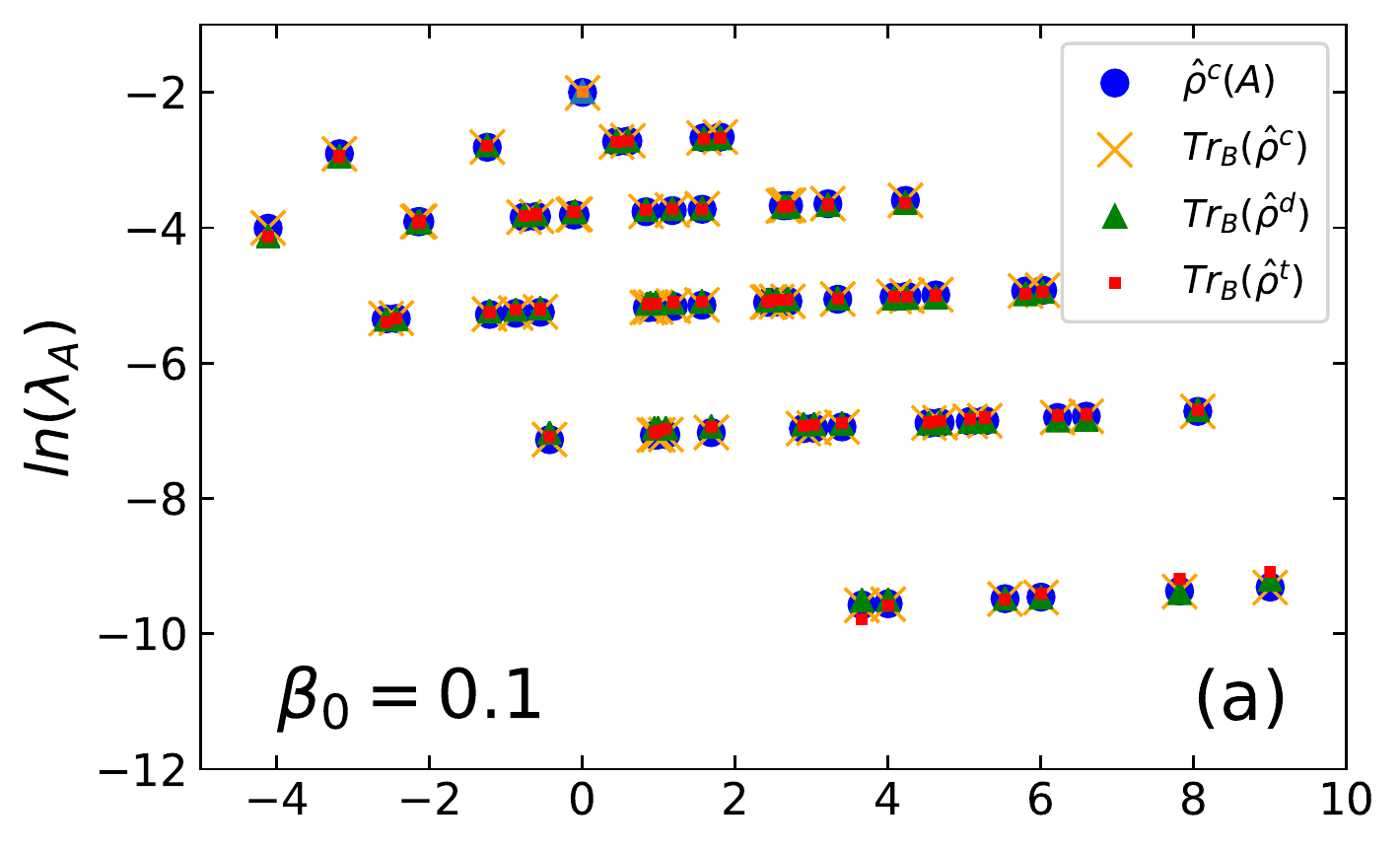}
    \includegraphics[width=0.48\textwidth]{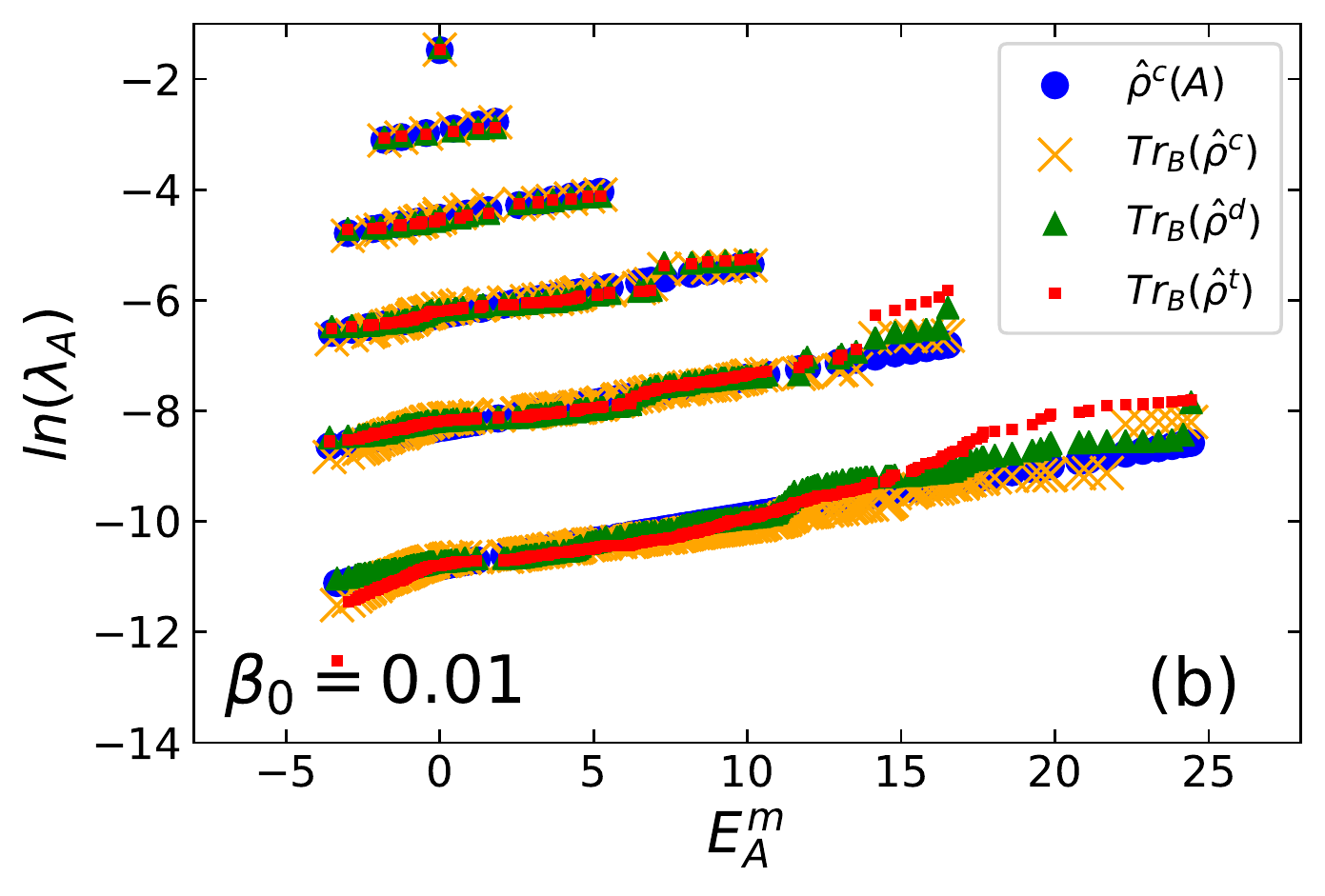}
    \caption{\label{fig:splfmreducedrho6} (Color online) Natural logarithms of eigenvalues of reduced density matrices for the canonical ensemble, the diagonal ensemble and the time-evolved state at $tJ=6000$, compared with the weights in the canonical ensemble constructed by the Hamiltonian for the subsystem $\hat{H}_A$. The subsystem contains the left six sites of the lattice. The results are for spinless fermions with the initial temperature $\beta_0 = 0.1$ (a) and bosons with the initial temperature $\beta_0 = 0.01$ (b). In each subplot, from top to bottom, the linearly arranged points represent zero, one, two, three, four and five particle states respectively.
    }
  \end{figure} 

\begin{figure}[t!]
  \centering
    \includegraphics[width=0.48\textwidth]{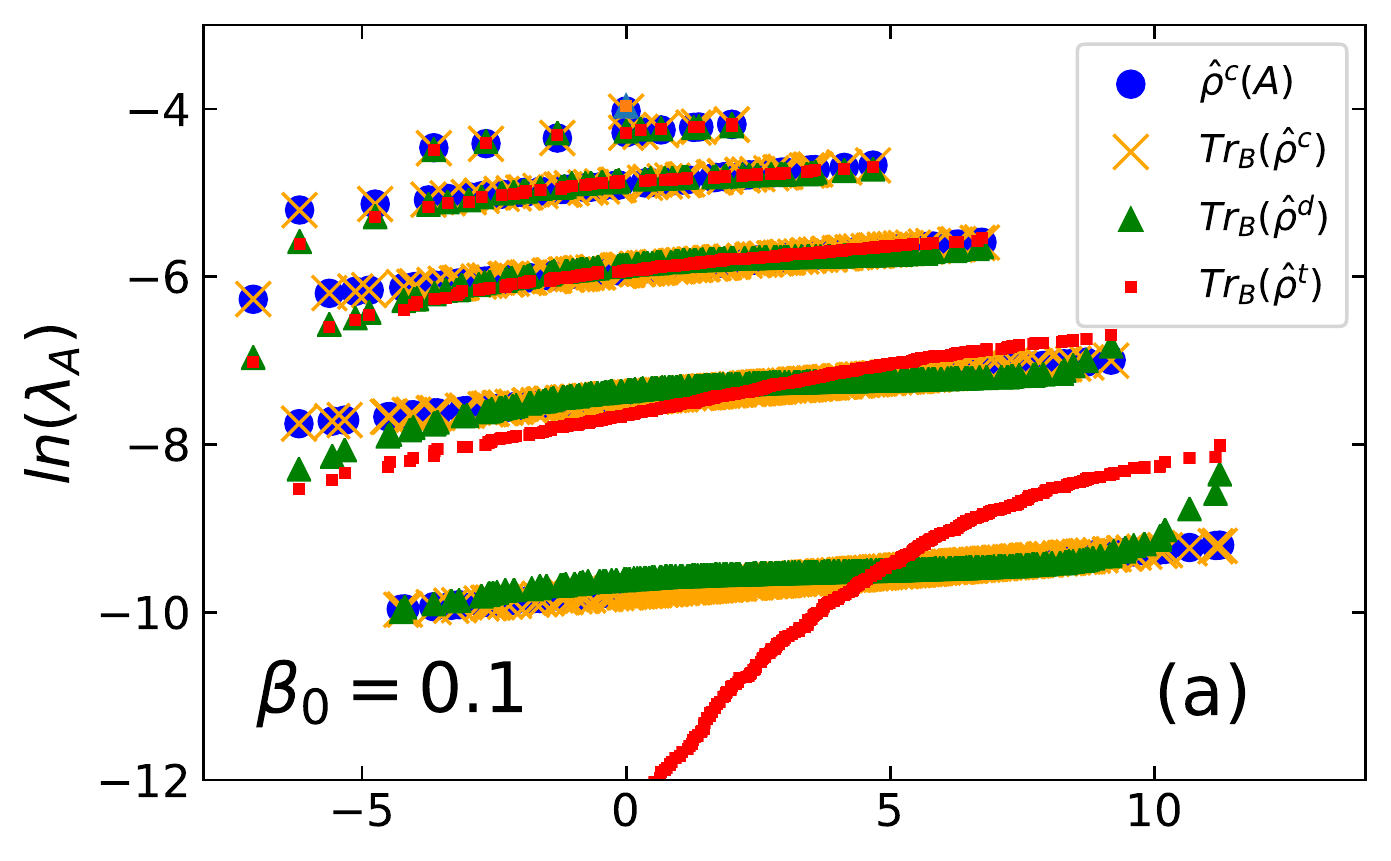}
    \includegraphics[width=0.48\textwidth]{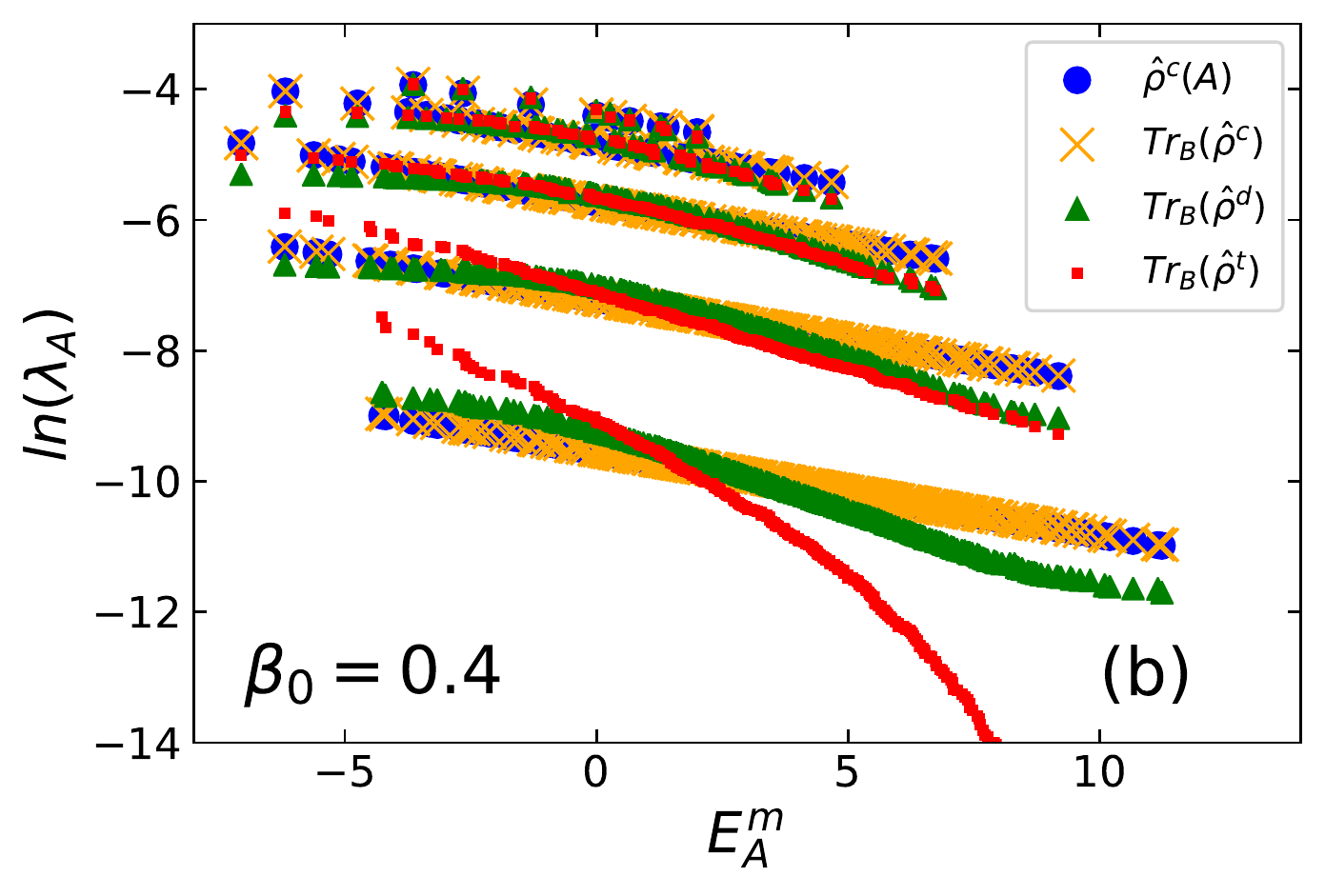}
    \caption{\label{fig:splfmreducedrho} (Color online) Same as Fig. \ref{fig:splfmreducedrho6} but only for spinless fermions. The subsystem contains the left ten sites of the lattice. The initial temperatures are $\beta_0 = 0.1$ (a) and $\beta_0 = 0.4$ (b). Note that in (b), the point for the zero-particle state is very close to the one-particle states line.
    }
  \end{figure} 
  
\begin{figure}[t!]
  \centering
    \includegraphics[width=0.48\textwidth]{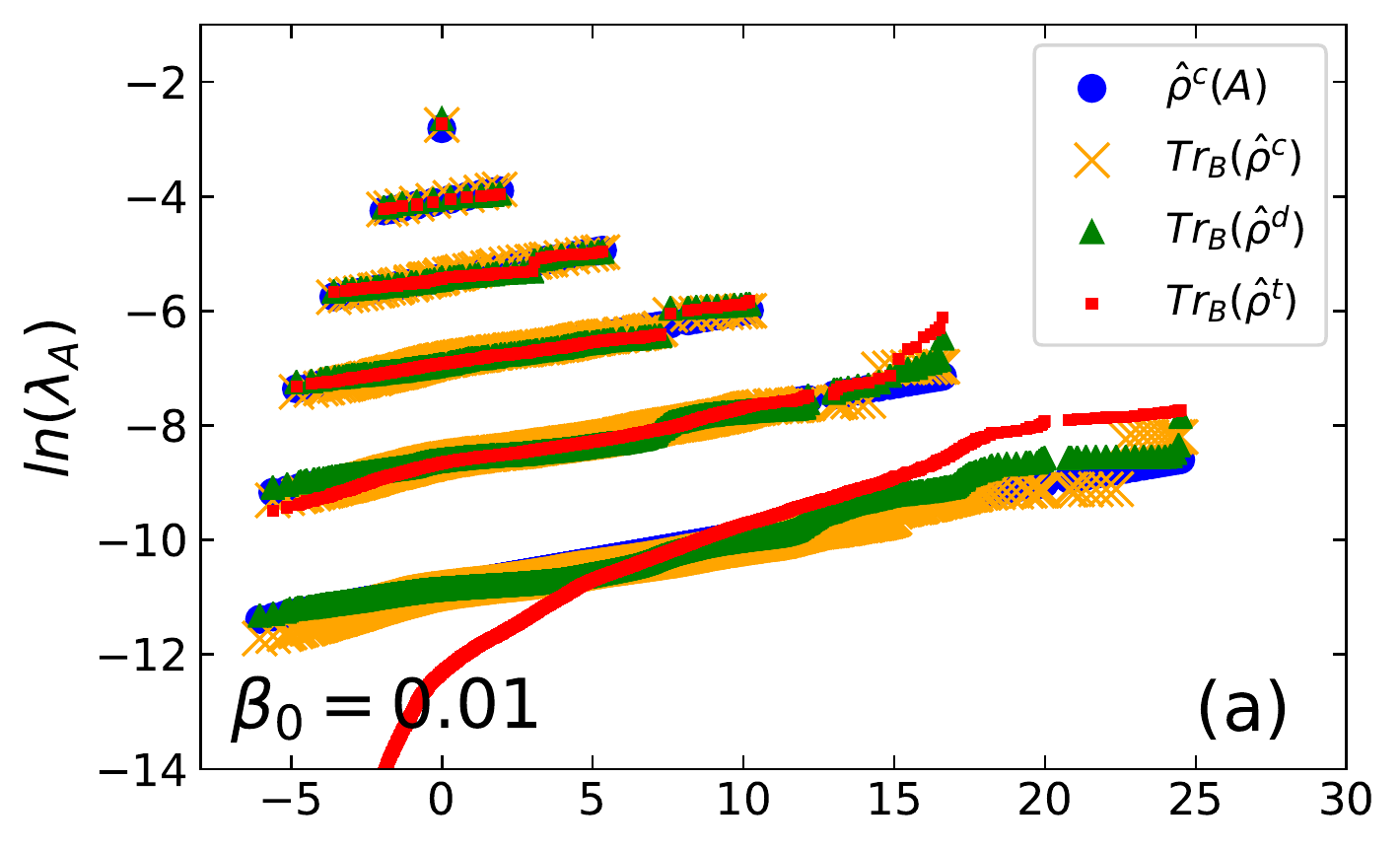}
    \includegraphics[width=0.48\textwidth]{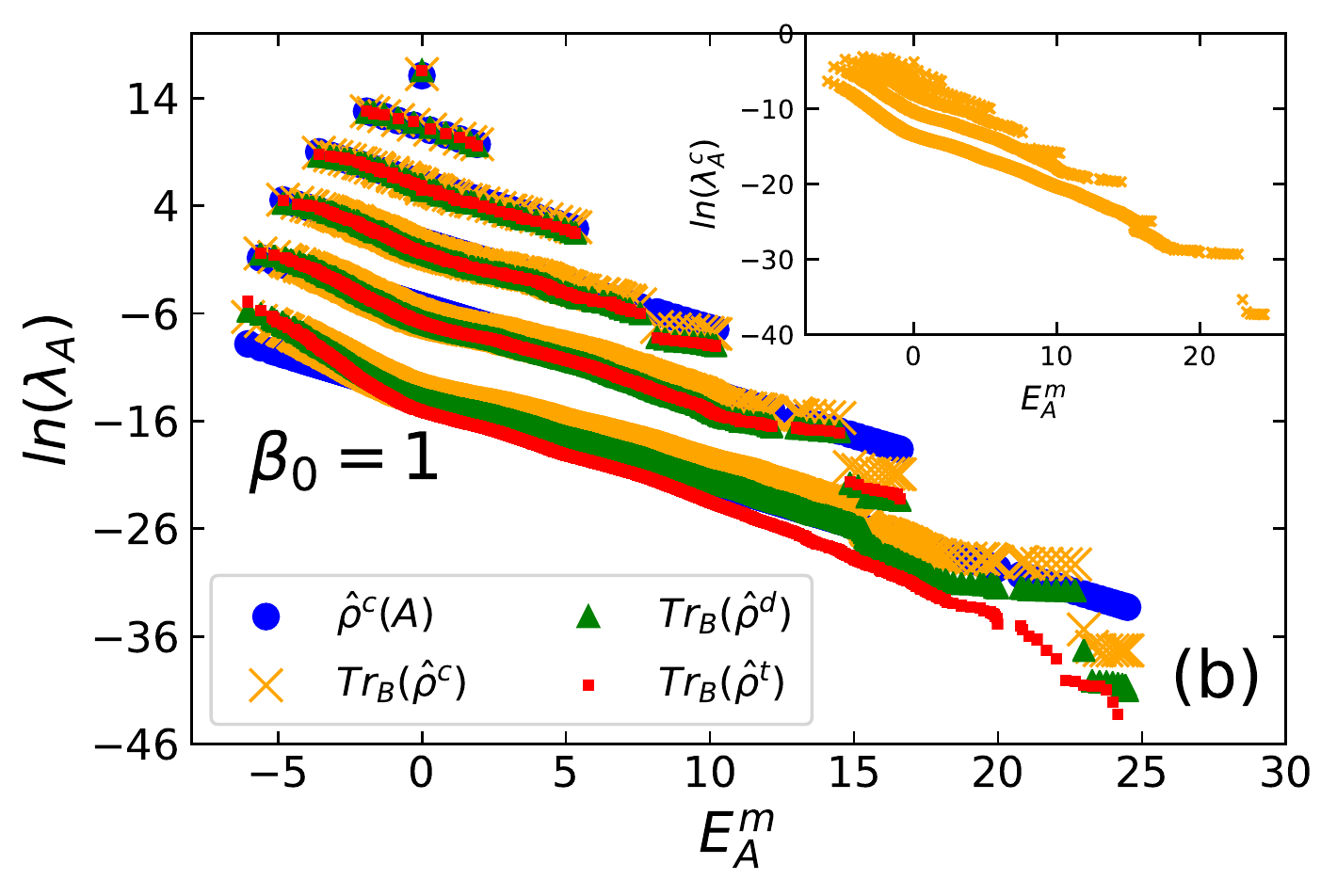}
    \caption{\label{fig:bhreducedrho} (Color online) Same as Fig. \ref{fig:splfmreducedrho}, but for bosons with initial temperatures $\beta_0=0.01$ (a), $\beta_0=1$ (b). In (b), those points for zero, one, two, three and four particle states are shifted up by constants for better view. The inset shows the actual eigenvalues for the reduced density matrix of canonical ensemble without shift.
    }
  \end{figure} 
  
\begin{figure}[t!]
  \centering
    \includegraphics[width=0.5\textwidth]{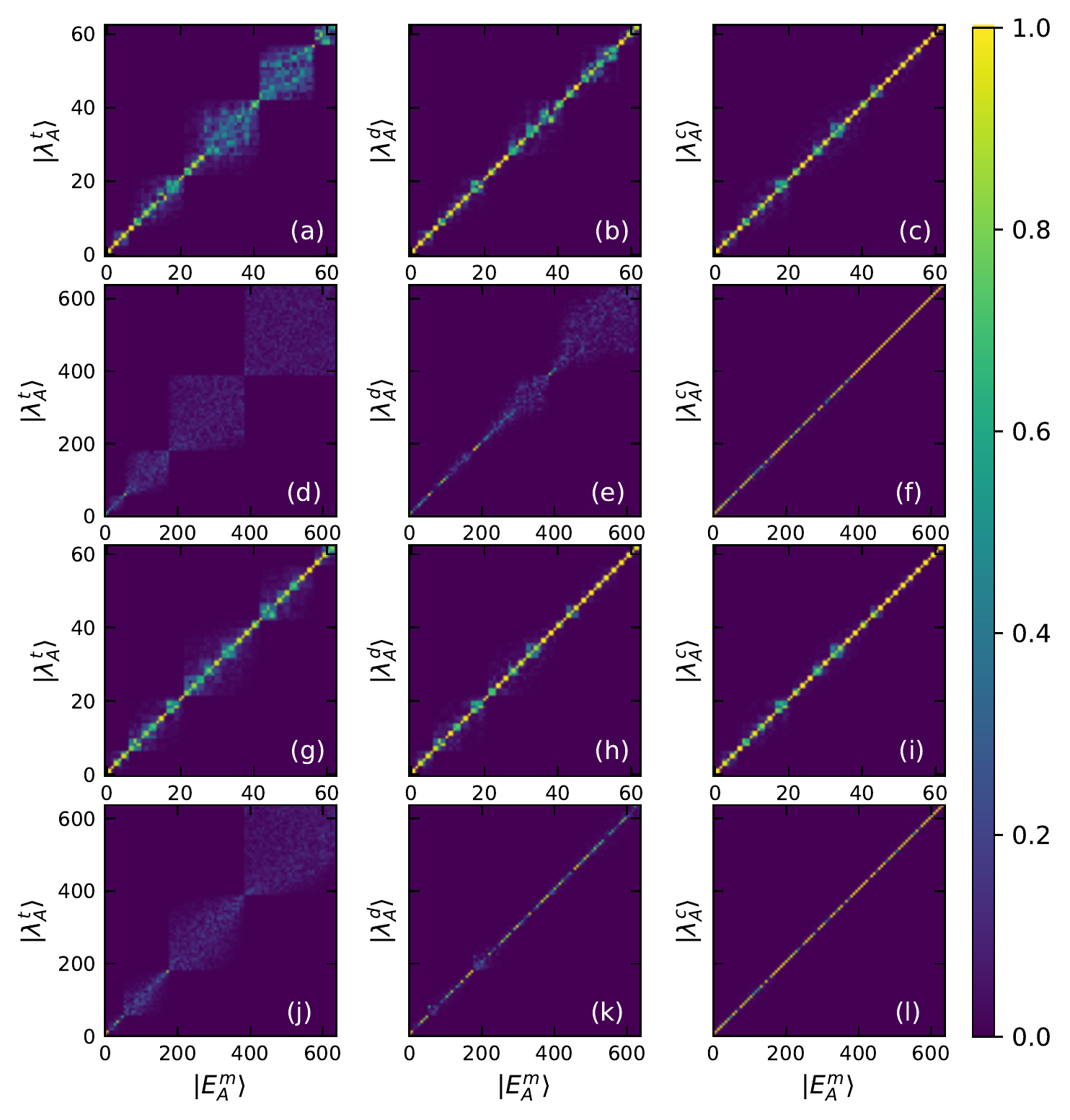}
    \caption{\label{fig:splfmreducedrho6eigenovlp} (Color online) Overlaps between eigenstates of reduced density matrices of $\hat{\rho}^t$ (a, d, g, j), $\hat{\rho}^d$ (b, e, h, k), $\hat{\rho}^c$ (c, f, i, l) and those of $\hat{H}_A$. The subsystem contains the left six sites in (a-c, g-i) and ten sites in (d-f, j-l). The results are for spinless fermions with initial inverse temperature $\beta_0 = 0.1$ (a-f) and $\beta_0 = 0.4$ (g-l).
    }
  \end{figure} 

The above observations indicate that the reduced density matrices of $\hat{\rho}^t$ at long time, $\hat{\rho}^d$ and $\hat{\rho}^c$ may be very similar, for not too large subsystems. In order to check this, we calculate the eigenvalues of the reduced density matrices $\lambda_A$ and plot $\ln{\lambda_A}$ as a function of the eigenenergy $E^m_A$ of the Hamiltonian associated with the subsystem $\hat{H}_A$. Another important question is to decide if these eigenvalues $\lambda_A$ are equal to the weights in thermal density matrix of the subsystem $\hat{\rho}^c(A) = e^{-\beta_{eff}\hat{H}_A} / Z_A$. In order to answer this question, we consider subsystems containing more than five sites,  for which the reduced density matrix can be divided into six sub-sectors which have zero, one, two, three, four and five particles respectively. If the sub-sector contains $p$ particles, the bath of the subsystem should have $5-p$ particles. The reduced density matrices of density matrices considered above are all mixtures of states with fixed number of particles in subsystem $A$. We need to construct the thermal state of subsystem $A$ with a correct mixture of thermal states in these six sub-sectors. Notice that the Hamiltonian can be written as $\hat{H} = \hat{H}_A + \hat{H}_B + \hat{H}_{AB}$. $\hat{H}_A$ and $\hat{H}_B$ are extensive quantities, while $\hat{H}_{AB}$ is a non-extensive local interaction with negligible contribution to the total energy. Then the product of each pair of eigenstates $\ket{E_A}, \ket{E_B}$ can approximate an eigenstate of $H$ with energy $E \simeq E_A + E_B$. A thermal state can then be approximated by a product of two thermal states associated with the subsystem and its complement. To be specific, let $\{\ket{E_A^{m_p}}\}$ be the eigenstates of $H_A^p$ with $p$ particles, $\{\ket{E_B^{m_{5-p}}}\}$ be the eigenstates of $H_B^{5-p}$ with $5-p$ particles, the whole thermal density matrix of two uncorrelated subsystems $A$ and $B$ at temperature $\beta_{eff}$ is
\begin{eqnarray}
\label{eq:ceabprod}
\nonumber \hat{\rho}^c(AB) =&& \sum_{p=0}^5 \sum_{m_p,m_{5-p}} \frac{e^{-\beta_{eff}(E_A^{m_p} + E_B^{m_{5-p}})}}{Z_{AB}} \\&& \ket{E_A^{m_p}} \otimes \ket{E_B^{m_{5-p}}} \bra{E_B^{m_{5-p}}} \otimes \bra{E_A^{m_{p}}} 
\end{eqnarray}
where $Z_{AB} = \sum_{p=0}^5 Z_A^p Z_B^{5-p}$ is the partition function of the whole system. So the thermal density matrix of $A$ is
\begin{eqnarray}
\label{eq:cereducedfromprod}
\nonumber \hat{\rho}^c(A) &=& \Tr_B \hat{\rho}^c(AB) \\ &=& \sum_{p=0}^5 \frac{Z_B^{5-p} Z_A^p}{Z_{AB}} \sum_{m_p} \frac{e^{-\beta_{eff}E_A^{m_p}}}{Z_{A}^p} \ket{E_A^{m_p}} \bra{E_A^{m_{p}}}.
\end{eqnarray}
Thus we have obtained the thermal state of subsystem $A$ as a weighted mixture of states with fixed number of particles. The weights can be obtained by computing the spectrum in subsystems $A$ and $B$ separately.

Fig. \ref{fig:splfmreducedrho6} shows the results for the subsystem containing left six sites. Fig. \ref{fig:splfmreducedrho6}(a) is for spinless fermions with $\beta_0 = 0.1$ and Fig. \ref{fig:splfmreducedrho6}(b) is for bosons with $\beta_0 = 0.01$. Fig. \ref{fig:splfmreducedrho} depicts the results for spinless fermions with $\beta_0 = 0.1$ [Fig. \ref{fig:splfmreducedrho}(a)] and $\beta_0 = 0.4$ [Fig. \ref{fig:splfmreducedrho}(b)], where the subsystem contains left ten sites. And Fig. \ref{fig:bhreducedrho} shows the results for bosons with $\beta_0 = 0.01$ [Fig. \ref{fig:bhreducedrho}(a)] and $\beta_0 = 1$ [Fig. \ref{fig:bhreducedrho}(b)], where the subsystem also contains left ten sites. Fig. \ref{fig:splfmreducedrho6}, Fig. \ref{fig:splfmreducedrho}(a) and Fig. \ref{fig:bhreducedrho}(a) have negative effective temperatures, so the slopes of the lines are positive, while the effective temperatures are positive for Fig. \ref{fig:splfmreducedrho}(b) and Fig. \ref{fig:bhreducedrho}(b), the slopes are negative. The points linearly arranged belong to the same sub-sector with a certain particle number. From top to bottom, they are zero, one, two, three, four, five particle states, respectively. First note that for spinless fermions the spectrum of reduced CE, $\Tr_B{(\hat{\rho}^c)}$, match perfectly with that of $\hat{\rho}^c(A)$ for all cases shown here. However, there are two main deviations for bosons. First, there are some ripples in the spectra of reduced CE around that of $\hat{\rho}^c(A)$ for sub-sectors containing more than two particles. This effect is more obvious for the lower temperature $\beta_0 = 1$ than for the higher temperature $\beta_0 = 0.01$ in the same subsystem, see Fig. \ref{fig:bhreducedrho}. Second, there are discontinuities near the top of the spectrum for sub-sectors containing more than two particles. The states near the top of the spectrum in those sub-sectors jump higher from the lines for the negative effective temperature but jump lower from the lines for the case of positive effective temperature. But overall they still match well. 

For spinless fermions, it can be seen that in Fig. \ref{fig:splfmreducedrho6}(a), the four density matrices have almost the same spectra, except that the line of the five-particle sector in the time-evolved state rotates anti-clockwise a little. In Fig. \ref{fig:splfmreducedrho}(a), the initial temperature is the same, but the subsystem is bigger, then only the first two sub-sectors containing zero and one particle have perfect match for all four density matrices. While for the sub-sector containing two particles, the eigenvalues of $\Tr_B{\hat{\rho}^d}$ and $\Tr_B{\hat{\rho}^t}$ become smaller near the bottom of the spectra, similar to Fig. \ref{fig:spinlessfdece} where the DE also has lower weights near the bottom of the spectra. For sub-sectors with more particles, the reduced DE can always stay around the reduced CE because the similarity of weights between DE and CE as shown in Fig. \ref{fig:spinlessfdece}, while $\Tr_B{\hat{\rho}^t}$ has bigger and bigger deviations. In Fig. \ref{fig:splfmreducedrho}(b) where the initial temperature is lower, the spectra in the first two sectors still match well. For the sub-sector with two particles, the eigenvalues for $\Tr_B{\hat{\rho}^d}$ and $\Tr_B{\hat{\rho}^t}$ are smaller near both the top and the bottom of the spectrum, also similar to the behavior of weights in Fig. \ref{fig:spinlessfdece}. The fact that the reduced DE have bigger deviations at lower initial temperature is also consistent with the result in Fig. \ref{fig:spinlessfdece} that the DE has bigger difference from the CE at lower initial temperature. The spectrum of $\Tr_B{\hat{\rho}^t}$ is also very different from the other three for the five-particle states. Note that the big deviations in the time evolved state is not due to time fluctuations. We have checked the results for other ten different times and they behave almost the same. So the eigenvalues of $\Tr_B{\hat{\rho}^t}$ still equilibrate but not to those of reduced DE.

For bosons, as shown in Fig. \ref{fig:bhreducedrho}, we can still see the ripples and discontinuities in $\Tr_B{(\hat{\rho}^d)}$ and $\Tr_B{(\hat{\rho}^t)}$ for the sub-sectors containing more than two particles. The spectrum for reduced DE match well to the reduced CE for all three cases considered here, because the DE and the CE for bosons are very similar for initial inverse temperature smaller than $1$, as shown in Fig. \ref{fig:bosondece}. For $\beta_0=1$, Fig. \ref{fig:bosondece}(c) shows that most of the weights in DE are a little smaller than those in CE, which also happens here for the reduced DE and the reduced CE. $\Tr_B{(\hat{\rho}^t)}$ have eigenvalues matching well to the other three for sub-sectors containing no more than four particles, but big differences appear for five-particle states in the subsystem with ten sites.

For all the cases considered here, most of the eigenvalues of all four density matrices in at least the first four sub-sectors with no more than three particles agree with each other. The contribution from these eigenvalues dominate the value of R\'enyi entropy, so the entropy of the subsystem containing the left ten sites or fewer all have indiscernible difference in Sec. \ref{sec:entropy}. Another question is if the corresponding eigenstates, $\ket{\lambda_A}$, are the same with the eigenstates of the Hamiltonian associated with the subsystem, $\ket{E_A^m}$. If it is, then any observables residing in this subsystem are expected to thermalize as long as they depend little on the high energy eigenstates belonging to the  sub-sectors with large number of particles. To check this, we plot the overlaps between eigenstates of the three density matrices and those of $\hat{H}_A$ in Fig. \ref{fig:splfmreducedrho6eigenovlp}. The results are for spinless fermions with $\beta_0 = 0.1$ [Fig. \ref{fig:splfmreducedrho6eigenovlp}(a-f)] and $\beta_0 = 0.4$ [Fig. \ref{fig:splfmreducedrho6eigenovlp}(g-l)]. Fig. \ref{fig:splfmreducedrho6eigenovlp}(a-c)(g-i) are for the subsystem containing the left six sites and Fig. \ref{fig:splfmreducedrho6eigenovlp}(d-f)(j-l) are for the subsystem containing the left ten sites. Fig. \ref{fig:splfmreducedrho6eigenovlp}(a,d,g,j) are for the reduced time-evolved states, Fig. \ref{fig:splfmreducedrho6eigenovlp}(b,e,h,k) for the reduced DE and Fig. \ref{fig:splfmreducedrho6eigenovlp}(c,f,i,l) for the reduced CE. For $\beta_0 = 0.1$, we see from Fig. \ref{fig:splfmreducedrho6eigenovlp}(c)(f) that most of the eigenstates of the reduced CE are the same with those of $\hat{H}_A$ (referred as ``good" eigenstates here). And we have shown that the eigenvalues are very close. So the reduced density matrix of a thermal state is still thermal with the same temperature. This is the intensive property of temperature in quantum mechanics. The reduced DE for the subsystem containing six sites still have many good eigenstates, most of which are in the sub-sectors with no more than two particles, while for the subsystem containing ten sites it has only a small portion of eigenstates that are good. As the expectation value of an observable depend on the diagonal ensemble, we conclude that nearly all obervables thermalize as long as they reside in small subsystems and few-particle sub-sectors. The reduced time-evolved state has much fewer good eigenstates. But it can be seen that for the subsystem containing six sites, the first two sub-sectors containing no more than one particle (first 7 states) still have good resutls. So for small subsystems and few-particle sub-sectors, the time-evolved states becomes thermal at long enough time. The results for the lower initial temperature $\beta_0 = 0.4$ are better than those for $\beta_0 = 0.1$, where it is clear that there are more good eigenstates in both the reduced time-evolved state and the reduced DE. Note that for the six-site subsystem, unlike the case with $\beta_0 = 0.1$ where most of the good eigenstates are in low particle number sub-sectors, $\beta_0 = 0.4$ has most of the good eigenstates in high particle number sub-sectors, which may due to the different signs of effective temeprature. We also investigate the results for $\beta_0 = 1$, which have similar behaviors to $\beta_0 = 0.4$ but are a little worse than $\beta_0 = 0.4$. Note that more good eigenstates does not mean that the density matrix is closer to thermal. As high particle number eigenstates have much lower weights, the reduced time-evolved states for higher initial temperatures may be still closer to the thermal states.

We conclude this section with a comment on the theory of equilibration and thermalization. Intuitively, few-particle sub-sectors have more degrees of freedom serving as their bath, so they should behave more like the corresponding thermal states. For the sub-sector containing five particles, the bath only has one degree of freedom. But there still are a large number of dephasing terms, which could result in the equilibration of eigenvalues of reduced time-evolved states but not to those of reduced DE, similar to the behaviors of R\'enyi entropy. In Ref. \cite{linden2009quantum}, a theorem states that the time averaged distance between the reduced time-evolved density matrix and the reduced DE density matrix is bounded by $1/2 \sqrt{d_A / d^{eff}(\Tr_A{(\hat{\rho}_d)})} \leq 1/2 \sqrt{d_A^2 / d^{eff}(\hat{\rho}_d)}$, where $d_A$ is the dimension of the subsystem, $d^{eff}(\Tr_A{(\hat{\rho}_d)})$ is the effective dimension of the reduced DE and $d^{eff}(\hat{\rho}_d)$ is the effective dimension of the DE. From the stronger bound we see that the subsystem must be smaller than half in the sense $l_A / L < 1/2$. For the weaker bound, the maximal $d^{eff}(\hat{\rho}_d)$ is the dimension of the  Hilbert space for the whole system when the weights in the DE are constant. In our case, the weights in the CE and the DE are very similar at high effective temperatures, so for $l_A / L < 1/2$, the subsystem equilibrates to the reduced DE density matrix in the thermodynamic limit. Ref.~\cite{linden2009quantum} also proves that the effective dimension of the diagonal ensemble is the order of the dimension of the Hilbert space (with exponentially small probability to be smaller). We can directly count the dimension of the whole system and the subsystem, and use Stirling formula to estimate the bound. When $l_A / L = 1/2$ and large $L$, the dimension of the whole system for spinless fermions $\ln{d^f_w} \approx (2\ln{2} - 3/4\ln{3})L$, and $\ln{d^f_A} \approx L/2 \ln{2}$. So $(d^f_A)^2 / d^f_w \approx e^{0.131 L}$ is exponentially large with the size of the system. A careful calculation shows that we need $l_A / L < 0.4175$ for the weaker bound to be exponentially small at high temperature. Similarly for bosons, $\ln{d^b_w} \approx (5/4 \ln{5} - 2\ln{2}) L$, $\ln{d^b_A} \approx (3/4 \ln{3} - 1/2 \ln{2}) L$, $(d^b_A)^2 / d^b_w \approx e^{0.33 L}$. And we need $l_A / L < 0.2045$ for the weaker bound to be exponentially small. The numeric results for $l_A / L = 1/2$ show good agreement between eigenvalues but not good agreement between eigenstates. Following the derivation in the Appendix A of Ref. \cite{linden2009quantum}, replacing the coefficients by the matrix elements considered here, the average distance is bounded by $1/2 \sqrt{d_A e^{-S_2^{\beta_0}} T_{BA}^o}$, where again $T_{BA}^o$ is a typical value of $T_{BA}^{mn}$. For $\beta_0 = 0$, $d_A e^{-S_2^{\beta_0}}$ is at most an algebraically increasing function of system size, while $T_{BA}^o$ is typically exponentially small. So in our case, it is possible for subsystems bigger than half of the system ($l_A / L > 1/2$) to equilibrate to the corresponding reduced DE density matrix in the thermodynamic limit.

Next, the DE is a weighted sum of eigenstate projectors. ETH in the strong sense states that every eigenstate is thermal for few-body observables \cite{kim2014testing}. And the similarity of reduced density matrices between the DE and the CE can be explained by the strong form of ETH which states that the reduced density matrix of a single finite energy density eigenstate of chaotic many-body quantum systems is equivalent to the thermal density matrix of the subsystem as long as the subsystem is much smaller than its complement and as long as $l_A / L < 1/2$ for many observables \cite{garrison2018does}. The subsystem ETH \cite{dymarsky2018subsystem} states that the norm of the ``off-diagonal" matrix $\hat{\rho}_A^{mn} = \Tr_B{(\ket{m}\bra{n})}$ ($m \neq n$) is exponentially small for subsystems with $l_A / L < 1/2$, which explains the similarity between reduced $\hat{\rho}^t$ and reduced DE. Inside the small window where the energy density is peaked, all the reduced density matrix of eigenstates give the same thermal states for the subsystem. We show that even the full density matrix of the DE is close to the CE at high temperatures for the models we study here. So at high temperatures, the size of the subsystem that thermalizes just depends on the size of the subsystem that equilibrates to the reduced DE.

\section{Momentum Distribution Function}\label{sec:momentum}
\begin{figure}[t!]
  \centering
    \includegraphics[width=0.5\textwidth]{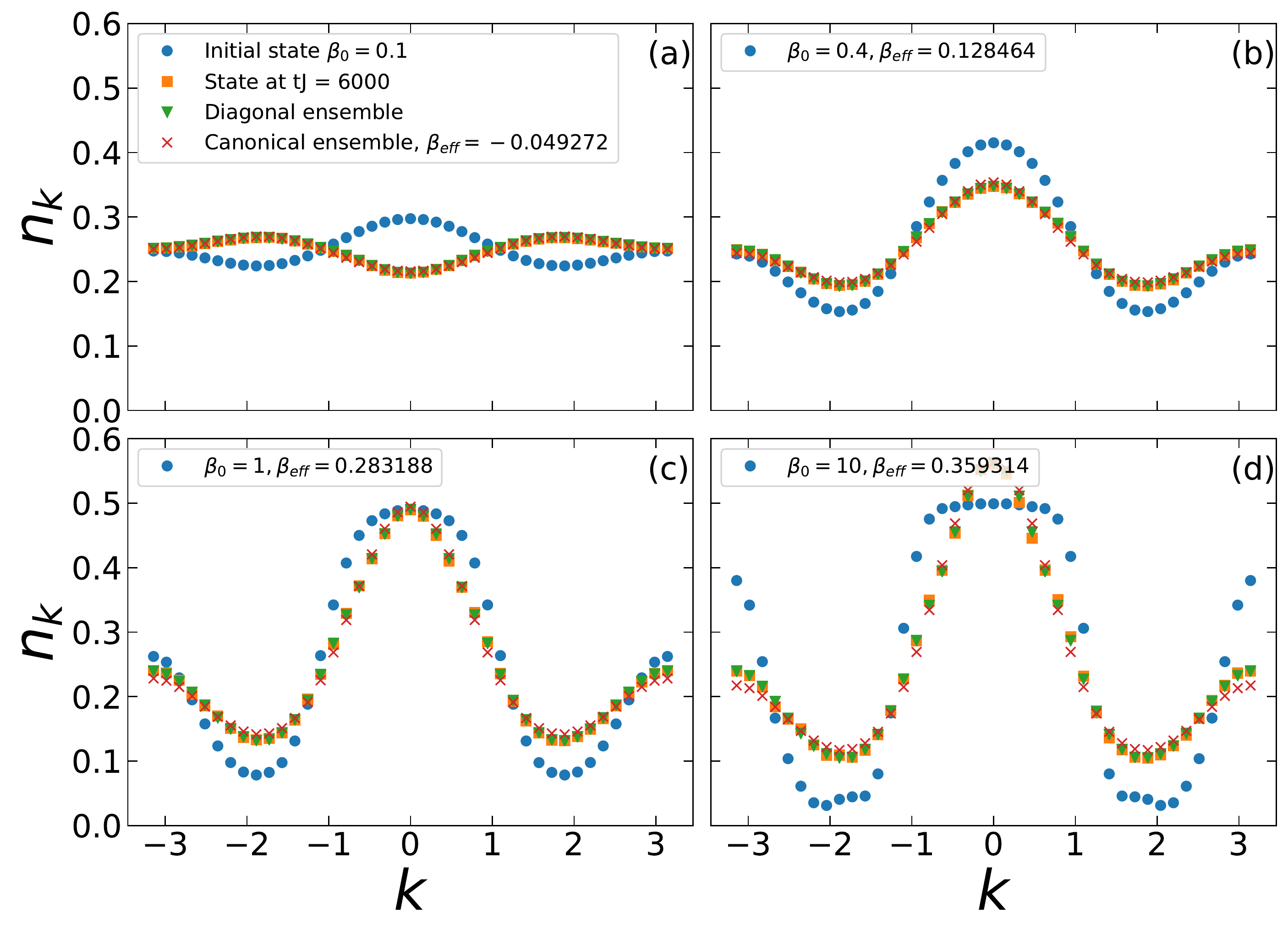}
    \caption{\label{fig:nkdissplfm20s5p} (Color online)  MDFs of the initial state, the time-evolved state at $tJ=6000$, the DE and the corresponding CE as a function of momentum $k \in (-\pi, \pi)$. The results are for spinless fermions, with different initial inverse temperatures $\beta_0 = 0.1$ (a), $\beta_0 = 0.4$ (b), $\beta_0 = 1$ (c) and $\beta_0 = 10$ (d).
    }
  \end{figure} 
\begin{figure}[t!]
  \centering
    \includegraphics[width=0.5\textwidth]{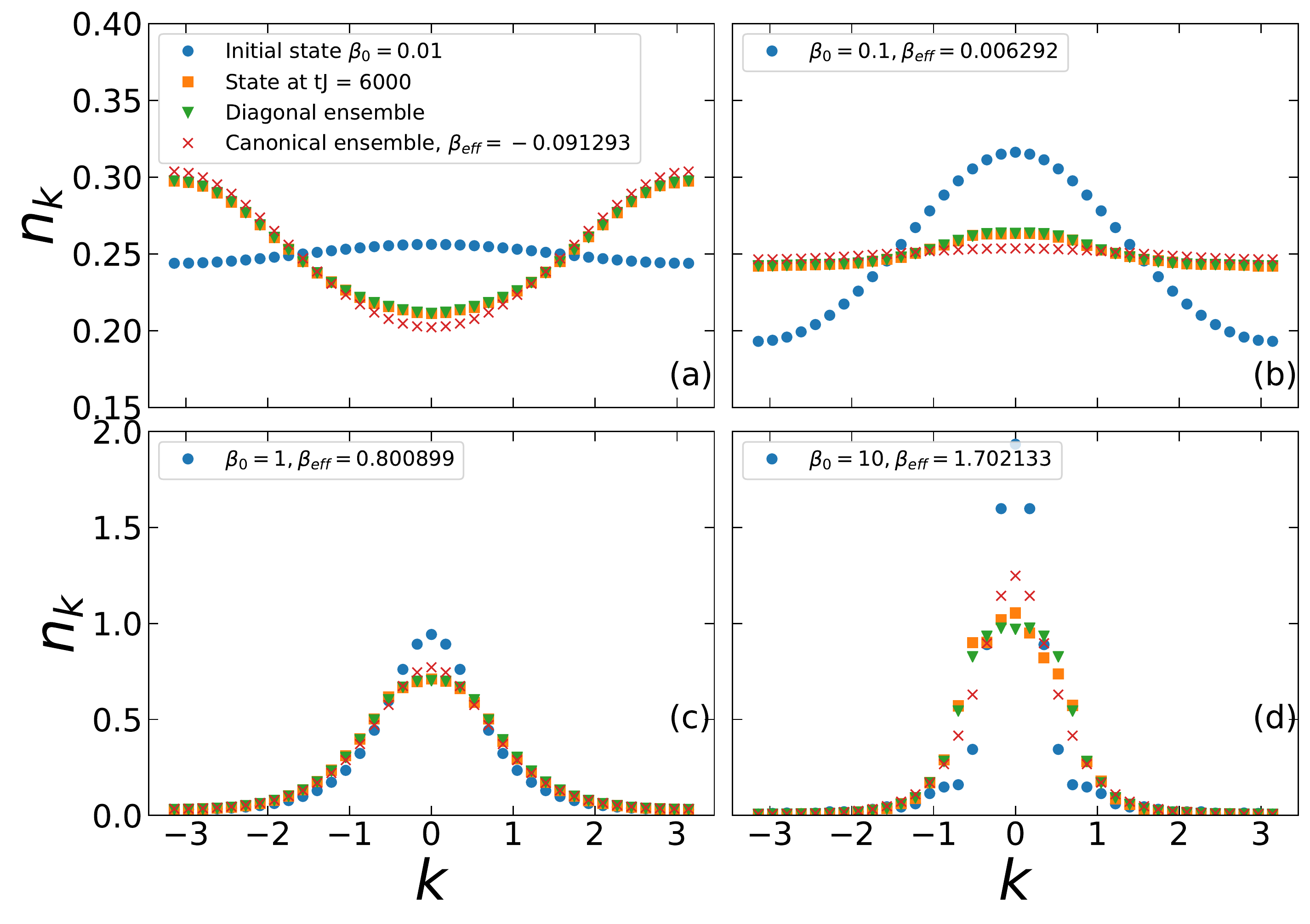}
    \caption{\label{fig:nkdisbh20s5p} (Color online) Same as Fig. \ref{fig:nkdissplfm20s5p} but for bosons. The initial inverse temperatures are $\beta_0 = 0.01$ (a), $\beta_0 = 0.1$ (b), $\beta_0 = 1$ (c) and $\beta_0 = 10$ (d).
    }
  \end{figure} 
In this section, we use the momentum distribution function (MDF) to test the above conclusions. The MDF for fermions is
\begin{eqnarray}
\label{eq:mdffermion}
n^f_k = \frac{1}{L} \sum_{p,q} e^{-ik(p-q)} \langle c^\dagger_p c_q \rangle,
\end{eqnarray}
where $p,q = 1,2,...,L$ are the real-space locations. And the MDF for bosons is
\begin{eqnarray}
n^b_k = \frac{1}{L} \sum_{p,q} e^{-ik(p-q)} \langle a^\dagger_p a_q \rangle.
\end{eqnarray}
The MDF is a few-body observable because it contains a summation of two-body observables, so it is expected to thermalize according to the above results. Regardless, we calculate the MDF for the initial state $\hat{\rho}^0$, the time-evolved state $\hat{\rho}^t$ at long time $tJ = 6000$, the DE density matrix $\hat{\rho}^d$ and the CE density matrix $\hat{\rho}^c$. If the MDF equilibrates, its expectation value using $\hat{\rho}^t$ should be the same as that using $\hat{\rho}^d$. And if the MDF thermalizes, all the expectation values, using $\hat{\rho}^t$, $\hat{\rho}^d$ and $\hat{\rho}^c$ should be the same.

Fig. \ref{fig:nkdissplfm20s5p} shows the results for spinless fermions with the same four initial temperatures as those in Fig. \ref{fig:splessfententropy4bs}. Let's first briefly describe the MDF in the initial state. At low initial temperature $\beta_0=10$ [Fig. \ref{fig:nkdissplfm20s5p}(d)], there are plateaus around $k = 0, \pm 1.8$. The momenta around $k=0, \pm \pi$ have the highest population, while the momenta around $k = \pm 1.8$ have the lowest population. These behaviors can be understood by Fourier transforming the kinetic terms in Hamiltonian Eq. (\ref{eq:SFhamiltonian}), $-2(\cos{k}+\cos{2k})$, which has one global minima $k = 0$, two local minima $k = \pm \pi$ and two global maxima $k \approx \pm 1.8$. At low temperature, the global minimum $k=0$ is occupied first, then the nearby levels, one particle per level, due to the Pauli exclusion principle. The height of the plateau is $0.5$ because of normalization by the full volume. If the surrounding levels have higher energy than the other two local minima $k = \pm \pi$ do, particles start to occupy $k = \pm \pi$ and their surrounding levels. Very few particles will occupy the two maxima $k = \pm 1.8$ and their surroundings. At higher temperatures (Fig. \ref{fig:nkdissplfm20s5p}(a)(b)(c)), the plateaus are smoothed out by thermal
fluctuations. Close to the infinite temperature limit, more and more particles occupy the high energy levels, and eventually the MDF becomes flat.

The equilibration and thermalization of the MDF in Fig. \ref{fig:nkdissplfm20s5p} are very clear. In Fig. \ref{fig:nkdissplfm20s5p}(a), $\beta_0 = 0.1$, the MDFs for $\hat{\rho}_t$, $\hat{\rho}_d$ and $\hat{\rho}_c$ do not have a visible difference, indicating strong thermalization of the MDF. The inversion of occupation is very interesting, and is a result of the  changing of sign of the temperature. The effective final temperature  we calculated for this case is indeed negative. The difference in three MDFs is still extremely small in Fig. \ref{fig:nkdissplfm20s5p}(b) where the initial temperature is lower $\beta_0=0.4$. Further decreasing the initial temperature results in bigger difference between the MDF for $\hat{\rho}_c$ and the other two MDFs. This difference is clear in Fig. \ref{fig:nkdissplfm20s5p}(d), where the MDF for $\hat{\rho}_c$ has lower occupation around $k = \pm \pi$, higher occupation around $k = \pm 1.8$, lower occupation for $k \in (-1.1, -0.7)$ and higher occupation for $k \in (-0.7, 0)$. These differences are probably finite size effects because they are much smaller than those in smaller systems with the same particle filling. Since these differences are overall relatively small, the MDF in the CE is still a good approximation for the real MDF at long times. It should be emphasized that although there are three slightly asymmetric points for $k \in (0, 0.5)$ in Fig. \ref{fig:nkdissplfm20s5p}(d), which is due to time fluctuations, the MDFs for DE and time-evolved state are the same for all four initial temperatures. 

Fig. \ref{fig:nkdisbh20s5p} shows the results for bosons. As the model we consider for bosons only has nearest-neighbor hopping, the hopping energy only have a single minimum at $k = 0$ and maxima at $k = \pm \pi$. So for positive effective temperatures [Fig. \ref{fig:nkdisbh20s5p}(b)(c)(d)], the occupation is peaked at $k=0$, while for negative effective temperatures [Fig. \ref{fig:nkdisbh20s5p}(a)], the occupation is the lowest at $k=0$. Note that the MDF for $\hat{\rho}^t$ and $\hat{\rho}^d$ has little difference in Fig. \ref{fig:nkdisbh20s5p}(a)(b)(c), and the small difference around $k=0$ in Fig. \ref{fig:nkdisbh20s5p}(d) comes from time fluctuations. While they have obvious deviations from that of the CE for all cases, we can see that the MDF of the CE is more peaked at $k=0$ for positive effective temperatures in Fig. \ref{fig:nkdisbh20s5p}(c)(d), while its value around $k=0$ is smaller than that of the DE for negative effective temperature in Fig. \ref{fig:nkdisbh20s5p}(a). In Fig. \ref{fig:nkdisbh20s5p}(b), the effective temperature is close to infinity, MDF of the CE is flatter than that of the DE. Again these deviations are much smaller than those in smaller systems, so we believe they are finite-size effects. 

From the above results we conclude that equilibration is much easier than thermalization. The intuitive reason is that the time-dependent terms are expected to cancel out due to dephasing. And it can also be partially understood from the similarity of $\Tr_B{(\hat{\rho}^d)}$ and $\Tr_B{(\hat{\rho}^t)}$ discussed in the last section. To see it more rigorously, we follow the derivation in Ref. \cite{dalessio2016, zhang2011quantum} and calculate the variance of an observable in time
\begin{eqnarray}
\label{eq:VarO} \triangle^2 O &=& \sum_{m \neq n} |\rho_{0, mn}|^2 |O_{mn}|^2 < M \sum_{m \neq n} |\rho_{0, mn}|^2,
\end{eqnarray}
where $M$ is the maximal off-diagonal matrix element of $\hat{O}$, then similar to Eq. \ref{eq:s2difference},
\begin{eqnarray}
\label{eq:sumrho2} \triangle^2 O < M \left( e^{-S_2^{\beta_0}} - e^{-S_2^d} \right)
\end{eqnarray}
From the same arguments for Eq. (\ref{eq:s2difference}) we conclude that as long as the initial state is at a finite temperature, the variance exponentially decays with the volume of the system, as long as the off diagonal elements are not exponentially large as $e^{S_2^{\beta_0}}$. Moreover, even at $T = 0$, ETH asserts that the off diagonal elements of few-body observables are exponentially small with the size of the system. So initial thermal states substantially decrease the fluctuations in time.\\

\section{The Effective Temperature}\label{sec:effectiveT}
\begin{figure}[t!]
  \centering
    \includegraphics[width=0.49\textwidth]{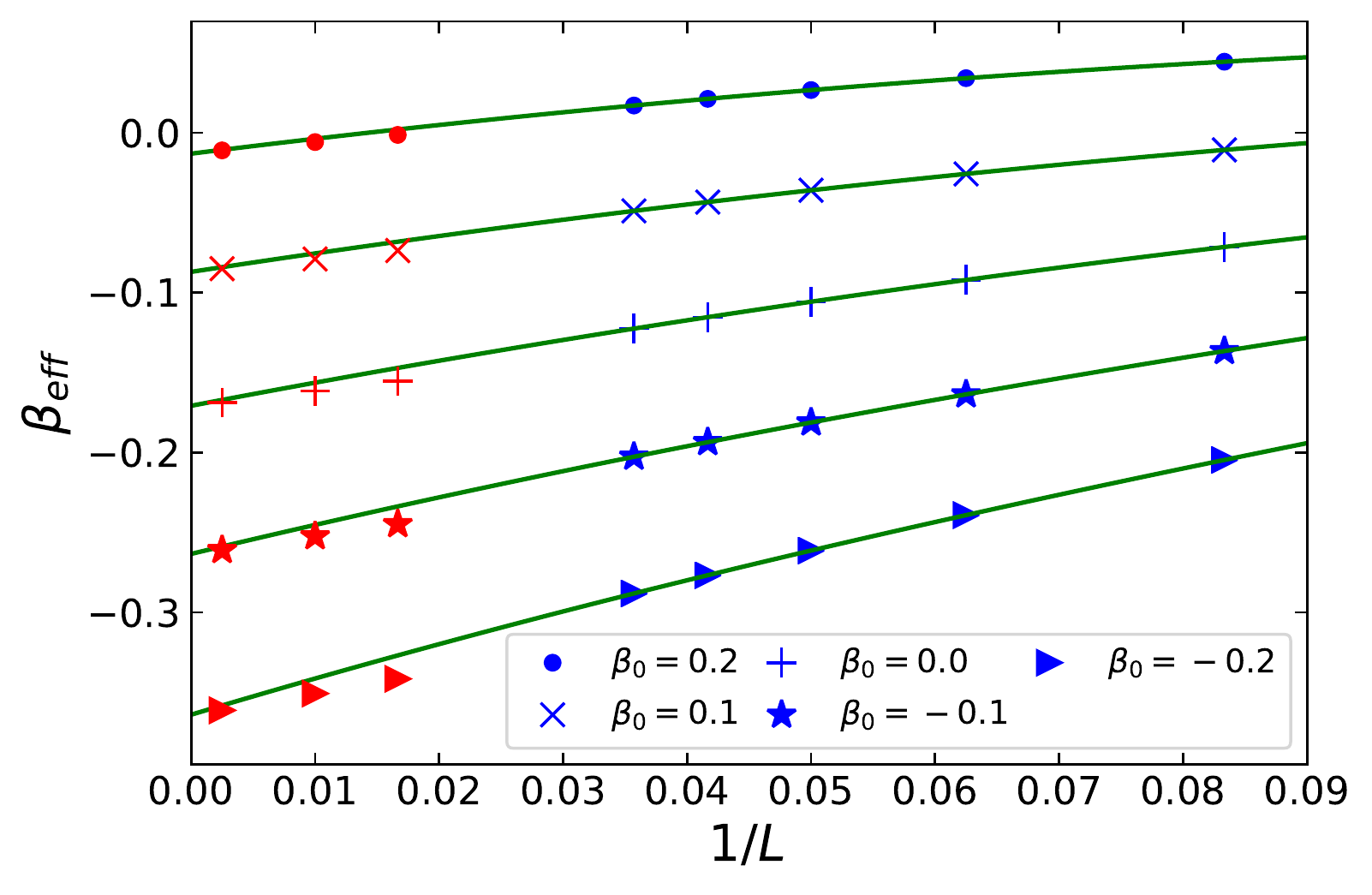}
    \caption{\label{fig:spfmbeteefffss} (Color online) The finite size scaling of the effective inverse temperature for spinless fermions. The lines from top to bottom, the initial inverse temperatures are $\beta_0 = 0.2, 0.1, 0, -0.1, -0.2$ respectively. The system sizes used for fitting are $L = 12, 16, 20, 24, 28$ (obtained with exact diagonalization, blue symbols online). The data points for larger $L$ ($L = 60, 100, 400$, red symbols online) are obtained by finite temperature DMRG with open boundary conditions.}
  \end{figure} 
  
\begin{figure}[t!]
  \centering
    \includegraphics[width=0.49\textwidth]{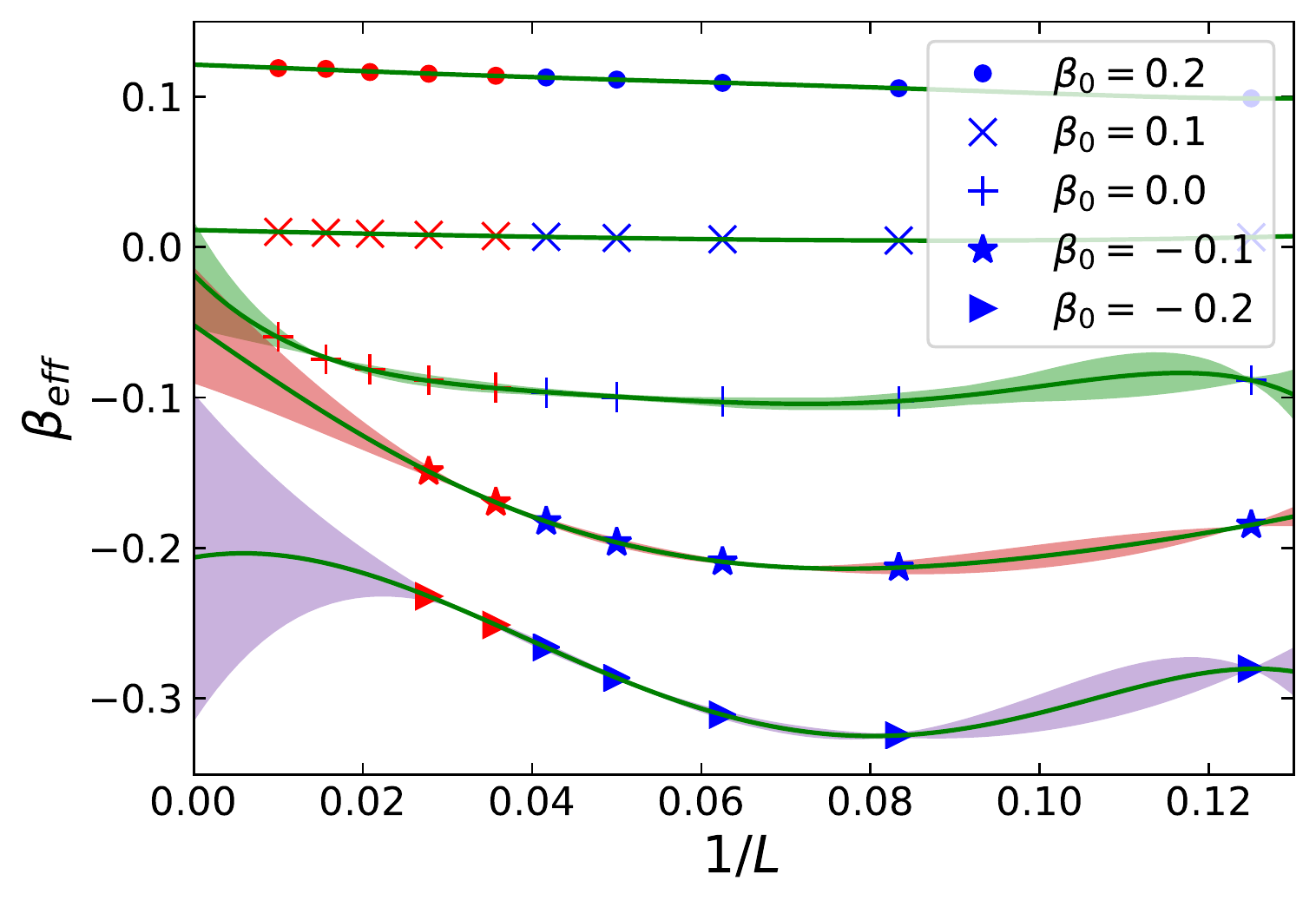}
    \caption{\label{fig:bhbeteefffss} (Color online) Same as Fig. \ref{fig:spfmbeteefffss} but for bosons. Data for smaller system sizes ($L = 8, 12, 16, 20, 24$, blue symbols online) are obtained with exact diagonalization and for larger sizes (red symbols online) are obtained with finite temperature DMRG. All the shown data is used for the curve fitting. The shaded regions are error estimates.}
  \end{figure} 

We have seen that the effective temperature obtained by Eq. \ref{eq:CETbyenergy} can be negative. It is interesting that the negative temperature state can be created simply by sudden expansion to a larger volume. The numerical results presented in the previous sections are for finite system sizes. An important question is whether this negative temperature survives in the thermodynamic limit. To answer this question, we do a finite size analysis of the effective inverse temperature $\beta_{eff}$. In order to go to larger system sizes, we consider a configuration where $L/4$ particles expand inside a ring of $L$ sites and carry out exact diagonalization of the final ($L$-site) Hamiltonian. Making use of the translation invariance \cite{sandvik2010computational}, we can then calculate the spectrum of the final Hamiltonian up to 28 sites with 7 particles for spinless fermions (the dimension of Hilbert space is $1,184,040$) and 24 sites with 6 particles for bosons (the dimension of Hilbert space is $475,020$). We also use finite temperature density matrix renormalization group (DMRG) \cite{PhysRevLett.93.207204, PhysRevB.72.220401, PhysRevB.94.115157} with approximated matrix product operators (MPO) applied to matrix product states (MPS) \cite{PhysRevLett.75.3537} for larger systems. We perform the calculations using the ITensor C++ Library \footnote{Version 3.0.0, http://itensor.org/}, approximating the time evolution operator with MPO to the second order \cite{PhysRevB.91.165112} and setting the imaginary time evolution step to be $\tau = 0.001$. We calculated systems with $L = 60, 100, 400$ for fermions and $L = 28, 36, 48, 64, 100$ for bosons. The data for bosons with $L = 28, 36, 48$ are calculated using MPO with periodic boundary conditions, while results of $L = 64, 100$ are calculated with open boundary conditions.
Convergence of finite temperature DMRG results is challenging for large bosonic systems at negative temperatures, and only data that has reliably converged is  shown in Fig.~\ref{fig:bhbeteefffss}. For fermionic systems, we use results from exact diagonalization ($L = 12, 16, 20, 24, 28$) in the curve fitting procedure and the DMRG data for larger systems for checking extrapolation ($L = 60, 100, 400$). These results are shown in Fig. \ref{fig:spfmbeteefffss}. We use all converged data to do curve fitting for bosonic systems. The boundary effects for $L = 64, 100$ (open boundary conditions) are negligible.

\begin{figure}[t!]
  \centering
    \includegraphics[width=0.49\textwidth]{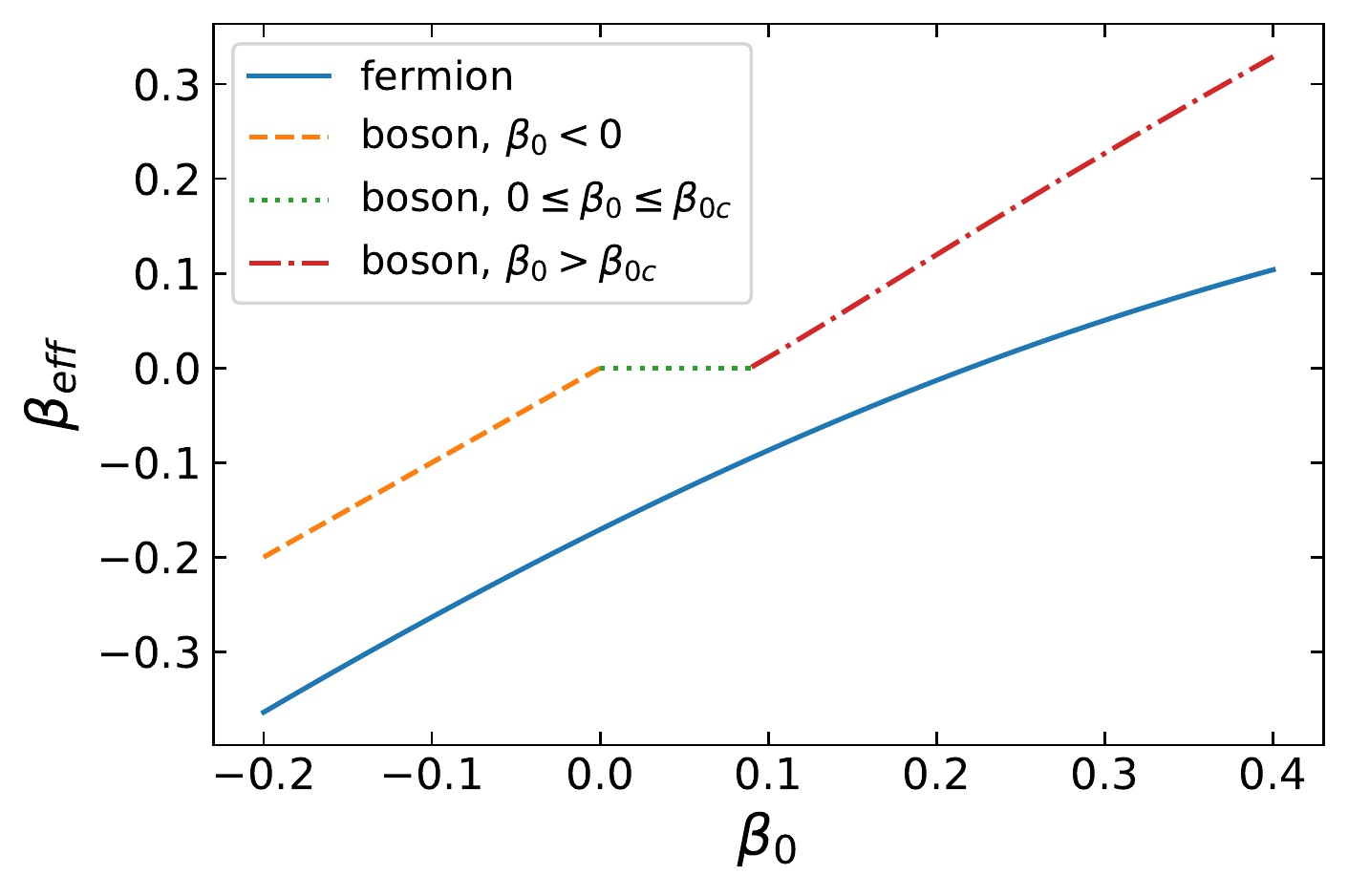}
    \caption{\label{fig:extrapolated} (Color online) The extrapolated $\beta_{eff}$ as a function of $\beta_0$ in the thermodynamic limit. For bosons, $\beta_{0c} = 0.0891(6)$ is the largest initial inverse temperature that has zero effective inverse temperature. Lines for $\beta_0 < \beta_{0c}$ are sketched based on the arguments ($\beta_{eff} = \beta_0$ for $\beta_0 < 0$ and $\beta_{eff} = 0$ for $0 \leq \beta_0 \leq \beta_{0c}$) explained in the main text.}
  \end{figure}

The results for spinless fermions are depicted in Fig. \ref{fig:spfmbeteefffss}. We see, for all initial temperatures shown in the figure, that $\beta_{eff} < \beta_0$ and $\beta_{eff}$ decreases as $L$ increases. The effective inverse temperature $\beta_{eff}$ is an approximated linear function of $1/L$. The lines in Fig. \ref{fig:spfmbeteefffss} are curve fits for data from exact diagonalization ($L = 12, 16, 20, 24, 28$) with third-degree polynomials. The curve fitting is stable to the degree of the polynomials and has negligible difference when higher degree polynomials are used. The curve fits can be extrapolated and the values of $\beta_{eff}$ in the thermodynamic limit can be extracted. An initial inverse temperature $\beta_0 = 0.2$ already results in a negative temperature state for $L \rightarrow \infty$, and higher initial temperatures result in more negative inverse temperature states after expansion. The validity of the extrapolation with data from exact diagonalization is confirmed by the data from finite temperature DMRG, which are very close and slightly below the extrapolation for $L = 60, 100$ due to boundary effects (we use open boundary conditions in the DMRG calculations) and coincide with the extrapolation for $L = 400$. By repeating the procedure of extrapolation from $\beta_0 = -0.2$ to $\beta_0 = 0.4$ with stepsize $0.01$, we show the effective inverse temperature as a function of initial inverse temperature in the thermodynamic limit in Fig.~\ref{fig:extrapolated}. It is a smooth function with slight concavity, from where we see that the initial inverse temperature that has zero effective inverse temperature is $0.22$.

Negative temperature occur in systems with bounded spectrum and therefore have a peak in the thermal entropy as function of the energy.
For energies above the value where the peak occurs, the thermal entropy decreases with energy and these high energy states have negative temperature. High energy states for the fermionic model correspond to particles occupying neighboring sites (so that the energy is in the form of high interaction energy), or occupying high-energy $k$-states (energy in the form of high kinetic energy), depending on which scale dominates. In either case, and even when the two scales are comparable, having high energy reduces the number of allowed microscopic  configurations and the thermal entropy decreases with increasing energy. The thermal entropy is maximum when all the eigenvalues of the density operators are equal, which is the case of infinite temperature in CE. This is the standard picture for thermal entropy and negative temperature. Now, in Sec. \ref{sec:entropy}, we have shown that the three density matrices (exact time-evolved at long times, diagonal, and canonical) all have very similar spectra for small subsystems, and the DE and the CE continue to agree for larger subsystems, even up to the entire system in some cases. So the entanglement entropy of small subsystems behaves the same way as the thermal entropy at negative temperature. Here, we extract the effective temperature by finding a CE with the same total energy as the initial system (Eq. \ref{eq:tevolexpE}), and thus find negative effective temperatures for high energies (high enough initial temperature). In order to draw contrast later to the case of bosons that we discuss below, we note also that even the highest eigenenergy of the fermionic model is extensive and the canonical ensemble is therefore well defined for negative temperature. So as long as the initial energy is higher than the infinite temperature energy of the final Hamiltonian, the effective temperature becomes negative after expansion.

\begin{figure}[t!]
  \centering
    \includegraphics[width=0.49\textwidth]{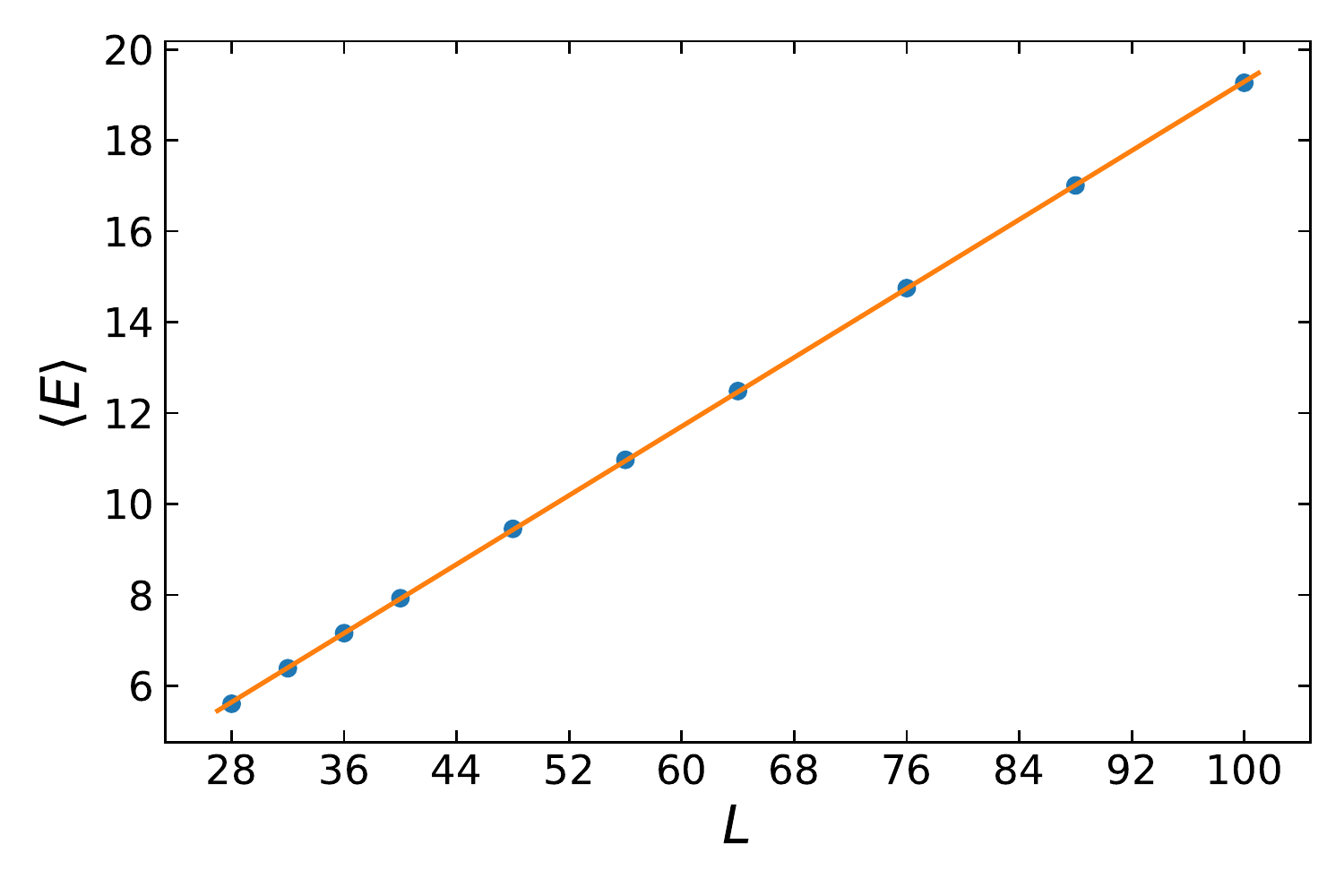}
    \caption{\label{fig:extensive} (Color online) Total energy of quarter-filled bosonic systems as a function of system size for inverse temperature $\beta = -1/L$. The data points correspond to $L = 28, 32, 36, 40, 48, 56, 64, 76, 88, 100$. The solid line is a linear fit.}
  \end{figure}

The same plot as Fig. \ref{fig:spfmbeteefffss} for fermions is shown in Fig. \ref{fig:bhbeteefffss} for bosons. As in the case of fermions we see that $\beta_{eff} < \beta_0$ but for bosons, $\beta_{eff}$ increases with the system size $L$ for large enough systems. Due to strong finite size effects, we use both exact diagonalization and DMRG data to do curve fitting in order to get an estimate for the error of the  extrapolated value of $\beta_{eff}$ in the thermodynamic limit. The solid lines in Fig. \ref{fig:bhbeteefffss} are fifth-degree polynomial fits and  the estimate for the errors are obtained by changing the degree of the polynomials. For $\beta_0 = 0.1, 0.2$, the curve fits are stable to both the degree of polynomials and number of data points included, and the error bars for the extrapolated values at $L \rightarrow \infty$ are too small to be seen in the plot ($< 10^{-3}$). This stability lasts down to the extrapolated initial inverse temperature $\beta_{0c} = 0.0891(6)$ that has zero effective inverse temperature. The extrapolated $\beta_{eff}$ as a function of $\beta_0$ ($\beta_{0c} < \beta_0 < 0.4$) in the thermodynamic limit is depicted in Fig.~\ref{fig:extrapolated}, which exhibits much less concavity compared to the case of fermions. For smaller initial inverse temperatures and negative $\beta_{eff}$, $\beta_{eff}$ decreases at first and then increases for larger systems. The curve fits are very sensitive to the degree of the polynomials used in the curve fitting, so there is large uncertainty in the extrapolated ($L\rightarrow \infty$) value of $\beta_{eff}$ for these cases. Finite temperature DMRG is difficult for large systems of bosons at negative temperature due to high onsite occupation and huge quadratic onsite energy, so we do not have reliable values of $\beta_{eff}$ for $L > 48$ with $\beta_0 = -0.1, -0.2$. We now discuss the upturn in $\beta_{eff}$ with increasing system size $L$ that is exhibited by the bosons but not the fermions. For the bosonic model, since $U/J=3$, the highest energy state corresponds  the bosons all occupying a single site, which has energy $\sim N_p(N_p-1)$, and the next highest energy state has energy $\sim (N_p-1)(N_p-2)$, where $N_p = L/4$ is the total number of bosons. The energy difference between these two states is linear in $N_p$. When the temperature is finite and negative and $L$ is large, the Boltzmann weights in CE for the highest energy state dominate \cite{braun2013}, all bosons stay at the same site and the energy is quadratic in $N_p$, and therefore quadratic in $L$. Initial states with non-negative or $-1/L$ like initial inverse temperature have extensive total energy, and we expect that the negative temperature should not survive in the thermodynamic limit for bosons with repulsive interactions. However, this needs to be reconciled with the previous argument, based on decreasing entropy with increasing energy, that the final temperature should be negative. The infinite temperature energy of $m$ bosons in $N$ sites is $U m(m-1) / (N+1)$. The energy of the initial Hamiltonian at infinite temperature is $U L(L-4) / 8(L+2)$ ($m = L/4, N = L/2$), twice that of the final Hamiltonian at infinite temperature $U L(L-4) / 16(L+1)$ ($m = L/4, N = L$). So when the initial state has extensive energy and the initial energy is higher than the energy of the final Hamiltonian at infinite temperature,  $\beta_{eff}$ should be negative. These two expectations can be conciliated by having $\beta_{eff} \sim -1/L$ when $\beta_0=0$, that is, while $\beta_{eff}$ does indeed become negative for finite systems, it tends back to zero (infinite temperature) in the thermodynamic limit.
The form $\beta_{eff} \sim -1/L$ returns the Boltzman weights to their conventional form and ensures that the extensive property of the total energy is maintained. In order to check this last remark, we calculated the total energy for quarter-filled bosonic systems with $L = 28,32,36,40,48,56,64,76,88,100$, and inverse temperature given by $\beta = -1/L$ using finite temperature DMRG. The results are plotted in Fig. \ref{fig:extensive} as a function of system size $L$. The linear fit demonstrates the extensive property of the total energy. For $\beta_0 = 0$, $\beta_{eff}$ is then expected to go to $0$ as $\beta_{eff} = - 1/L$ for very large systems. Larger scale computations is needed to confirm this numerically, but the statement is consistent with our numerical results within the error bars (Fig. \ref{fig:bhbeteefffss}). The statement holds for any $0 \leq \beta_0 \leq \beta_{0c}$ in the thermodynamic limit where the initial energy is extensive and exceeds the infinite temperature energy of the final Hamiltonian. Finally, if the initial temperature is already negative, for large $L$ all bosons already form a single particle that moves together, and the total energy is already non-extensive. The effective temperature should then be the same as the initial temperature in the thermodynamic limit. This is consistent with the extrapolation in Fig.~\ref{fig:bhbeteefffss}, within error bars, for $\beta_0 = -0.2$. In summary, $\beta_{eff} = \beta_0$ for $\beta_0 < 0$ and $\beta_{eff} = 0$ for $0 \leq \beta_0 \leq \beta_{0c}$, as shown in Fig.~\ref{fig:extrapolated}.

\section{Conclusion}\label{sec:conclusion}

We studied the Joule expansion for  one-dimensional nonintegrable quantum lattice systems containing bosons or spinless fermions. We  calculated and compared the weights of the eigenstates in the DE and the CE, for different initial temperatures, for the two models. The effective final temperature for the CE is obtained by matching the total energy of the system before and after the expansion. The effective temperature can be negative when the initial temperature is high enough, where the weights of the DE and the CE can match very well. As the DE determines the equilibrated expectation values of observables, it is expected that extensive observables can thermalize at high initial temperatures. We analyzed the von Neumann and second order R\'enyi entropies in the DE, the CE and for the exact numerical time-evolved state at long times. The agreement between the DE and the CE can extend to the whole system for high initial temperatures because of the similarity in weights. And the agreement between the time-evolved states and the other two can extend beyond $l_A = L/2$ at high temperatures. We provided analytical arguments and numerical confirmation that $\Delta S_{2A}$, the difference between the second-order Renyi entropy calculated in the DE and the fully time-evolved time-averaged one at long times, around $l_A = L/2$ is exponentially small in system size $L$ (Fig. \ref{fig:sffssdREvsL}). We examined the eigenvalues and the eigenstates of the reduced density matrices directly and found that the reduced time-evolved density matrix at long times can be thermal for small subsystems. The momentum distribution functions have strong thermalizations at high initial temperature, and show population inversion after the expansion if the effective temperature is negative.

Note again that for initial temperatures larger than a certain value, the system thermalizes in a negative-temperature state. Negative temperature states have been created in cold atoms in spin systems\cite{medley2011} and for motional degrees of freedom of a fermionic system\cite{braun2013}, and are of interest for a number of exotic phenomena that can be realized with them\cite{mandt2011}. We propose that negative temperature states can be created by sudden expansion to a larger system size. In this paper we only considered repulsive interactions. For fermions, the negative temperature is a feature that remains in the thermodynamic limit. For bosons with repulsive interactions, there can be no negative temperature state in the thermodynamic limit since the energy spectrum is unbounded. For finite, experimentally realistic system sizes, there can be negative-temperature states and these can also be obtained through Joule expansion.

There are still a number of unanswered questions. For example, what is the maximal size of the subsystem whose R\'enyi entropy can equilibrate to those of the reduced DE and the reduced CE for a certain initial temperature? And what is the biggest subsystem whose reduced time-evolved density matrix is thermal for a given initial temperature? Why do the eigenvalues of reduced time-evolved states in high-particle-number sub-sectors equilibrate to different values than the reduced DE? Answering these questions needs further detailed study of the diagonal and off-diagonal matrix elements of initial states in the basis of final Hamiltonian and the reduced density matrix of projector operators $\ket{m}\bra{m}$ and ``off-diagonal" operators $\ket{m}\bra{n}$ ($m \neq n$), which can be the subject of future work.

In terms of experimental realization of the models studied, the bosonic model is a simple Bose Hubbard model that can be created with cold atoms on optical lattices. Creating the wall that initially confines the particles to the left half of the system requires  local manipulation of the optical potential and 
may be realized using holographic techniques \cite{islam2015measuring}. The thermalization dynamics from quantum quench for the simple one dimensional Bose Hubbard model with $U/J = 0.38$ and effective temperature $\beta J = 0.09$ has been studied experimentally \cite{Kaufman794}. The spinless fermions model we studied contains nearest-neighbor and next-nearest-neighbor hoppings and interactions. As discussed in Sec. \ref{sec:model}, it is equivalent to a zig-zag two-leg ladder with only nearest-neighbor terms (inter and intra-leg). Nearest-neighbor interactions may be created and tuned via Rydberg-dressed potentials\cite{Henkel2010,Pupillo2010,Zeiher2016}. The inter and intra-leg nearest-neighbor interactions and hopping can be tuned to different values by varying the lattice spacing on the rung and along the legs to different values and exploiting the range and shape of the Rydberg-dressed potential, as was done in Ref. \cite{zhang2018prl}.  
 Multi-mode cavity photon-mediated
interactions \cite{Vaidya:2018fp} may also be used to create the needed interactions.

\vskip20pt
\begin{acknowledgments}
This work was supported in part by the National Science Foundation (NSF) under Grant No. DMR-1411345 (SWT) and by the U.S. Department of Energy (DOE) under Award Number DE-SC0019139 (YM). 
\end{acknowledgments}


\begin{thebibliography}{101}%
\makeatletter
\providecommand \@ifxundefined [1]{%
 \@ifx{#1\undefined}
}%
\providecommand \@ifnum [1]{%
 \ifnum #1\expandafter \@firstoftwo
 \else \expandafter \@secondoftwo
 \fi
}%
\providecommand \@ifx [1]{%
 \ifx #1\expandafter \@firstoftwo
 \else \expandafter \@secondoftwo
 \fi
}%
\providecommand \natexlab [1]{#1}%
\providecommand \enquote  [1]{``#1''}%
\providecommand \bibnamefont  [1]{#1}%
\providecommand \bibfnamefont [1]{#1}%
\providecommand \citenamefont [1]{#1}%
\providecommand \href@noop [0]{\@secondoftwo}%
\providecommand \href [0]{\begingroup \@sanitize@url \@href}%
\providecommand \@href[1]{\@@startlink{#1}\@@href}%
\providecommand \@@href[1]{\endgroup#1\@@endlink}%
\providecommand \@sanitize@url [0]{\catcode `\\12\catcode `\$12\catcode
  `\&12\catcode `\#12\catcode `\^12\catcode `\_12\catcode `\%12\relax}%
\providecommand \@@startlink[1]{}%
\providecommand \@@endlink[0]{}%
\providecommand \url  [0]{\begingroup\@sanitize@url \@url }%
\providecommand \@url [1]{\endgroup\@href {#1}{\urlprefix }}%
\providecommand \urlprefix  [0]{URL }%
\providecommand \Eprint [0]{\href }%
\providecommand \doibase [0]{http://dx.doi.org/}%
\providecommand \selectlanguage [0]{\@gobble}%
\providecommand \bibinfo  [0]{\@secondoftwo}%
\providecommand \bibfield  [0]{\@secondoftwo}%
\providecommand \translation [1]{[#1]}%
\providecommand \BibitemOpen [0]{}%
\providecommand \bibitemStop [0]{}%
\providecommand \bibitemNoStop [0]{.\EOS\space}%
\providecommand \EOS [0]{\spacefactor3000\relax}%
\providecommand \BibitemShut  [1]{\csname bibitem#1\endcsname}%
\let\auto@bib@innerbib\@empty
\bibitem [{\citenamefont {von Neumann}(2010)}]{von2010proof}%
  \BibitemOpen
  \bibfield  {author} {\bibinfo {author} {\bibfnamefont {J.}~\bibnamefont {von
  Neumann}},\ }\href {\doibase 10.1140/epjh/e2010-00008-5} {\bibfield
  {journal} {\bibinfo  {journal} {The European Physical Journal H}\ }\textbf
  {\bibinfo {volume} {35}},\ \bibinfo {pages} {201} (\bibinfo {year}
  {2010})}\BibitemShut {NoStop}%
\bibitem [{\citenamefont {Jensen}\ and\ \citenamefont
  {Shankar}(1985)}]{jensen1985}%
  \BibitemOpen
  \bibfield  {author} {\bibinfo {author} {\bibfnamefont {R.~V.}\ \bibnamefont
  {Jensen}}\ and\ \bibinfo {author} {\bibfnamefont {R.}~\bibnamefont
  {Shankar}},\ }\href {\doibase 10.1103/PhysRevLett.54.1879} {\bibfield
  {journal} {\bibinfo  {journal} {Phys. Rev. Lett.}\ }\textbf {\bibinfo
  {volume} {54}},\ \bibinfo {pages} {1879} (\bibinfo {year}
  {1985})}\BibitemShut {NoStop}%
\bibitem [{\citenamefont {Deutsch}(1991)}]{deutsch1991quantum}%
  \BibitemOpen
  \bibfield  {author} {\bibinfo {author} {\bibfnamefont {J.~M.}\ \bibnamefont
  {Deutsch}},\ }\href {\doibase 10.1103/PhysRevA.43.2046} {\bibfield  {journal}
  {\bibinfo  {journal} {Phys. Rev. A}\ }\textbf {\bibinfo {volume} {43}},\
  \bibinfo {pages} {2046} (\bibinfo {year} {1991})}\BibitemShut {NoStop}%
\bibitem [{\citenamefont {Goldstein}\ \emph {et~al.}(2010)\citenamefont
  {Goldstein}, \citenamefont {Lebowitz}, \citenamefont {Tumulka},\ and\
  \citenamefont {Zangh{\`i}}}]{goldstein2010long}%
  \BibitemOpen
  \bibfield  {author} {\bibinfo {author} {\bibfnamefont {S.}~\bibnamefont
  {Goldstein}}, \bibinfo {author} {\bibfnamefont {J.~L.}\ \bibnamefont
  {Lebowitz}}, \bibinfo {author} {\bibfnamefont {R.}~\bibnamefont {Tumulka}}, \
  and\ \bibinfo {author} {\bibfnamefont {N.}~\bibnamefont {Zangh{\`i}}},\
  }\href {\doibase 10.1140/epjh/e2010-00007-7} {\bibfield  {journal} {\bibinfo
  {journal} {The European Physical Journal H}\ }\textbf {\bibinfo {volume}
  {35}},\ \bibinfo {pages} {173} (\bibinfo {year} {2010})}\BibitemShut
  {NoStop}%
\bibitem [{\citenamefont {Mehta}(2004)}]{mehta2004random}%
  \BibitemOpen
  \bibfield  {author} {\bibinfo {author} {\bibfnamefont {M.~L.}\ \bibnamefont
  {Mehta}},\ }\href@noop {} {\emph {\bibinfo {title} {Random matrices}}},\
  Vol.\ \bibinfo {volume} {142}\ (\bibinfo  {publisher} {Elsevier},\ \bibinfo
  {year} {2004})\BibitemShut {NoStop}%
\bibitem [{\citenamefont {Guhr}\ \emph {et~al.}(1998)\citenamefont {Guhr},
  \citenamefont {Müller–Groeling},\ and\ \citenamefont
  {Weidenmüller}}]{guhr1998random}%
  \BibitemOpen
  \bibfield  {author} {\bibinfo {author} {\bibfnamefont {T.}~\bibnamefont
  {Guhr}}, \bibinfo {author} {\bibfnamefont {A.}~\bibnamefont
  {Müller–Groeling}}, \ and\ \bibinfo {author} {\bibfnamefont {H.~A.}\
  \bibnamefont {Weidenmüller}},\ }\href {\doibase
  https://doi.org/10.1016/S0370-1573(97)00088-4} {\bibfield  {journal}
  {\bibinfo  {journal} {Physics Reports}\ }\textbf {\bibinfo {volume} {299}},\
  \bibinfo {pages} {189 } (\bibinfo {year} {1998})}\BibitemShut {NoStop}%
\bibitem [{\citenamefont {Srednicki}(1999)}]{srednicki1999approach}%
  \BibitemOpen
  \bibfield  {author} {\bibinfo {author} {\bibfnamefont {M.}~\bibnamefont
  {Srednicki}},\ }\href {\doibase 10.1088/0305-4470/32/7/007} {\bibfield
  {journal} {\bibinfo  {journal} {Journal of Physics A: Mathematical and
  General}\ }\textbf {\bibinfo {volume} {32}},\ \bibinfo {pages} {1163}
  (\bibinfo {year} {1999})}\BibitemShut {NoStop}%
\bibitem [{\citenamefont {Rigol}\ and\ \citenamefont
  {Srednicki}(2012)}]{rigol2012alternatives}%
  \BibitemOpen
  \bibfield  {author} {\bibinfo {author} {\bibfnamefont {M.}~\bibnamefont
  {Rigol}}\ and\ \bibinfo {author} {\bibfnamefont {M.}~\bibnamefont
  {Srednicki}},\ }\href {\doibase 10.1103/PhysRevLett.108.110601} {\bibfield
  {journal} {\bibinfo  {journal} {Phys. Rev. Lett.}\ }\textbf {\bibinfo
  {volume} {108}},\ \bibinfo {pages} {110601} (\bibinfo {year}
  {2012})}\BibitemShut {NoStop}%
\bibitem [{\citenamefont {Santos}\ and\ \citenamefont
  {Rigol}(2010{\natexlab{a}})}]{santos2010onset}%
  \BibitemOpen
  \bibfield  {author} {\bibinfo {author} {\bibfnamefont {L.~F.}\ \bibnamefont
  {Santos}}\ and\ \bibinfo {author} {\bibfnamefont {M.}~\bibnamefont {Rigol}},\
  }\href {\doibase 10.1103/PhysRevE.81.036206} {\bibfield  {journal} {\bibinfo
  {journal} {Phys. Rev. E}\ }\textbf {\bibinfo {volume} {81}},\ \bibinfo
  {pages} {036206} (\bibinfo {year} {2010}{\natexlab{a}})}\BibitemShut
  {NoStop}%
\bibitem [{\citenamefont {Kim}\ \emph {et~al.}(2014)\citenamefont {Kim},
  \citenamefont {Ikeda},\ and\ \citenamefont {Huse}}]{kim2014testing}%
  \BibitemOpen
  \bibfield  {author} {\bibinfo {author} {\bibfnamefont {H.}~\bibnamefont
  {Kim}}, \bibinfo {author} {\bibfnamefont {T.~N.}\ \bibnamefont {Ikeda}}, \
  and\ \bibinfo {author} {\bibfnamefont {D.~A.}\ \bibnamefont {Huse}},\ }\href
  {\doibase 10.1103/PhysRevE.90.052105} {\bibfield  {journal} {\bibinfo
  {journal} {Phys. Rev. E}\ }\textbf {\bibinfo {volume} {90}},\ \bibinfo
  {pages} {052105} (\bibinfo {year} {2014})}\BibitemShut {NoStop}%
\bibitem [{\citenamefont {Yoshizawa}\ \emph {et~al.}(2018)\citenamefont
  {Yoshizawa}, \citenamefont {Iyoda},\ and\ \citenamefont
  {Sagawa}}]{yoshizawa2018numerical}%
  \BibitemOpen
  \bibfield  {author} {\bibinfo {author} {\bibfnamefont {T.}~\bibnamefont
  {Yoshizawa}}, \bibinfo {author} {\bibfnamefont {E.}~\bibnamefont {Iyoda}}, \
  and\ \bibinfo {author} {\bibfnamefont {T.}~\bibnamefont {Sagawa}},\ }\href
  {\doibase 10.1103/PhysRevLett.120.200604} {\bibfield  {journal} {\bibinfo
  {journal} {Phys. Rev. Lett.}\ }\textbf {\bibinfo {volume} {120}},\ \bibinfo
  {pages} {200604} (\bibinfo {year} {2018})}\BibitemShut {NoStop}%
\bibitem [{\citenamefont {Rigol}\ \emph {et~al.}(2008)\citenamefont {Rigol},
  \citenamefont {Dunjko},\ and\ \citenamefont
  {Olshanii}}]{rigol2008thermalization}%
  \BibitemOpen
  \bibfield  {author} {\bibinfo {author} {\bibfnamefont {M.}~\bibnamefont
  {Rigol}}, \bibinfo {author} {\bibfnamefont {V.}~\bibnamefont {Dunjko}}, \
  and\ \bibinfo {author} {\bibfnamefont {M.}~\bibnamefont {Olshanii}},\ }\href
  {https://doi.org/10.1038/nature06838 http://10.0.4.14/nature06838
  https://www.nature.com/articles/nature06838{\#}supplementary-information}
  {\bibfield  {journal} {\bibinfo  {journal} {Nature}\ }\textbf {\bibinfo
  {volume} {452}},\ \bibinfo {pages} {854} (\bibinfo {year}
  {2008})}\BibitemShut {NoStop}%
\bibitem [{\citenamefont {Rigol}(2009)}]{rigol2009quantum}%
  \BibitemOpen
  \bibfield  {author} {\bibinfo {author} {\bibfnamefont {M.}~\bibnamefont
  {Rigol}},\ }\href {\doibase 10.1103/PhysRevA.80.053607} {\bibfield  {journal}
  {\bibinfo  {journal} {Phys. Rev. A}\ }\textbf {\bibinfo {volume} {80}},\
  \bibinfo {pages} {053607} (\bibinfo {year} {2009})}\BibitemShut {NoStop}%
\bibitem [{\citenamefont {Santos}\ and\ \citenamefont
  {Rigol}(2010{\natexlab{b}})}]{santos2010localization}%
  \BibitemOpen
  \bibfield  {author} {\bibinfo {author} {\bibfnamefont {L.~F.}\ \bibnamefont
  {Santos}}\ and\ \bibinfo {author} {\bibfnamefont {M.}~\bibnamefont {Rigol}},\
  }\href {\doibase 10.1103/PhysRevE.82.031130} {\bibfield  {journal} {\bibinfo
  {journal} {Phys. Rev. E}\ }\textbf {\bibinfo {volume} {82}},\ \bibinfo
  {pages} {031130} (\bibinfo {year} {2010}{\natexlab{b}})}\BibitemShut
  {NoStop}%
\bibitem [{\citenamefont {Biroli}\ \emph {et~al.}(2010)\citenamefont {Biroli},
  \citenamefont {Kollath},\ and\ \citenamefont {L\"auchli}}]{biroli2010effect}%
  \BibitemOpen
  \bibfield  {author} {\bibinfo {author} {\bibfnamefont {G.}~\bibnamefont
  {Biroli}}, \bibinfo {author} {\bibfnamefont {C.}~\bibnamefont {Kollath}}, \
  and\ \bibinfo {author} {\bibfnamefont {A.~M.}\ \bibnamefont {L\"auchli}},\
  }\href {\doibase 10.1103/PhysRevLett.105.250401} {\bibfield  {journal}
  {\bibinfo  {journal} {Phys. Rev. Lett.}\ }\textbf {\bibinfo {volume} {105}},\
  \bibinfo {pages} {250401} (\bibinfo {year} {2010})}\BibitemShut {NoStop}%
\bibitem [{\citenamefont {Genway}\ \emph {et~al.}(2012)\citenamefont {Genway},
  \citenamefont {Ho},\ and\ \citenamefont {Lee}}]{genway2012thermalization}%
  \BibitemOpen
  \bibfield  {author} {\bibinfo {author} {\bibfnamefont {S.}~\bibnamefont
  {Genway}}, \bibinfo {author} {\bibfnamefont {A.~F.}\ \bibnamefont {Ho}}, \
  and\ \bibinfo {author} {\bibfnamefont {D.~K.~K.}\ \bibnamefont {Lee}},\
  }\href {\doibase 10.1103/PhysRevA.86.023609} {\bibfield  {journal} {\bibinfo
  {journal} {Phys. Rev. A}\ }\textbf {\bibinfo {volume} {86}},\ \bibinfo
  {pages} {023609} (\bibinfo {year} {2012})}\BibitemShut {NoStop}%
\bibitem [{\citenamefont {Khatami}\ \emph {et~al.}(2012)\citenamefont
  {Khatami}, \citenamefont {Rigol}, \citenamefont {Rela\~no},\ and\
  \citenamefont {Garc\'{\i}a-Garc\'{\i}a}}]{khatami2012quantum}%
  \BibitemOpen
  \bibfield  {author} {\bibinfo {author} {\bibfnamefont {E.}~\bibnamefont
  {Khatami}}, \bibinfo {author} {\bibfnamefont {M.}~\bibnamefont {Rigol}},
  \bibinfo {author} {\bibfnamefont {A.}~\bibnamefont {Rela\~no}}, \ and\
  \bibinfo {author} {\bibfnamefont {A.~M.}\ \bibnamefont
  {Garc\'{\i}a-Garc\'{\i}a}},\ }\href {\doibase 10.1103/PhysRevE.85.050102}
  {\bibfield  {journal} {\bibinfo  {journal} {Phys. Rev. E}\ }\textbf {\bibinfo
  {volume} {85}},\ \bibinfo {pages} {050102(R)} (\bibinfo {year}
  {2012})}\BibitemShut {NoStop}%
\bibitem [{\citenamefont {Neuenhahn}\ and\ \citenamefont
  {Marquardt}(2012)}]{neuenhahn2012thermalization}%
  \BibitemOpen
  \bibfield  {author} {\bibinfo {author} {\bibfnamefont {C.}~\bibnamefont
  {Neuenhahn}}\ and\ \bibinfo {author} {\bibfnamefont {F.}~\bibnamefont
  {Marquardt}},\ }\href {\doibase 10.1103/PhysRevE.85.060101} {\bibfield
  {journal} {\bibinfo  {journal} {Phys. Rev. E}\ }\textbf {\bibinfo {volume}
  {85}},\ \bibinfo {pages} {060101(R)} (\bibinfo {year} {2012})}\BibitemShut
  {NoStop}%
\bibitem [{\citenamefont {Khatami}\ \emph {et~al.}(2013)\citenamefont
  {Khatami}, \citenamefont {Pupillo}, \citenamefont {Srednicki},\ and\
  \citenamefont {Rigol}}]{khatami2013fluctuation}%
  \BibitemOpen
  \bibfield  {author} {\bibinfo {author} {\bibfnamefont {E.}~\bibnamefont
  {Khatami}}, \bibinfo {author} {\bibfnamefont {G.}~\bibnamefont {Pupillo}},
  \bibinfo {author} {\bibfnamefont {M.}~\bibnamefont {Srednicki}}, \ and\
  \bibinfo {author} {\bibfnamefont {M.}~\bibnamefont {Rigol}},\ }\href
  {\doibase 10.1103/PhysRevLett.111.050403} {\bibfield  {journal} {\bibinfo
  {journal} {Phys. Rev. Lett.}\ }\textbf {\bibinfo {volume} {111}},\ \bibinfo
  {pages} {050403} (\bibinfo {year} {2013})}\BibitemShut {NoStop}%
\bibitem [{\citenamefont {Ikeda}\ \emph {et~al.}(2013)\citenamefont {Ikeda},
  \citenamefont {Watanabe},\ and\ \citenamefont {Ueda}}]{ikeda2013finite}%
  \BibitemOpen
  \bibfield  {author} {\bibinfo {author} {\bibfnamefont {T.~N.}\ \bibnamefont
  {Ikeda}}, \bibinfo {author} {\bibfnamefont {Y.}~\bibnamefont {Watanabe}}, \
  and\ \bibinfo {author} {\bibfnamefont {M.}~\bibnamefont {Ueda}},\ }\href
  {\doibase 10.1103/PhysRevE.87.012125} {\bibfield  {journal} {\bibinfo
  {journal} {Phys. Rev. E}\ }\textbf {\bibinfo {volume} {87}},\ \bibinfo
  {pages} {012125} (\bibinfo {year} {2013})}\BibitemShut {NoStop}%
\bibitem [{\citenamefont {Beugeling}\ \emph {et~al.}(2014)\citenamefont
  {Beugeling}, \citenamefont {Moessner},\ and\ \citenamefont
  {Haque}}]{beugeling2014finite}%
  \BibitemOpen
  \bibfield  {author} {\bibinfo {author} {\bibfnamefont {W.}~\bibnamefont
  {Beugeling}}, \bibinfo {author} {\bibfnamefont {R.}~\bibnamefont {Moessner}},
  \ and\ \bibinfo {author} {\bibfnamefont {M.}~\bibnamefont {Haque}},\ }\href
  {\doibase 10.1103/PhysRevE.89.042112} {\bibfield  {journal} {\bibinfo
  {journal} {Phys. Rev. E}\ }\textbf {\bibinfo {volume} {89}},\ \bibinfo
  {pages} {042112} (\bibinfo {year} {2014})}\BibitemShut {NoStop}%
\bibitem [{\citenamefont {Sorg}\ \emph {et~al.}(2014)\citenamefont {Sorg},
  \citenamefont {Vidmar}, \citenamefont {Pollet},\ and\ \citenamefont
  {Heidrich-Meisner}}]{sorg2014relaxation}%
  \BibitemOpen
  \bibfield  {author} {\bibinfo {author} {\bibfnamefont {S.}~\bibnamefont
  {Sorg}}, \bibinfo {author} {\bibfnamefont {L.}~\bibnamefont {Vidmar}},
  \bibinfo {author} {\bibfnamefont {L.}~\bibnamefont {Pollet}}, \ and\ \bibinfo
  {author} {\bibfnamefont {F.}~\bibnamefont {Heidrich-Meisner}},\ }\href
  {\doibase 10.1103/PhysRevA.90.033606} {\bibfield  {journal} {\bibinfo
  {journal} {Phys. Rev. A}\ }\textbf {\bibinfo {volume} {90}},\ \bibinfo
  {pages} {033606} (\bibinfo {year} {2014})}\BibitemShut {NoStop}%
\bibitem [{\citenamefont {Beugeling}\ \emph {et~al.}(2015)\citenamefont
  {Beugeling}, \citenamefont {Moessner},\ and\ \citenamefont
  {Haque}}]{beugeling2015off}%
  \BibitemOpen
  \bibfield  {author} {\bibinfo {author} {\bibfnamefont {W.}~\bibnamefont
  {Beugeling}}, \bibinfo {author} {\bibfnamefont {R.}~\bibnamefont {Moessner}},
  \ and\ \bibinfo {author} {\bibfnamefont {M.}~\bibnamefont {Haque}},\ }\href
  {\doibase 10.1103/PhysRevE.91.012144} {\bibfield  {journal} {\bibinfo
  {journal} {Phys. Rev. E}\ }\textbf {\bibinfo {volume} {91}},\ \bibinfo
  {pages} {012144} (\bibinfo {year} {2015})}\BibitemShut {NoStop}%
\bibitem [{\citenamefont {Mondaini}\ \emph {et~al.}(2016)\citenamefont
  {Mondaini}, \citenamefont {Fratus}, \citenamefont {Srednicki},\ and\
  \citenamefont {Rigol}}]{mondaini2016eigenstate}%
  \BibitemOpen
  \bibfield  {author} {\bibinfo {author} {\bibfnamefont {R.}~\bibnamefont
  {Mondaini}}, \bibinfo {author} {\bibfnamefont {K.~R.}\ \bibnamefont
  {Fratus}}, \bibinfo {author} {\bibfnamefont {M.}~\bibnamefont {Srednicki}}, \
  and\ \bibinfo {author} {\bibfnamefont {M.}~\bibnamefont {Rigol}},\ }\href
  {\doibase 10.1103/PhysRevE.93.032104} {\bibfield  {journal} {\bibinfo
  {journal} {Phys. Rev. E}\ }\textbf {\bibinfo {volume} {93}},\ \bibinfo
  {pages} {032104} (\bibinfo {year} {2016})}\BibitemShut {NoStop}%
\bibitem [{\citenamefont {Mondaini}\ and\ \citenamefont
  {Rigol}(2017)}]{mondaini2017eigenstate}%
  \BibitemOpen
  \bibfield  {author} {\bibinfo {author} {\bibfnamefont {R.}~\bibnamefont
  {Mondaini}}\ and\ \bibinfo {author} {\bibfnamefont {M.}~\bibnamefont
  {Rigol}},\ }\href {\doibase 10.1103/PhysRevE.96.012157} {\bibfield  {journal}
  {\bibinfo  {journal} {Phys. Rev. E}\ }\textbf {\bibinfo {volume} {96}},\
  \bibinfo {pages} {012157} (\bibinfo {year} {2017})}\BibitemShut {NoStop}%
\bibitem [{\citenamefont {Hikida}\ \emph {et~al.}(2018)\citenamefont {Hikida},
  \citenamefont {Kusuki},\ and\ \citenamefont
  {Takayanagi}}]{hikida2018eigenstate}%
  \BibitemOpen
  \bibfield  {author} {\bibinfo {author} {\bibfnamefont {Y.}~\bibnamefont
  {Hikida}}, \bibinfo {author} {\bibfnamefont {Y.}~\bibnamefont {Kusuki}}, \
  and\ \bibinfo {author} {\bibfnamefont {T.}~\bibnamefont {Takayanagi}},\
  }\href {\doibase 10.1103/PhysRevD.98.026003} {\bibfield  {journal} {\bibinfo
  {journal} {Phys. Rev. D}\ }\textbf {\bibinfo {volume} {98}},\ \bibinfo
  {pages} {026003} (\bibinfo {year} {2018})}\BibitemShut {NoStop}%
\bibitem [{\citenamefont {Gogolin}\ \emph {et~al.}(2011)\citenamefont
  {Gogolin}, \citenamefont {M\"uller},\ and\ \citenamefont
  {Eisert}}]{gogolin2011absence}%
  \BibitemOpen
  \bibfield  {author} {\bibinfo {author} {\bibfnamefont {C.}~\bibnamefont
  {Gogolin}}, \bibinfo {author} {\bibfnamefont {M.~P.}\ \bibnamefont
  {M\"uller}}, \ and\ \bibinfo {author} {\bibfnamefont {J.}~\bibnamefont
  {Eisert}},\ }\href {\doibase 10.1103/PhysRevLett.106.040401} {\bibfield
  {journal} {\bibinfo  {journal} {Phys. Rev. Lett.}\ }\textbf {\bibinfo
  {volume} {106}},\ \bibinfo {pages} {040401} (\bibinfo {year}
  {2011})}\BibitemShut {NoStop}%
\bibitem [{\citenamefont {Luitz}\ and\ \citenamefont
  {Bar~Lev}(2016)}]{luitz2016anomalous}%
  \BibitemOpen
  \bibfield  {author} {\bibinfo {author} {\bibfnamefont {D.~J.}\ \bibnamefont
  {Luitz}}\ and\ \bibinfo {author} {\bibfnamefont {Y.}~\bibnamefont
  {Bar~Lev}},\ }\href {\doibase 10.1103/PhysRevLett.117.170404} {\bibfield
  {journal} {\bibinfo  {journal} {Phys. Rev. Lett.}\ }\textbf {\bibinfo
  {volume} {117}},\ \bibinfo {pages} {170404} (\bibinfo {year}
  {2016})}\BibitemShut {NoStop}%
\bibitem [{\citenamefont {Garrison}\ and\ \citenamefont
  {Grover}(2018)}]{garrison2018does}%
  \BibitemOpen
  \bibfield  {author} {\bibinfo {author} {\bibfnamefont {J.~R.}\ \bibnamefont
  {Garrison}}\ and\ \bibinfo {author} {\bibfnamefont {T.}~\bibnamefont
  {Grover}},\ }\href {\doibase 10.1103/PhysRevX.8.021026} {\bibfield  {journal}
  {\bibinfo  {journal} {Phys. Rev. X}\ }\textbf {\bibinfo {volume} {8}},\
  \bibinfo {pages} {021026} (\bibinfo {year} {2018})}\BibitemShut {NoStop}%
\bibitem [{\citenamefont {Dymarsky}\ \emph {et~al.}(2018)\citenamefont
  {Dymarsky}, \citenamefont {Lashkari},\ and\ \citenamefont
  {Liu}}]{dymarsky2018subsystem}%
  \BibitemOpen
  \bibfield  {author} {\bibinfo {author} {\bibfnamefont {A.}~\bibnamefont
  {Dymarsky}}, \bibinfo {author} {\bibfnamefont {N.}~\bibnamefont {Lashkari}},
  \ and\ \bibinfo {author} {\bibfnamefont {H.}~\bibnamefont {Liu}},\ }\href
  {\doibase 10.1103/PhysRevE.97.012140} {\bibfield  {journal} {\bibinfo
  {journal} {Phys. Rev. E}\ }\textbf {\bibinfo {volume} {97}},\ \bibinfo
  {pages} {012140} (\bibinfo {year} {2018})}\BibitemShut {NoStop}%
\bibitem [{\citenamefont {Calabrese}\ and\ \citenamefont
  {Cardy}(2006)}]{calabrese2006time}%
  \BibitemOpen
  \bibfield  {author} {\bibinfo {author} {\bibfnamefont {P.}~\bibnamefont
  {Calabrese}}\ and\ \bibinfo {author} {\bibfnamefont {J.}~\bibnamefont
  {Cardy}},\ }\href {\doibase 10.1103/PhysRevLett.96.136801} {\bibfield
  {journal} {\bibinfo  {journal} {Phys. Rev. Lett.}\ }\textbf {\bibinfo
  {volume} {96}},\ \bibinfo {pages} {136801} (\bibinfo {year}
  {2006})}\BibitemShut {NoStop}%
\bibitem [{\citenamefont {Calabrese}\ and\ \citenamefont
  {Cardy}(2007{\natexlab{a}})}]{calabrese2007quantum}%
  \BibitemOpen
  \bibfield  {author} {\bibinfo {author} {\bibfnamefont {P.}~\bibnamefont
  {Calabrese}}\ and\ \bibinfo {author} {\bibfnamefont {J.}~\bibnamefont
  {Cardy}},\ }\href {\doibase 10.1088/1742-5468/2007/06/p06008} {\bibfield
  {journal} {\bibinfo  {journal} {Journal of Statistical Mechanics: Theory and
  Experiment}\ }\textbf {\bibinfo {volume} {2007}},\ \bibinfo {pages} {P06008}
  (\bibinfo {year} {2007}{\natexlab{a}})}\BibitemShut {NoStop}%
\bibitem [{\citenamefont {Sotiriadis}\ \emph {et~al.}(2009)\citenamefont
  {Sotiriadis}, \citenamefont {Calabrese},\ and\ \citenamefont
  {Cardy}}]{sotiriadis2009quantum}%
  \BibitemOpen
  \bibfield  {author} {\bibinfo {author} {\bibfnamefont {S.}~\bibnamefont
  {Sotiriadis}}, \bibinfo {author} {\bibfnamefont {P.}~\bibnamefont
  {Calabrese}}, \ and\ \bibinfo {author} {\bibfnamefont {J.}~\bibnamefont
  {Cardy}},\ }\href {\doibase 10.1209/0295-5075/87/20002} {\bibfield  {journal}
  {\bibinfo  {journal} {{EPL} (Europhysics Letters)}\ }\textbf {\bibinfo
  {volume} {87}},\ \bibinfo {pages} {20002} (\bibinfo {year}
  {2009})}\BibitemShut {NoStop}%
\bibitem [{\citenamefont {Rigol}\ and\ \citenamefont
  {Fitzpatrick}(2011)}]{rigol2011initial}%
  \BibitemOpen
  \bibfield  {author} {\bibinfo {author} {\bibfnamefont {M.}~\bibnamefont
  {Rigol}}\ and\ \bibinfo {author} {\bibfnamefont {M.}~\bibnamefont
  {Fitzpatrick}},\ }\href {\doibase 10.1103/PhysRevA.84.033640} {\bibfield
  {journal} {\bibinfo  {journal} {Phys. Rev. A}\ }\textbf {\bibinfo {volume}
  {84}},\ \bibinfo {pages} {033640} (\bibinfo {year} {2011})}\BibitemShut
  {NoStop}%
\bibitem [{\citenamefont {He}\ and\ \citenamefont
  {Rigol}(2012)}]{he2012initial}%
  \BibitemOpen
  \bibfield  {author} {\bibinfo {author} {\bibfnamefont {K.}~\bibnamefont
  {He}}\ and\ \bibinfo {author} {\bibfnamefont {M.}~\bibnamefont {Rigol}},\
  }\href {\doibase 10.1103/PhysRevA.85.063609} {\bibfield  {journal} {\bibinfo
  {journal} {Phys. Rev. A}\ }\textbf {\bibinfo {volume} {85}},\ \bibinfo
  {pages} {063609} (\bibinfo {year} {2012})}\BibitemShut {NoStop}%
\bibitem [{\citenamefont {He}\ and\ \citenamefont
  {Rigol}(2013)}]{he2013initial}%
  \BibitemOpen
  \bibfield  {author} {\bibinfo {author} {\bibfnamefont {K.}~\bibnamefont
  {He}}\ and\ \bibinfo {author} {\bibfnamefont {M.}~\bibnamefont {Rigol}},\
  }\href {\doibase 10.1103/PhysRevA.87.043615} {\bibfield  {journal} {\bibinfo
  {journal} {Phys. Rev. A}\ }\textbf {\bibinfo {volume} {87}},\ \bibinfo
  {pages} {043615} (\bibinfo {year} {2013})}\BibitemShut {NoStop}%
\bibitem [{\citenamefont {Mistakidis}\ \emph {et~al.}(2014)\citenamefont
  {Mistakidis}, \citenamefont {Cao},\ and\ \citenamefont
  {Schmelcher}}]{Mistakidis_2014}%
  \BibitemOpen
  \bibfield  {author} {\bibinfo {author} {\bibfnamefont {S.~I.}\ \bibnamefont
  {Mistakidis}}, \bibinfo {author} {\bibfnamefont {L.}~\bibnamefont {Cao}}, \
  and\ \bibinfo {author} {\bibfnamefont {P.}~\bibnamefont {Schmelcher}},\
  }\href {\doibase 10.1088/0953-4075/47/22/225303} {\bibfield  {journal}
  {\bibinfo  {journal} {Journal of Physics B: Atomic, Molecular and Optical
  Physics}\ }\textbf {\bibinfo {volume} {47}},\ \bibinfo {pages} {225303}
  (\bibinfo {year} {2014})}\BibitemShut {NoStop}%
\bibitem [{\citenamefont {Mistakidis}\ \emph {et~al.}(2015)\citenamefont
  {Mistakidis}, \citenamefont {Cao},\ and\ \citenamefont
  {Schmelcher}}]{PhysRevA.91.033611}%
  \BibitemOpen
  \bibfield  {author} {\bibinfo {author} {\bibfnamefont {S.~I.}\ \bibnamefont
  {Mistakidis}}, \bibinfo {author} {\bibfnamefont {L.}~\bibnamefont {Cao}}, \
  and\ \bibinfo {author} {\bibfnamefont {P.}~\bibnamefont {Schmelcher}},\
  }\href {\doibase 10.1103/PhysRevA.91.033611} {\bibfield  {journal} {\bibinfo
  {journal} {Phys. Rev. A}\ }\textbf {\bibinfo {volume} {91}},\ \bibinfo
  {pages} {033611} (\bibinfo {year} {2015})}\BibitemShut {NoStop}%
\bibitem [{\citenamefont {Cardy}(2016)}]{cardy2016quantum}%
  \BibitemOpen
  \bibfield  {author} {\bibinfo {author} {\bibfnamefont {J.}~\bibnamefont
  {Cardy}},\ }\href {\doibase 10.1088/1742-5468/2016/02/023103} {\bibfield
  {journal} {\bibinfo  {journal} {Journal of Statistical Mechanics: Theory and
  Experiment}\ }\textbf {\bibinfo {volume} {2016}},\ \bibinfo {pages} {023103}
  (\bibinfo {year} {2016})}\BibitemShut {NoStop}%
\bibitem [{\citenamefont {Neuhaus-Steinmetz}\ \emph {et~al.}(2017)\citenamefont
  {Neuhaus-Steinmetz}, \citenamefont {Mistakidis},\ and\ \citenamefont
  {Schmelcher}}]{PhysRevA.95.053610}%
  \BibitemOpen
  \bibfield  {author} {\bibinfo {author} {\bibfnamefont {J.}~\bibnamefont
  {Neuhaus-Steinmetz}}, \bibinfo {author} {\bibfnamefont {S.~I.}\ \bibnamefont
  {Mistakidis}}, \ and\ \bibinfo {author} {\bibfnamefont {P.}~\bibnamefont
  {Schmelcher}},\ }\href {\doibase 10.1103/PhysRevA.95.053610} {\bibfield
  {journal} {\bibinfo  {journal} {Phys. Rev. A}\ }\textbf {\bibinfo {volume}
  {95}},\ \bibinfo {pages} {053610} (\bibinfo {year} {2017})}\BibitemShut
  {NoStop}%
\bibitem [{\citenamefont {Mistakidis}\ and\ \citenamefont
  {Schmelcher}(2017)}]{PhysRevA.95.013625}%
  \BibitemOpen
  \bibfield  {author} {\bibinfo {author} {\bibfnamefont {S.~I.}\ \bibnamefont
  {Mistakidis}}\ and\ \bibinfo {author} {\bibfnamefont {P.}~\bibnamefont
  {Schmelcher}},\ }\href {\doibase 10.1103/PhysRevA.95.013625} {\bibfield
  {journal} {\bibinfo  {journal} {Phys. Rev. A}\ }\textbf {\bibinfo {volume}
  {95}},\ \bibinfo {pages} {013625} (\bibinfo {year} {2017})}\BibitemShut
  {NoStop}%
\bibitem [{\citenamefont {Pla{\ss}mann}\ \emph {et~al.}(2018)\citenamefont
  {Pla{\ss}mann}, \citenamefont {Mistakidis},\ and\ \citenamefont
  {Schmelcher}}]{Pla_mann_2018}%
  \BibitemOpen
  \bibfield  {author} {\bibinfo {author} {\bibfnamefont {T.}~\bibnamefont
  {Pla{\ss}mann}}, \bibinfo {author} {\bibfnamefont {S.~I.}\ \bibnamefont
  {Mistakidis}}, \ and\ \bibinfo {author} {\bibfnamefont {P.}~\bibnamefont
  {Schmelcher}},\ }\href {\doibase 10.1088/1361-6455/aae57a} {\bibfield
  {journal} {\bibinfo  {journal} {Journal of Physics B: Atomic, Molecular and
  Optical Physics}\ }\textbf {\bibinfo {volume} {51}},\ \bibinfo {pages}
  {225001} (\bibinfo {year} {2018})}\BibitemShut {NoStop}%
\bibitem [{\citenamefont {Mistakidis}\ \emph {et~al.}(2018)\citenamefont
  {Mistakidis}, \citenamefont {Koutentakis},\ and\ \citenamefont
  {Schmelcher}}]{MISTAKIDIS2018106}%
  \BibitemOpen
  \bibfield  {author} {\bibinfo {author} {\bibfnamefont {S.}~\bibnamefont
  {Mistakidis}}, \bibinfo {author} {\bibfnamefont {G.}~\bibnamefont
  {Koutentakis}}, \ and\ \bibinfo {author} {\bibfnamefont {P.}~\bibnamefont
  {Schmelcher}},\ }\href {\doibase
  https://doi.org/10.1016/j.chemphys.2017.11.022} {\bibfield  {journal}
  {\bibinfo  {journal} {Chemical Physics}\ }\textbf {\bibinfo {volume} {509}},\
  \bibinfo {pages} {106 } (\bibinfo {year} {2018})},\ \bibinfo {note}
  {high-dimensional quantum dynamics (on the occasion of the 70th birthday of
  Hans-Dieter Meyer)}\BibitemShut {NoStop}%
\bibitem [{\citenamefont {Calabrese}\ and\ \citenamefont
  {Cardy}(2007{\natexlab{b}})}]{calabrese2007entanglement}%
  \BibitemOpen
  \bibfield  {author} {\bibinfo {author} {\bibfnamefont {P.}~\bibnamefont
  {Calabrese}}\ and\ \bibinfo {author} {\bibfnamefont {J.}~\bibnamefont
  {Cardy}},\ }\href {\doibase 10.1088/1742-5468/2007/10/p10004} {\bibfield
  {journal} {\bibinfo  {journal} {Journal of Statistical Mechanics: Theory and
  Experiment}\ }\textbf {\bibinfo {volume} {2007}},\ \bibinfo {pages} {P10004}
  (\bibinfo {year} {2007}{\natexlab{b}})}\BibitemShut {NoStop}%
\bibitem [{\citenamefont {Eisler}\ and\ \citenamefont
  {Peschel}(2007)}]{eisler2007evolution}%
  \BibitemOpen
  \bibfield  {author} {\bibinfo {author} {\bibfnamefont {V.}~\bibnamefont
  {Eisler}}\ and\ \bibinfo {author} {\bibfnamefont {I.}~\bibnamefont
  {Peschel}},\ }\href {\doibase 10.1088/1742-5468/2007/06/p06005} {\bibfield
  {journal} {\bibinfo  {journal} {Journal of Statistical Mechanics: Theory and
  Experiment}\ }\textbf {\bibinfo {volume} {2007}},\ \bibinfo {pages} {P06005}
  (\bibinfo {year} {2007})}\BibitemShut {NoStop}%
\bibitem [{\citenamefont {Rigol}\ and\ \citenamefont
  {Muramatsu}(2004)}]{rigol2004emergence}%
  \BibitemOpen
  \bibfield  {author} {\bibinfo {author} {\bibfnamefont {M.}~\bibnamefont
  {Rigol}}\ and\ \bibinfo {author} {\bibfnamefont {A.}~\bibnamefont
  {Muramatsu}},\ }\href {\doibase 10.1103/PhysRevLett.93.230404} {\bibfield
  {journal} {\bibinfo  {journal} {Phys. Rev. Lett.}\ }\textbf {\bibinfo
  {volume} {93}},\ \bibinfo {pages} {230404} (\bibinfo {year}
  {2004})}\BibitemShut {NoStop}%
\bibitem [{\citenamefont {Minguzzi}\ and\ \citenamefont
  {Gangardt}(2005)}]{minguzzi2005exact}%
  \BibitemOpen
  \bibfield  {author} {\bibinfo {author} {\bibfnamefont {A.}~\bibnamefont
  {Minguzzi}}\ and\ \bibinfo {author} {\bibfnamefont {D.~M.}\ \bibnamefont
  {Gangardt}},\ }\href {\doibase 10.1103/PhysRevLett.94.240404} {\bibfield
  {journal} {\bibinfo  {journal} {Phys. Rev. Lett.}\ }\textbf {\bibinfo
  {volume} {94}},\ \bibinfo {pages} {240404} (\bibinfo {year}
  {2005})}\BibitemShut {NoStop}%
\bibitem [{\citenamefont {Rigol}\ and\ \citenamefont
  {Muramatsu}(2005{\natexlab{a}})}]{rigol2005fermionization}%
  \BibitemOpen
  \bibfield  {author} {\bibinfo {author} {\bibfnamefont {M.}~\bibnamefont
  {Rigol}}\ and\ \bibinfo {author} {\bibfnamefont {A.}~\bibnamefont
  {Muramatsu}},\ }\href {\doibase 10.1103/PhysRevLett.94.240403} {\bibfield
  {journal} {\bibinfo  {journal} {Phys. Rev. Lett.}\ }\textbf {\bibinfo
  {volume} {94}},\ \bibinfo {pages} {240403} (\bibinfo {year}
  {2005}{\natexlab{a}})}\BibitemShut {NoStop}%
\bibitem [{\citenamefont {Rigol}\ and\ \citenamefont
  {Muramatsu}(2005{\natexlab{b}})}]{rigol2005free}%
  \BibitemOpen
  \bibfield  {author} {\bibinfo {author} {\bibfnamefont {M.}~\bibnamefont
  {Rigol}}\ and\ \bibinfo {author} {\bibfnamefont {A.}~\bibnamefont
  {Muramatsu}},\ }\href {\doibase 10.1142/S0217984905008876} {\bibfield
  {journal} {\bibinfo  {journal} {Modern Physics Letters B}\ }\textbf {\bibinfo
  {volume} {19}},\ \bibinfo {pages} {861} (\bibinfo {year}
  {2005}{\natexlab{b}})},\ \Eprint
  {http://arxiv.org/abs/https://doi.org/10.1142/S0217984905008876}
  {https://doi.org/10.1142/S0217984905008876} \BibitemShut {NoStop}%
\bibitem [{\citenamefont {Camalet}(2008)}]{camalet2008joule}%
  \BibitemOpen
  \bibfield  {author} {\bibinfo {author} {\bibfnamefont {S.}~\bibnamefont
  {Camalet}},\ }\href {\doibase 10.1103/PhysRevLett.100.180401} {\bibfield
  {journal} {\bibinfo  {journal} {Phys. Rev. Lett.}\ }\textbf {\bibinfo
  {volume} {100}},\ \bibinfo {pages} {180401} (\bibinfo {year}
  {2008})}\BibitemShut {NoStop}%
\bibitem [{\citenamefont {Heidrich-Meisner}\ \emph {et~al.}(2009)\citenamefont
  {Heidrich-Meisner}, \citenamefont {Manmana}, \citenamefont {Rigol},
  \citenamefont {Muramatsu}, \citenamefont {Feiguin},\ and\ \citenamefont
  {Dagotto}}]{heidrich2009quantum}%
  \BibitemOpen
  \bibfield  {author} {\bibinfo {author} {\bibfnamefont {F.}~\bibnamefont
  {Heidrich-Meisner}}, \bibinfo {author} {\bibfnamefont {S.~R.}\ \bibnamefont
  {Manmana}}, \bibinfo {author} {\bibfnamefont {M.}~\bibnamefont {Rigol}},
  \bibinfo {author} {\bibfnamefont {A.}~\bibnamefont {Muramatsu}}, \bibinfo
  {author} {\bibfnamefont {A.~E.}\ \bibnamefont {Feiguin}}, \ and\ \bibinfo
  {author} {\bibfnamefont {E.}~\bibnamefont {Dagotto}},\ }\href {\doibase
  10.1103/PhysRevA.80.041603} {\bibfield  {journal} {\bibinfo  {journal} {Phys.
  Rev. A}\ }\textbf {\bibinfo {volume} {80}},\ \bibinfo {pages} {041603(R)}
  (\bibinfo {year} {2009})}\BibitemShut {NoStop}%
\bibitem [{\citenamefont {Kajala}\ \emph {et~al.}(2011)\citenamefont {Kajala},
  \citenamefont {Massel},\ and\ \citenamefont
  {T\"orm\"a}}]{kajala2011expansion}%
  \BibitemOpen
  \bibfield  {author} {\bibinfo {author} {\bibfnamefont {J.}~\bibnamefont
  {Kajala}}, \bibinfo {author} {\bibfnamefont {F.}~\bibnamefont {Massel}}, \
  and\ \bibinfo {author} {\bibfnamefont {P.}~\bibnamefont {T\"orm\"a}},\ }\href
  {\doibase 10.1103/PhysRevLett.106.206401} {\bibfield  {journal} {\bibinfo
  {journal} {Phys. Rev. Lett.}\ }\textbf {\bibinfo {volume} {106}},\ \bibinfo
  {pages} {206401} (\bibinfo {year} {2011})}\BibitemShut {NoStop}%
\bibitem [{\citenamefont {Vidmar}\ \emph {et~al.}(2013)\citenamefont {Vidmar},
  \citenamefont {Langer}, \citenamefont {McCulloch}, \citenamefont {Schneider},
  \citenamefont {Schollw\"ock},\ and\ \citenamefont
  {Heidrich-Meisner}}]{vidmar2013sudden}%
  \BibitemOpen
  \bibfield  {author} {\bibinfo {author} {\bibfnamefont {L.}~\bibnamefont
  {Vidmar}}, \bibinfo {author} {\bibfnamefont {S.}~\bibnamefont {Langer}},
  \bibinfo {author} {\bibfnamefont {I.~P.}\ \bibnamefont {McCulloch}}, \bibinfo
  {author} {\bibfnamefont {U.}~\bibnamefont {Schneider}}, \bibinfo {author}
  {\bibfnamefont {U.}~\bibnamefont {Schollw\"ock}}, \ and\ \bibinfo {author}
  {\bibfnamefont {F.}~\bibnamefont {Heidrich-Meisner}},\ }\href {\doibase
  10.1103/PhysRevB.88.235117} {\bibfield  {journal} {\bibinfo  {journal} {Phys.
  Rev. B}\ }\textbf {\bibinfo {volume} {88}},\ \bibinfo {pages} {235117}
  (\bibinfo {year} {2013})}\BibitemShut {NoStop}%
\bibitem [{\citenamefont {Ronzheimer}\ \emph {et~al.}(2013)\citenamefont
  {Ronzheimer}, \citenamefont {Schreiber}, \citenamefont {Braun}, \citenamefont
  {Hodgman}, \citenamefont {Langer}, \citenamefont {McCulloch}, \citenamefont
  {Heidrich-Meisner}, \citenamefont {Bloch},\ and\ \citenamefont
  {Schneider}}]{ronzheimer2013expansion}%
  \BibitemOpen
  \bibfield  {author} {\bibinfo {author} {\bibfnamefont {J.~P.}\ \bibnamefont
  {Ronzheimer}}, \bibinfo {author} {\bibfnamefont {M.}~\bibnamefont
  {Schreiber}}, \bibinfo {author} {\bibfnamefont {S.}~\bibnamefont {Braun}},
  \bibinfo {author} {\bibfnamefont {S.~S.}\ \bibnamefont {Hodgman}}, \bibinfo
  {author} {\bibfnamefont {S.}~\bibnamefont {Langer}}, \bibinfo {author}
  {\bibfnamefont {I.~P.}\ \bibnamefont {McCulloch}}, \bibinfo {author}
  {\bibfnamefont {F.}~\bibnamefont {Heidrich-Meisner}}, \bibinfo {author}
  {\bibfnamefont {I.}~\bibnamefont {Bloch}}, \ and\ \bibinfo {author}
  {\bibfnamefont {U.}~\bibnamefont {Schneider}},\ }\href {\doibase
  10.1103/PhysRevLett.110.205301} {\bibfield  {journal} {\bibinfo  {journal}
  {Phys. Rev. Lett.}\ }\textbf {\bibinfo {volume} {110}},\ \bibinfo {pages}
  {205301} (\bibinfo {year} {2013})}\BibitemShut {NoStop}%
\bibitem [{\citenamefont {Xia}\ \emph {et~al.}(2015)\citenamefont {Xia},
  \citenamefont {Zundel}, \citenamefont {Carrasquilla}, \citenamefont
  {Reinhard}, \citenamefont {Wilson}, \citenamefont {Rigol},\ and\
  \citenamefont {Weiss}}]{xia2015quantum}%
  \BibitemOpen
  \bibfield  {author} {\bibinfo {author} {\bibfnamefont {L.}~\bibnamefont
  {Xia}}, \bibinfo {author} {\bibfnamefont {L.~A.}\ \bibnamefont {Zundel}},
  \bibinfo {author} {\bibfnamefont {J.}~\bibnamefont {Carrasquilla}}, \bibinfo
  {author} {\bibfnamefont {A.}~\bibnamefont {Reinhard}}, \bibinfo {author}
  {\bibfnamefont {J.~M.}\ \bibnamefont {Wilson}}, \bibinfo {author}
  {\bibfnamefont {M.}~\bibnamefont {Rigol}}, \ and\ \bibinfo {author}
  {\bibfnamefont {D.~S.}\ \bibnamefont {Weiss}},\ }\href
  {https://doi.org/10.1038/nphys3244 http://10.0.4.14/nphys3244
  https://www.nature.com/articles/nphys3244{\#}supplementary-information}
  {\bibfield  {journal} {\bibinfo  {journal} {Nature Physics}\ }\textbf
  {\bibinfo {volume} {11}},\ \bibinfo {pages} {316} (\bibinfo {year}
  {2015})}\BibitemShut {NoStop}%
\bibitem [{\citenamefont {Campbell}\ \emph {et~al.}(2015)\citenamefont
  {Campbell}, \citenamefont {Gangardt},\ and\ \citenamefont
  {Kheruntsyan}}]{campbell2015sudden}%
  \BibitemOpen
  \bibfield  {author} {\bibinfo {author} {\bibfnamefont {A.~S.}\ \bibnamefont
  {Campbell}}, \bibinfo {author} {\bibfnamefont {D.~M.}\ \bibnamefont
  {Gangardt}}, \ and\ \bibinfo {author} {\bibfnamefont {K.~V.}\ \bibnamefont
  {Kheruntsyan}},\ }\href {\doibase 10.1103/PhysRevLett.114.125302} {\bibfield
  {journal} {\bibinfo  {journal} {Phys. Rev. Lett.}\ }\textbf {\bibinfo
  {volume} {114}},\ \bibinfo {pages} {125302} (\bibinfo {year}
  {2015})}\BibitemShut {NoStop}%
\bibitem [{\citenamefont {Vidmar}\ \emph {et~al.}(2015)\citenamefont {Vidmar},
  \citenamefont {Ronzheimer}, \citenamefont {Schreiber}, \citenamefont {Braun},
  \citenamefont {Hodgman}, \citenamefont {Langer}, \citenamefont
  {Heidrich-Meisner}, \citenamefont {Bloch},\ and\ \citenamefont
  {Schneider}}]{vidmar2015dynamical}%
  \BibitemOpen
  \bibfield  {author} {\bibinfo {author} {\bibfnamefont {L.}~\bibnamefont
  {Vidmar}}, \bibinfo {author} {\bibfnamefont {J.~P.}\ \bibnamefont
  {Ronzheimer}}, \bibinfo {author} {\bibfnamefont {M.}~\bibnamefont
  {Schreiber}}, \bibinfo {author} {\bibfnamefont {S.}~\bibnamefont {Braun}},
  \bibinfo {author} {\bibfnamefont {S.~S.}\ \bibnamefont {Hodgman}}, \bibinfo
  {author} {\bibfnamefont {S.}~\bibnamefont {Langer}}, \bibinfo {author}
  {\bibfnamefont {F.}~\bibnamefont {Heidrich-Meisner}}, \bibinfo {author}
  {\bibfnamefont {I.}~\bibnamefont {Bloch}}, \ and\ \bibinfo {author}
  {\bibfnamefont {U.}~\bibnamefont {Schneider}},\ }\href {\doibase
  10.1103/PhysRevLett.115.175301} {\bibfield  {journal} {\bibinfo  {journal}
  {Phys. Rev. Lett.}\ }\textbf {\bibinfo {volume} {115}},\ \bibinfo {pages}
  {175301} (\bibinfo {year} {2015})}\BibitemShut {NoStop}%
\bibitem [{\citenamefont {Vidmar}\ \emph
  {et~al.}(2017{\natexlab{a}})\citenamefont {Vidmar}, \citenamefont {Iyer},\
  and\ \citenamefont {Rigol}}]{vidmar2017emergent}%
  \BibitemOpen
  \bibfield  {author} {\bibinfo {author} {\bibfnamefont {L.}~\bibnamefont
  {Vidmar}}, \bibinfo {author} {\bibfnamefont {D.}~\bibnamefont {Iyer}}, \ and\
  \bibinfo {author} {\bibfnamefont {M.}~\bibnamefont {Rigol}},\ }\href
  {\doibase 10.1103/PhysRevX.7.021012} {\bibfield  {journal} {\bibinfo
  {journal} {Phys. Rev. X}\ }\textbf {\bibinfo {volume} {7}},\ \bibinfo {pages}
  {021012} (\bibinfo {year} {2017}{\natexlab{a}})}\BibitemShut {NoStop}%
\bibitem [{\citenamefont {Vidmar}\ \emph
  {et~al.}(2017{\natexlab{b}})\citenamefont {Vidmar}, \citenamefont {Xu},\ and\
  \citenamefont {Rigol}}]{vidmar2017emergentpra}%
  \BibitemOpen
  \bibfield  {author} {\bibinfo {author} {\bibfnamefont {L.}~\bibnamefont
  {Vidmar}}, \bibinfo {author} {\bibfnamefont {W.}~\bibnamefont {Xu}}, \ and\
  \bibinfo {author} {\bibfnamefont {M.}~\bibnamefont {Rigol}},\ }\href
  {\doibase 10.1103/PhysRevA.96.013608} {\bibfield  {journal} {\bibinfo
  {journal} {Phys. Rev. A}\ }\textbf {\bibinfo {volume} {96}},\ \bibinfo
  {pages} {013608} (\bibinfo {year} {2017}{\natexlab{b}})}\BibitemShut
  {NoStop}%
\bibitem [{\citenamefont {Xu}\ and\ \citenamefont
  {Rigol}(2017)}]{xu2017expansion}%
  \BibitemOpen
  \bibfield  {author} {\bibinfo {author} {\bibfnamefont {W.}~\bibnamefont
  {Xu}}\ and\ \bibinfo {author} {\bibfnamefont {M.}~\bibnamefont {Rigol}},\
  }\href {\doibase 10.1103/PhysRevA.95.033617} {\bibfield  {journal} {\bibinfo
  {journal} {Phys. Rev. A}\ }\textbf {\bibinfo {volume} {95}},\ \bibinfo
  {pages} {033617} (\bibinfo {year} {2017})}\BibitemShut {NoStop}%
\bibitem [{\citenamefont {Herbrych}\ \emph {et~al.}(2017)\citenamefont
  {Herbrych}, \citenamefont {Feiguin}, \citenamefont {Dagotto},\ and\
  \citenamefont {Heidrich-Meisner}}]{herbrych2017efficiency}%
  \BibitemOpen
  \bibfield  {author} {\bibinfo {author} {\bibfnamefont {J.}~\bibnamefont
  {Herbrych}}, \bibinfo {author} {\bibfnamefont {A.~E.}\ \bibnamefont
  {Feiguin}}, \bibinfo {author} {\bibfnamefont {E.}~\bibnamefont {Dagotto}}, \
  and\ \bibinfo {author} {\bibfnamefont {F.}~\bibnamefont {Heidrich-Meisner}},\
  }\href {\doibase 10.1103/PhysRevA.96.033617} {\bibfield  {journal} {\bibinfo
  {journal} {Phys. Rev. A}\ }\textbf {\bibinfo {volume} {96}},\ \bibinfo
  {pages} {033617} (\bibinfo {year} {2017})}\BibitemShut {NoStop}%
\bibitem [{\citenamefont {Koutentakis}\ \emph {et~al.}(2017)\citenamefont
  {Koutentakis}, \citenamefont {Mistakidis},\ and\ \citenamefont
  {Schmelcher}}]{PhysRevA.95.013617}%
  \BibitemOpen
  \bibfield  {author} {\bibinfo {author} {\bibfnamefont {G.~M.}\ \bibnamefont
  {Koutentakis}}, \bibinfo {author} {\bibfnamefont {S.~I.}\ \bibnamefont
  {Mistakidis}}, \ and\ \bibinfo {author} {\bibfnamefont {P.}~\bibnamefont
  {Schmelcher}},\ }\href {\doibase 10.1103/PhysRevA.95.013617} {\bibfield
  {journal} {\bibinfo  {journal} {Phys. Rev. A}\ }\textbf {\bibinfo {volume}
  {95}},\ \bibinfo {pages} {013617} (\bibinfo {year} {2017})}\BibitemShut
  {NoStop}%
\bibitem [{\citenamefont {{Noh}}\ \emph {et~al.}(2018)\citenamefont {{Noh}},
  \citenamefont {{Iyoda}},\ and\ \citenamefont {{Sagawa}}}]{noh2018heating}%
  \BibitemOpen
  \bibfield  {author} {\bibinfo {author} {\bibfnamefont {J.~D.}\ \bibnamefont
  {{Noh}}}, \bibinfo {author} {\bibfnamefont {E.}~\bibnamefont {{Iyoda}}}, \
  and\ \bibinfo {author} {\bibfnamefont {T.}~\bibnamefont {{Sagawa}}},\
  }\href@noop {} {\bibfield  {journal} {\bibinfo  {journal} {arXiv e-prints}\
  ,\ \bibinfo {eid} {arXiv:1811.10051}} (\bibinfo {year} {2018})},\ \Eprint
  {http://arxiv.org/abs/1811.10051} {arXiv:1811.10051 [cond-mat.stat-mech]}
  \BibitemShut {NoStop}%
\bibitem [{\citenamefont {Scherg}\ \emph {et~al.}(2018)\citenamefont {Scherg},
  \citenamefont {Kohlert}, \citenamefont {Herbrych}, \citenamefont {Stolpp},
  \citenamefont {Bordia}, \citenamefont {Schneider}, \citenamefont
  {Heidrich-Meisner}, \citenamefont {Bloch},\ and\ \citenamefont
  {Aidelsburger}}]{scherg2018nonequilibrium}%
  \BibitemOpen
  \bibfield  {author} {\bibinfo {author} {\bibfnamefont {S.}~\bibnamefont
  {Scherg}}, \bibinfo {author} {\bibfnamefont {T.}~\bibnamefont {Kohlert}},
  \bibinfo {author} {\bibfnamefont {J.}~\bibnamefont {Herbrych}}, \bibinfo
  {author} {\bibfnamefont {J.}~\bibnamefont {Stolpp}}, \bibinfo {author}
  {\bibfnamefont {P.}~\bibnamefont {Bordia}}, \bibinfo {author} {\bibfnamefont
  {U.}~\bibnamefont {Schneider}}, \bibinfo {author} {\bibfnamefont
  {F.}~\bibnamefont {Heidrich-Meisner}}, \bibinfo {author} {\bibfnamefont
  {I.}~\bibnamefont {Bloch}}, \ and\ \bibinfo {author} {\bibfnamefont
  {M.}~\bibnamefont {Aidelsburger}},\ }\href {\doibase
  10.1103/PhysRevLett.121.130402} {\bibfield  {journal} {\bibinfo  {journal}
  {Phys. Rev. Lett.}\ }\textbf {\bibinfo {volume} {121}},\ \bibinfo {pages}
  {130402} (\bibinfo {year} {2018})}\BibitemShut {NoStop}%
\bibitem [{\citenamefont {Siegl}\ \emph {et~al.}(2018)\citenamefont {Siegl},
  \citenamefont {Mistakidis},\ and\ \citenamefont
  {Schmelcher}}]{PhysRevA.97.053626}%
  \BibitemOpen
  \bibfield  {author} {\bibinfo {author} {\bibfnamefont {P.}~\bibnamefont
  {Siegl}}, \bibinfo {author} {\bibfnamefont {S.~I.}\ \bibnamefont
  {Mistakidis}}, \ and\ \bibinfo {author} {\bibfnamefont {P.}~\bibnamefont
  {Schmelcher}},\ }\href {\doibase 10.1103/PhysRevA.97.053626} {\bibfield
  {journal} {\bibinfo  {journal} {Phys. Rev. A}\ }\textbf {\bibinfo {volume}
  {97}},\ \bibinfo {pages} {053626} (\bibinfo {year} {2018})}\BibitemShut
  {NoStop}%
\bibitem [{\citenamefont {Kaminishi}\ \emph {et~al.}(2015)\citenamefont
  {Kaminishi}, \citenamefont {Sato},\ and\ \citenamefont
  {Deguchi}}]{kaminishi2015recurrence}%
  \BibitemOpen
  \bibfield  {author} {\bibinfo {author} {\bibfnamefont {E.}~\bibnamefont
  {Kaminishi}}, \bibinfo {author} {\bibfnamefont {J.}~\bibnamefont {Sato}}, \
  and\ \bibinfo {author} {\bibfnamefont {T.}~\bibnamefont {Deguchi}},\ }\href
  {\doibase 10.7566/JPSJ.84.064002} {\bibfield  {journal} {\bibinfo  {journal}
  {Journal of the Physical Society of Japan}\ }\textbf {\bibinfo {volume}
  {84}},\ \bibinfo {pages} {064002} (\bibinfo {year} {2015})},\ \Eprint
  {http://arxiv.org/abs/https://doi.org/10.7566/JPSJ.84.064002}
  {https://doi.org/10.7566/JPSJ.84.064002} \BibitemShut {NoStop}%
\bibitem [{\citenamefont {Bocchieri}\ and\ \citenamefont
  {Loinger}(1957)}]{bocchieri1957quantum}%
  \BibitemOpen
  \bibfield  {author} {\bibinfo {author} {\bibfnamefont {P.}~\bibnamefont
  {Bocchieri}}\ and\ \bibinfo {author} {\bibfnamefont {A.}~\bibnamefont
  {Loinger}},\ }\href {\doibase 10.1103/PhysRev.107.337} {\bibfield  {journal}
  {\bibinfo  {journal} {Phys. Rev.}\ }\textbf {\bibinfo {volume} {107}},\
  \bibinfo {pages} {337} (\bibinfo {year} {1957})}\BibitemShut {NoStop}%
\bibitem [{\citenamefont {Mori}\ \emph {et~al.}(2018)\citenamefont {Mori},
  \citenamefont {Ikeda}, \citenamefont {Kaminishi},\ and\ \citenamefont
  {Ueda}}]{mori2018thermalization}%
  \BibitemOpen
  \bibfield  {author} {\bibinfo {author} {\bibfnamefont {T.}~\bibnamefont
  {Mori}}, \bibinfo {author} {\bibfnamefont {T.~N.}\ \bibnamefont {Ikeda}},
  \bibinfo {author} {\bibfnamefont {E.}~\bibnamefont {Kaminishi}}, \ and\
  \bibinfo {author} {\bibfnamefont {M.}~\bibnamefont {Ueda}},\ }\href {\doibase
  10.1088/1361-6455/aabcdf} {\bibfield  {journal} {\bibinfo  {journal} {Journal
  of Physics B: Atomic, Molecular and Optical Physics}\ }\textbf {\bibinfo
  {volume} {51}},\ \bibinfo {pages} {112001} (\bibinfo {year}
  {2018})}\BibitemShut {NoStop}%
\bibitem [{\citenamefont {Kollath}\ \emph {et~al.}(2010)\citenamefont
  {Kollath}, \citenamefont {Roux}, \citenamefont {Biroli},\ and\ \citenamefont
  {Läuchli}}]{kollath2010statistical}%
  \BibitemOpen
  \bibfield  {author} {\bibinfo {author} {\bibfnamefont {C.}~\bibnamefont
  {Kollath}}, \bibinfo {author} {\bibfnamefont {G.}~\bibnamefont {Roux}},
  \bibinfo {author} {\bibfnamefont {G.}~\bibnamefont {Biroli}}, \ and\ \bibinfo
  {author} {\bibfnamefont {A.~M.}\ \bibnamefont {Läuchli}},\ }\href
  {\doibase 10.1088/1742-5468/2010/08/p08011} {\bibfield  {journal} {\bibinfo
  {journal} {Journal of Statistical Mechanics: Theory and Experiment}\ }\textbf
  {\bibinfo {volume} {2010}},\ \bibinfo {pages} {P08011} (\bibinfo {year}
  {2010})}\BibitemShut {NoStop}%
\bibitem [{\citenamefont {Bohigas}\ \emph {et~al.}(1984)\citenamefont
  {Bohigas}, \citenamefont {Giannoni},\ and\ \citenamefont
  {Schmit}}]{bohigas1984characterization}%
  \BibitemOpen
  \bibfield  {author} {\bibinfo {author} {\bibfnamefont {O.}~\bibnamefont
  {Bohigas}}, \bibinfo {author} {\bibfnamefont {M.~J.}\ \bibnamefont
  {Giannoni}}, \ and\ \bibinfo {author} {\bibfnamefont {C.}~\bibnamefont
  {Schmit}},\ }\href {\doibase 10.1103/PhysRevLett.52.1} {\bibfield  {journal}
  {\bibinfo  {journal} {Phys. Rev. Lett.}\ }\textbf {\bibinfo {volume} {52}},\
  \bibinfo {pages} {1} (\bibinfo {year} {1984})}\BibitemShut {NoStop}%
\bibitem [{\citenamefont {D'Alessio}\ \emph {et~al.}(2016)\citenamefont
  {D'Alessio}, \citenamefont {Kafri}, \citenamefont {Polkovnikov},\ and\
  \citenamefont {Rigol}}]{dalessio2016}%
  \BibitemOpen
  \bibfield  {author} {\bibinfo {author} {\bibfnamefont {L.}~\bibnamefont
  {D'Alessio}}, \bibinfo {author} {\bibfnamefont {Y.}~\bibnamefont {Kafri}},
  \bibinfo {author} {\bibfnamefont {A.}~\bibnamefont {Polkovnikov}}, \ and\
  \bibinfo {author} {\bibfnamefont {M.}~\bibnamefont {Rigol}},\ }\href
  {\doibase 10.1080/00018732.2016.1198134} {\bibfield  {journal} {\bibinfo
  {journal} {Advances in Physics}\ }\textbf {\bibinfo {volume} {65}},\ \bibinfo
  {pages} {239} (\bibinfo {year} {2016})},\ \Eprint
  {http://arxiv.org/abs/https://doi.org/10.1080/00018732.2016.1198134}
  {https://doi.org/10.1080/00018732.2016.1198134} \BibitemShut {NoStop}%
\bibitem [{\citenamefont {Holzhey}\ \emph {et~al.}(1994)\citenamefont
  {Holzhey}, \citenamefont {Larsen},\ and\ \citenamefont
  {Wilczek}}]{holzhey1994geometric}%
  \BibitemOpen
  \bibfield  {author} {\bibinfo {author} {\bibfnamefont {C.}~\bibnamefont
  {Holzhey}}, \bibinfo {author} {\bibfnamefont {F.}~\bibnamefont {Larsen}}, \
  and\ \bibinfo {author} {\bibfnamefont {F.}~\bibnamefont {Wilczek}},\ }\href
  {\doibase https://doi.org/10.1016/0550-3213(94)90402-2} {\bibfield  {journal}
  {\bibinfo  {journal} {Nuclear Physics B}\ }\textbf {\bibinfo {volume}
  {424}},\ \bibinfo {pages} {443 } (\bibinfo {year} {1994})}\BibitemShut
  {NoStop}%
\bibitem [{\citenamefont {Vidal}\ \emph {et~al.}(2003)\citenamefont {Vidal},
  \citenamefont {Latorre}, \citenamefont {Rico},\ and\ \citenamefont
  {Kitaev}}]{vidal2003entanglement}%
  \BibitemOpen
  \bibfield  {author} {\bibinfo {author} {\bibfnamefont {G.}~\bibnamefont
  {Vidal}}, \bibinfo {author} {\bibfnamefont {J.~I.}\ \bibnamefont {Latorre}},
  \bibinfo {author} {\bibfnamefont {E.}~\bibnamefont {Rico}}, \ and\ \bibinfo
  {author} {\bibfnamefont {A.}~\bibnamefont {Kitaev}},\ }\href {\doibase
  10.1103/PhysRevLett.90.227902} {\bibfield  {journal} {\bibinfo  {journal}
  {Phys. Rev. Lett.}\ }\textbf {\bibinfo {volume} {90}},\ \bibinfo {pages}
  {227902} (\bibinfo {year} {2003})}\BibitemShut {NoStop}%
\bibitem [{\citenamefont {Jin}\ and\ \citenamefont
  {Korepin}(2004)}]{jin2004quantum}%
  \BibitemOpen
  \bibfield  {author} {\bibinfo {author} {\bibfnamefont {B.-Q.}\ \bibnamefont
  {Jin}}\ and\ \bibinfo {author} {\bibfnamefont {V.~E.}\ \bibnamefont
  {Korepin}},\ }\href {\doibase 10.1023/B:JOSS.0000037230.37166.42} {\bibfield
  {journal} {\bibinfo  {journal} {Journal of Statistical Physics}\ }\textbf
  {\bibinfo {volume} {116}},\ \bibinfo {pages} {79} (\bibinfo {year}
  {2004})}\BibitemShut {NoStop}%
\bibitem [{\citenamefont {Calabrese}\ and\ \citenamefont
  {Cardy}(2004)}]{calabrese2004entanglement}%
  \BibitemOpen
  \bibfield  {author} {\bibinfo {author} {\bibfnamefont {P.}~\bibnamefont
  {Calabrese}}\ and\ \bibinfo {author} {\bibfnamefont {J.}~\bibnamefont
  {Cardy}},\ }\href {\doibase 10.1088/1742-5468/2004/06/p06002} {\bibfield
  {journal} {\bibinfo  {journal} {Journal of Statistical Mechanics: Theory and
  Experiment}\ }\textbf {\bibinfo {volume} {2004}},\ \bibinfo {pages} {P06002}
  (\bibinfo {year} {2004})}\BibitemShut {NoStop}%
\bibitem [{\citenamefont {Calabrese}\ and\ \citenamefont
  {Cardy}(2009)}]{calabrese2009entanglement}%
  \BibitemOpen
  \bibfield  {author} {\bibinfo {author} {\bibfnamefont {P.}~\bibnamefont
  {Calabrese}}\ and\ \bibinfo {author} {\bibfnamefont {J.}~\bibnamefont
  {Cardy}},\ }\href {\doibase 10.1088/1751-8113/42/50/504005} {\bibfield
  {journal} {\bibinfo  {journal} {Journal of Physics A: Mathematical and
  Theoretical}\ }\textbf {\bibinfo {volume} {42}},\ \bibinfo {pages} {504005}
  (\bibinfo {year} {2009})}\BibitemShut {NoStop}%
\bibitem [{\citenamefont {Lu}\ and\ \citenamefont
  {Grover}(2019)}]{lu2017renyi}%
  \BibitemOpen
  \bibfield  {author} {\bibinfo {author} {\bibfnamefont {T.-C.}\ \bibnamefont
  {Lu}}\ and\ \bibinfo {author} {\bibfnamefont {T.}~\bibnamefont {Grover}},\
  }\href {\doibase 10.1103/PhysRevE.99.032111} {\bibfield  {journal} {\bibinfo
  {journal} {Phys. Rev. E}\ }\textbf {\bibinfo {volume} {99}},\ \bibinfo
  {pages} {032111} (\bibinfo {year} {2019})}\BibitemShut {NoStop}%
\bibitem [{\citenamefont {Korepin}(2004)}]{korepin2004universality}%
  \BibitemOpen
  \bibfield  {author} {\bibinfo {author} {\bibfnamefont {V.~E.}\ \bibnamefont
  {Korepin}},\ }\href {\doibase 10.1103/PhysRevLett.92.096402} {\bibfield
  {journal} {\bibinfo  {journal} {Phys. Rev. Lett.}\ }\textbf {\bibinfo
  {volume} {92}},\ \bibinfo {pages} {096402} (\bibinfo {year}
  {2004})}\BibitemShut {NoStop}%
\bibitem [{\citenamefont {Unmuth-Yockey}\ \emph {et~al.}(2017)\citenamefont
  {Unmuth-Yockey}, \citenamefont {Zhang}, \citenamefont {Preiss}, \citenamefont
  {Yang}, \citenamefont {Tsai},\ and\ \citenamefont
  {Meurice}}]{unmuth2017probing}%
  \BibitemOpen
  \bibfield  {author} {\bibinfo {author} {\bibfnamefont {J.}~\bibnamefont
  {Unmuth-Yockey}}, \bibinfo {author} {\bibfnamefont {J.}~\bibnamefont
  {Zhang}}, \bibinfo {author} {\bibfnamefont {P.~M.}\ \bibnamefont {Preiss}},
  \bibinfo {author} {\bibfnamefont {L.-P.}\ \bibnamefont {Yang}}, \bibinfo
  {author} {\bibfnamefont {S.-W.}\ \bibnamefont {Tsai}}, \ and\ \bibinfo
  {author} {\bibfnamefont {Y.}~\bibnamefont {Meurice}},\ }\href {\doibase
  10.1103/PhysRevA.96.023603} {\bibfield  {journal} {\bibinfo  {journal} {Phys.
  Rev. A}\ }\textbf {\bibinfo {volume} {96}},\ \bibinfo {pages} {023603}
  (\bibinfo {year} {2017})}\BibitemShut {NoStop}%
\bibitem [{\citenamefont {Zhang}\ \emph {et~al.}(2011)\citenamefont {Zhang},
  \citenamefont {Shen},\ and\ \citenamefont {Liu}}]{zhang2011quantum}%
  \BibitemOpen
  \bibfield  {author} {\bibinfo {author} {\bibfnamefont {J.~M.}\ \bibnamefont
  {Zhang}}, \bibinfo {author} {\bibfnamefont {C.}~\bibnamefont {Shen}}, \ and\
  \bibinfo {author} {\bibfnamefont {W.~M.}\ \bibnamefont {Liu}},\ }\href
  {\doibase 10.1103/PhysRevA.83.063622} {\bibfield  {journal} {\bibinfo
  {journal} {Phys. Rev. A}\ }\textbf {\bibinfo {volume} {83}},\ \bibinfo
  {pages} {063622} (\bibinfo {year} {2011})}\BibitemShut {NoStop}%
\bibitem [{\citenamefont {Ba\~nuls}\ \emph {et~al.}(2011)\citenamefont
  {Ba\~nuls}, \citenamefont {Cirac},\ and\ \citenamefont
  {Hastings}}]{banuls2011strong}%
  \BibitemOpen
  \bibfield  {author} {\bibinfo {author} {\bibfnamefont {M.~C.}\ \bibnamefont
  {Ba\~nuls}}, \bibinfo {author} {\bibfnamefont {J.~I.}\ \bibnamefont {Cirac}},
  \ and\ \bibinfo {author} {\bibfnamefont {M.~B.}\ \bibnamefont {Hastings}},\
  }\href {\doibase 10.1103/PhysRevLett.106.050405} {\bibfield  {journal}
  {\bibinfo  {journal} {Phys. Rev. Lett.}\ }\textbf {\bibinfo {volume} {106}},\
  \bibinfo {pages} {050405} (\bibinfo {year} {2011})}\BibitemShut {NoStop}%
\bibitem [{\citenamefont {Bonnes}\ \emph {et~al.}(2013)\citenamefont {Bonnes},
  \citenamefont {Pichler},\ and\ \citenamefont
  {L\"auchli}}]{bonnes2013entropy}%
  \BibitemOpen
  \bibfield  {author} {\bibinfo {author} {\bibfnamefont {L.}~\bibnamefont
  {Bonnes}}, \bibinfo {author} {\bibfnamefont {H.}~\bibnamefont {Pichler}}, \
  and\ \bibinfo {author} {\bibfnamefont {A.~M.}\ \bibnamefont {L\"auchli}},\
  }\href {\doibase 10.1103/PhysRevB.88.155103} {\bibfield  {journal} {\bibinfo
  {journal} {Phys. Rev. B}\ }\textbf {\bibinfo {volume} {88}},\ \bibinfo
  {pages} {155103} (\bibinfo {year} {2013})}\BibitemShut {NoStop}%
\bibitem [{\citenamefont {Daley}\ \emph {et~al.}(2012)\citenamefont {Daley},
  \citenamefont {Pichler}, \citenamefont {Schachenmayer},\ and\ \citenamefont
  {Zoller}}]{daley2012measuring}%
  \BibitemOpen
  \bibfield  {author} {\bibinfo {author} {\bibfnamefont {A.~J.}\ \bibnamefont
  {Daley}}, \bibinfo {author} {\bibfnamefont {H.}~\bibnamefont {Pichler}},
  \bibinfo {author} {\bibfnamefont {J.}~\bibnamefont {Schachenmayer}}, \ and\
  \bibinfo {author} {\bibfnamefont {P.}~\bibnamefont {Zoller}},\ }\href
  {\doibase 10.1103/PhysRevLett.109.020505} {\bibfield  {journal} {\bibinfo
  {journal} {Phys. Rev. Lett.}\ }\textbf {\bibinfo {volume} {109}},\ \bibinfo
  {pages} {020505} (\bibinfo {year} {2012})}\BibitemShut {NoStop}%
\bibitem [{\citenamefont {Islam}\ \emph {et~al.}(2015)\citenamefont {Islam},
  \citenamefont {Ma}, \citenamefont {Preiss}, \citenamefont {{Eric Tai}},
  \citenamefont {Lukin}, \citenamefont {Rispoli},\ and\ \citenamefont
  {Greiner}}]{islam2015measuring}%
  \BibitemOpen
  \bibfield  {author} {\bibinfo {author} {\bibfnamefont {R.}~\bibnamefont
  {Islam}}, \bibinfo {author} {\bibfnamefont {R.}~\bibnamefont {Ma}}, \bibinfo
  {author} {\bibfnamefont {P.~M.}\ \bibnamefont {Preiss}}, \bibinfo {author}
  {\bibfnamefont {M.}~\bibnamefont {{Eric Tai}}}, \bibinfo {author}
  {\bibfnamefont {A.}~\bibnamefont {Lukin}}, \bibinfo {author} {\bibfnamefont
  {M.}~\bibnamefont {Rispoli}}, \ and\ \bibinfo {author} {\bibfnamefont
  {M.}~\bibnamefont {Greiner}},\ }\href {https://doi.org/10.1038/nature15750
  http://10.0.4.14/nature15750
  https://www.nature.com/articles/nature15750{\#}supplementary-information}
  {\bibfield  {journal} {\bibinfo  {journal} {Nature}\ }\textbf {\bibinfo
  {volume} {528}},\ \bibinfo {pages} {77} (\bibinfo {year} {2015})}\BibitemShut
  {NoStop}%
\bibitem [{\citenamefont {Linden}\ \emph {et~al.}(2009)\citenamefont {Linden},
  \citenamefont {Popescu}, \citenamefont {Short},\ and\ \citenamefont
  {Winter}}]{linden2009quantum}%
  \BibitemOpen
  \bibfield  {author} {\bibinfo {author} {\bibfnamefont {N.}~\bibnamefont
  {Linden}}, \bibinfo {author} {\bibfnamefont {S.}~\bibnamefont {Popescu}},
  \bibinfo {author} {\bibfnamefont {A.~J.}\ \bibnamefont {Short}}, \ and\
  \bibinfo {author} {\bibfnamefont {A.}~\bibnamefont {Winter}},\ }\href
  {\doibase 10.1103/PhysRevE.79.061103} {\bibfield  {journal} {\bibinfo
  {journal} {Phys. Rev. E}\ }\textbf {\bibinfo {volume} {79}},\ \bibinfo
  {pages} {061103} (\bibinfo {year} {2009})}\BibitemShut {NoStop}%
\bibitem [{\citenamefont {Sandvik}(2010)}]{sandvik2010computational}%
  \BibitemOpen
  \bibfield  {author} {\bibinfo {author} {\bibfnamefont {A.~W.}\ \bibnamefont
  {Sandvik}},\ }\href {\doibase 10.1063/1.3518900} {\bibfield  {journal}
  {\bibinfo  {journal} {AIP Conference Proceedings}\ }\textbf {\bibinfo
  {volume} {1297}},\ \bibinfo {pages} {135} (\bibinfo {year} {2010})},\ \Eprint
  {http://arxiv.org/abs/https://aip.scitation.org/doi/pdf/10.1063/1.3518900}
  {https://aip.scitation.org/doi/pdf/10.1063/1.3518900} \BibitemShut {NoStop}%
\bibitem [{\citenamefont {Verstraete}\ \emph {et~al.}(2004)\citenamefont
  {Verstraete}, \citenamefont {Garc\'{\i}a-Ripoll},\ and\ \citenamefont
  {Cirac}}]{PhysRevLett.93.207204}%
  \BibitemOpen
  \bibfield  {author} {\bibinfo {author} {\bibfnamefont {F.}~\bibnamefont
  {Verstraete}}, \bibinfo {author} {\bibfnamefont {J.~J.}\ \bibnamefont
  {Garc\'{\i}a-Ripoll}}, \ and\ \bibinfo {author} {\bibfnamefont {J.~I.}\
  \bibnamefont {Cirac}},\ }\href {\doibase 10.1103/PhysRevLett.93.207204}
  {\bibfield  {journal} {\bibinfo  {journal} {Phys. Rev. Lett.}\ }\textbf
  {\bibinfo {volume} {93}},\ \bibinfo {pages} {207204} (\bibinfo {year}
  {2004})}\BibitemShut {NoStop}%
\bibitem [{\citenamefont {Feiguin}\ and\ \citenamefont
  {White}(2005)}]{PhysRevB.72.220401}%
  \BibitemOpen
  \bibfield  {author} {\bibinfo {author} {\bibfnamefont {A.~E.}\ \bibnamefont
  {Feiguin}}\ and\ \bibinfo {author} {\bibfnamefont {S.~R.}\ \bibnamefont
  {White}},\ }\href {\doibase 10.1103/PhysRevB.72.220401} {\bibfield  {journal}
  {\bibinfo  {journal} {Phys. Rev. B}\ }\textbf {\bibinfo {volume} {72}},\
  \bibinfo {pages} {220401(R)} (\bibinfo {year} {2005})}\BibitemShut {NoStop}%
\bibitem [{\citenamefont {Barthel}(2016)}]{PhysRevB.94.115157}%
  \BibitemOpen
  \bibfield  {author} {\bibinfo {author} {\bibfnamefont {T.}~\bibnamefont
  {Barthel}},\ }\href {\doibase 10.1103/PhysRevB.94.115157} {\bibfield
  {journal} {\bibinfo  {journal} {Phys. Rev. B}\ }\textbf {\bibinfo {volume}
  {94}},\ \bibinfo {pages} {115157} (\bibinfo {year} {2016})}\BibitemShut
  {NoStop}%
\bibitem [{\citenamefont {\"Ostlund}\ and\ \citenamefont
  {Rommer}(1995)}]{PhysRevLett.75.3537}%
  \BibitemOpen
  \bibfield  {author} {\bibinfo {author} {\bibfnamefont {S.}~\bibnamefont
  {\"Ostlund}}\ and\ \bibinfo {author} {\bibfnamefont {S.}~\bibnamefont
  {Rommer}},\ }\href {\doibase 10.1103/PhysRevLett.75.3537} {\bibfield
  {journal} {\bibinfo  {journal} {Phys. Rev. Lett.}\ }\textbf {\bibinfo
  {volume} {75}},\ \bibinfo {pages} {3537} (\bibinfo {year}
  {1995})}\BibitemShut {NoStop}%
\bibitem [{Note1()}]{Note1}%
  \BibitemOpen
  \bibinfo {note} {Version 3.0.0, http://itensor.org/}\BibitemShut {NoStop}%
\bibitem [{\citenamefont {Zaletel}\ \emph {et~al.}(2015)\citenamefont
  {Zaletel}, \citenamefont {Mong}, \citenamefont {Karrasch}, \citenamefont
  {Moore},\ and\ \citenamefont {Pollmann}}]{PhysRevB.91.165112}%
  \BibitemOpen
  \bibfield  {author} {\bibinfo {author} {\bibfnamefont {M.~P.}\ \bibnamefont
  {Zaletel}}, \bibinfo {author} {\bibfnamefont {R.~S.~K.}\ \bibnamefont
  {Mong}}, \bibinfo {author} {\bibfnamefont {C.}~\bibnamefont {Karrasch}},
  \bibinfo {author} {\bibfnamefont {J.~E.}\ \bibnamefont {Moore}}, \ and\
  \bibinfo {author} {\bibfnamefont {F.}~\bibnamefont {Pollmann}},\ }\href
  {\doibase 10.1103/PhysRevB.91.165112} {\bibfield  {journal} {\bibinfo
  {journal} {Phys. Rev. B}\ }\textbf {\bibinfo {volume} {91}},\ \bibinfo
  {pages} {165112} (\bibinfo {year} {2015})}\BibitemShut {NoStop}%
\bibitem [{\citenamefont {Braun}\ \emph {et~al.}(2013)\citenamefont {Braun},
  \citenamefont {Ronzheimer}, \citenamefont {Schreiber}, \citenamefont
  {Hodgman}, \citenamefont {Rom}, \citenamefont {Bloch},\ and\ \citenamefont
  {Schneider}}]{braun2013}%
  \BibitemOpen
  \bibfield  {author} {\bibinfo {author} {\bibfnamefont {S.}~\bibnamefont
  {Braun}}, \bibinfo {author} {\bibfnamefont {J.~P.}\ \bibnamefont
  {Ronzheimer}}, \bibinfo {author} {\bibfnamefont {M.}~\bibnamefont
  {Schreiber}}, \bibinfo {author} {\bibfnamefont {S.~S.}\ \bibnamefont
  {Hodgman}}, \bibinfo {author} {\bibfnamefont {T.}~\bibnamefont {Rom}},
  \bibinfo {author} {\bibfnamefont {I.}~\bibnamefont {Bloch}}, \ and\ \bibinfo
  {author} {\bibfnamefont {U.}~\bibnamefont {Schneider}},\ }\href {\doibase
  10.1126/science.1227831} {\bibfield  {journal} {\bibinfo  {journal}
  {Science}\ }\textbf {\bibinfo {volume} {339}},\ \bibinfo {pages} {52}
  (\bibinfo {year} {2013})},\ \Eprint
  {http://arxiv.org/abs/https://science.sciencemag.org/content/339/6115/52.full.pdf}
  {https://science.sciencemag.org/content/339/6115/52.full.pdf} \BibitemShut
  {NoStop}%
\bibitem [{\citenamefont {Medley}\ \emph {et~al.}(2011)\citenamefont {Medley},
  \citenamefont {Weld}, \citenamefont {Miyake}, \citenamefont {Pritchard},\
  and\ \citenamefont {Ketterle}}]{medley2011}%
  \BibitemOpen
  \bibfield  {author} {\bibinfo {author} {\bibfnamefont {P.}~\bibnamefont
  {Medley}}, \bibinfo {author} {\bibfnamefont {D.~M.}\ \bibnamefont {Weld}},
  \bibinfo {author} {\bibfnamefont {H.}~\bibnamefont {Miyake}}, \bibinfo
  {author} {\bibfnamefont {D.~E.}\ \bibnamefont {Pritchard}}, \ and\ \bibinfo
  {author} {\bibfnamefont {W.}~\bibnamefont {Ketterle}},\ }\href {\doibase
  10.1103/PhysRevLett.106.195301} {\bibfield  {journal} {\bibinfo  {journal}
  {Phys. Rev. Lett.}\ }\textbf {\bibinfo {volume} {106}},\ \bibinfo {pages}
  {195301} (\bibinfo {year} {2011})}\BibitemShut {NoStop}%
\bibitem [{\citenamefont {Mandt}\ \emph {et~al.}(2011)\citenamefont {Mandt},
  \citenamefont {Rapp},\ and\ \citenamefont {Rosch}}]{mandt2011}%
  \BibitemOpen
  \bibfield  {author} {\bibinfo {author} {\bibfnamefont {S.}~\bibnamefont
  {Mandt}}, \bibinfo {author} {\bibfnamefont {A.}~\bibnamefont {Rapp}}, \ and\
  \bibinfo {author} {\bibfnamefont {A.}~\bibnamefont {Rosch}},\ }\href
  {\doibase 10.1103/PhysRevLett.106.250602} {\bibfield  {journal} {\bibinfo
  {journal} {Physical Review Letters}\ }\textbf {\bibinfo {volume} {106}},\
  \bibinfo {pages} {250602} (\bibinfo {year} {2011})}\BibitemShut {NoStop}%
\bibitem [{\citenamefont {Kaufman}\ \emph {et~al.}(2016)\citenamefont
  {Kaufman}, \citenamefont {Tai}, \citenamefont {Lukin}, \citenamefont
  {Rispoli}, \citenamefont {Schittko}, \citenamefont {Preiss},\ and\
  \citenamefont {Greiner}}]{Kaufman794}%
  \BibitemOpen
  \bibfield  {author} {\bibinfo {author} {\bibfnamefont {A.~M.}\ \bibnamefont
  {Kaufman}}, \bibinfo {author} {\bibfnamefont {M.~E.}\ \bibnamefont {Tai}},
  \bibinfo {author} {\bibfnamefont {A.}~\bibnamefont {Lukin}}, \bibinfo
  {author} {\bibfnamefont {M.}~\bibnamefont {Rispoli}}, \bibinfo {author}
  {\bibfnamefont {R.}~\bibnamefont {Schittko}}, \bibinfo {author}
  {\bibfnamefont {P.~M.}\ \bibnamefont {Preiss}}, \ and\ \bibinfo {author}
  {\bibfnamefont {M.}~\bibnamefont {Greiner}},\ }\href {\doibase
  10.1126/science.aaf6725} {\bibfield  {journal} {\bibinfo  {journal}
  {Science}\ }\textbf {\bibinfo {volume} {353}},\ \bibinfo {pages} {794}
  (\bibinfo {year} {2016})},\ \Eprint
  {http://arxiv.org/abs/https://science.sciencemag.org/content/353/6301/794.full.pdf}
  {https://science.sciencemag.org/content/353/6301/794.full.pdf} \BibitemShut
  {NoStop}%
\bibitem [{\citenamefont {Henkel}\ \emph {et~al.}(2010)\citenamefont {Henkel},
  \citenamefont {Nath},\ and\ \citenamefont {Pohl}}]{Henkel2010}%
  \BibitemOpen
  \bibfield  {author} {\bibinfo {author} {\bibfnamefont {N.}~\bibnamefont
  {Henkel}}, \bibinfo {author} {\bibfnamefont {R.}~\bibnamefont {Nath}}, \ and\
  \bibinfo {author} {\bibfnamefont {T.}~\bibnamefont {Pohl}},\ }\href {\doibase
  10.1103/PhysRevLett.104.195302} {\bibfield  {journal} {\bibinfo  {journal}
  {Phys. Rev. Lett.}\ }\textbf {\bibinfo {volume} {104}},\ \bibinfo {pages}
  {195302} (\bibinfo {year} {2010})}\BibitemShut {NoStop}%
\bibitem [{\citenamefont {Pupillo}\ \emph {et~al.}(2010)\citenamefont
  {Pupillo}, \citenamefont {Micheli}, \citenamefont {Boninsegni}, \citenamefont
  {Lesanovsky},\ and\ \citenamefont {Zoller}}]{Pupillo2010}%
  \BibitemOpen
  \bibfield  {author} {\bibinfo {author} {\bibfnamefont {G.}~\bibnamefont
  {Pupillo}}, \bibinfo {author} {\bibfnamefont {A.}~\bibnamefont {Micheli}},
  \bibinfo {author} {\bibfnamefont {M.}~\bibnamefont {Boninsegni}}, \bibinfo
  {author} {\bibfnamefont {I.}~\bibnamefont {Lesanovsky}}, \ and\ \bibinfo
  {author} {\bibfnamefont {P.}~\bibnamefont {Zoller}},\ }\href {\doibase
  10.1103/PhysRevLett.104.223002} {\bibfield  {journal} {\bibinfo  {journal}
  {Phys. Rev. Lett.}\ }\textbf {\bibinfo {volume} {104}},\ \bibinfo {pages}
  {223002} (\bibinfo {year} {2010})}\BibitemShut {NoStop}%
\bibitem [{\citenamefont {Zeiher}\ \emph {et~al.}(2016)\citenamefont {Zeiher},
  \citenamefont {van Bijnen}, \citenamefont {Schau{\ss}}, \citenamefont {Hild},
  \citenamefont {Choi}, \citenamefont {Pohl}, \citenamefont {Bloch},\ and\
  \citenamefont {Gross}}]{Zeiher2016}%
  \BibitemOpen
  \bibfield  {author} {\bibinfo {author} {\bibfnamefont {J.}~\bibnamefont
  {Zeiher}}, \bibinfo {author} {\bibfnamefont {R.}~\bibnamefont {van Bijnen}},
  \bibinfo {author} {\bibfnamefont {P.}~\bibnamefont {Schau{\ss}}}, \bibinfo
  {author} {\bibfnamefont {S.}~\bibnamefont {Hild}}, \bibinfo {author}
  {\bibfnamefont {J.-y.}\ \bibnamefont {Choi}}, \bibinfo {author}
  {\bibfnamefont {T.}~\bibnamefont {Pohl}}, \bibinfo {author} {\bibfnamefont
  {I.}~\bibnamefont {Bloch}}, \ and\ \bibinfo {author} {\bibfnamefont
  {C.}~\bibnamefont {Gross}},\ }\href {https://doi.org/10.1038/nphys3835
  http://10.0.4.14/nphys3835
  https://www.nature.com/articles/nphys3835{\#}supplementary-information}
  {\bibfield  {journal} {\bibinfo  {journal} {Nature Physics}\ }\textbf
  {\bibinfo {volume} {12}},\ \bibinfo {pages} {1095} (\bibinfo {year}
  {2016})}\BibitemShut {NoStop}%
\bibitem [{\citenamefont {Zhang}\ \emph {et~al.}(2018)\citenamefont {Zhang},
  \citenamefont {Unmuth-Yockey}, \citenamefont {Zeiher}, \citenamefont
  {Bazavov}, \citenamefont {Tsai},\ and\ \citenamefont
  {Meurice}}]{zhang2018prl}%
  \BibitemOpen
  \bibfield  {author} {\bibinfo {author} {\bibfnamefont {J.}~\bibnamefont
  {Zhang}}, \bibinfo {author} {\bibfnamefont {J.}~\bibnamefont
  {Unmuth-Yockey}}, \bibinfo {author} {\bibfnamefont {J.}~\bibnamefont
  {Zeiher}}, \bibinfo {author} {\bibfnamefont {A.}~\bibnamefont {Bazavov}},
  \bibinfo {author} {\bibfnamefont {S.-W.}\ \bibnamefont {Tsai}}, \ and\
  \bibinfo {author} {\bibfnamefont {Y.}~\bibnamefont {Meurice}},\ }\href
  {\doibase 10.1103/PhysRevLett.121.223201} {\bibfield  {journal} {\bibinfo
  {journal} {Phys. Rev. Lett.}\ }\textbf {\bibinfo {volume} {121}},\ \bibinfo
  {pages} {223201} (\bibinfo {year} {2018})}\BibitemShut {NoStop}%
\bibitem [{\citenamefont {Vaidya}\ \emph {et~al.}(2018)\citenamefont {Vaidya},
  \citenamefont {Guo}, \citenamefont {Kroeze}, \citenamefont {Ballantine},
  \citenamefont {Koll\'ar}, \citenamefont {Keeling},\ and\ \citenamefont
  {Lev}}]{Vaidya:2018fp}%
  \BibitemOpen
  \bibfield  {author} {\bibinfo {author} {\bibfnamefont {V.~D.}\ \bibnamefont
  {Vaidya}}, \bibinfo {author} {\bibfnamefont {Y.}~\bibnamefont {Guo}},
  \bibinfo {author} {\bibfnamefont {R.~M.}\ \bibnamefont {Kroeze}}, \bibinfo
  {author} {\bibfnamefont {K.~E.}\ \bibnamefont {Ballantine}}, \bibinfo
  {author} {\bibfnamefont {A.~J.}\ \bibnamefont {Koll\'ar}}, \bibinfo {author}
  {\bibfnamefont {J.}~\bibnamefont {Keeling}}, \ and\ \bibinfo {author}
  {\bibfnamefont {B.~L.}\ \bibnamefont {Lev}},\ }\href {\doibase
  10.1103/PhysRevX.8.011002} {\bibfield  {journal} {\bibinfo  {journal} {Phys.
  Rev. X}\ }\textbf {\bibinfo {volume} {8}},\ \bibinfo {pages} {011002}
  (\bibinfo {year} {2018})}\BibitemShut {NoStop}%
\end{thebibliography}
\end{document}